\documentclass[letterpaper,twocolumn,10pt]{article}
\usepackage{usenix}

\usepackage{epsfig,endnotes,xspace,url,diagbox}

\usepackage{amsmath, amsthm, amsfonts}
\usepackage{bbm}
\usepackage{enumerate}
\usepackage{varioref}
\usepackage{graphicx}
\usepackage{subfigure}
\usepackage{hyperref}
\usepackage{pifont}

\usepackage{caption}
\usepackage{float}
\usepackage{cleveref}
\usepackage{thmtools}
\usepackage{thm-restate}

\usepackage{epstopdf}
\usepackage{xcolor}
\usepackage{colortbl} 

\usepackage{multirow}
\usepackage{comment}
\usepackage{color}
\usepackage[utf8]{inputenc}
\usepackage{mathtools}

\usepackage{wrapfig}
\usepackage{tabularx}
\usepackage{xspace}
\usepackage{algorithm}
\usepackage{algorithmic}
\usepackage{cite}
\usepackage{booktabs}
\usepackage{tcolorbox}
\usepackage{enumitem}
\usepackage{url}
\theoremstyle{plain}
\newtheorem{definition}{Definition}

\setlist[itemize]{leftmargin=*,noitemsep, topsep=0pt}

\newcommand{\argmax}{\mathop{\mathrm{argmax}}}

\newcommand{\ie}{\emph{i.e., }}
\newcommand{\eg}{\emph{e.g., }}

\newcommand{\mypara}[1]{\smallskip\noindent\textbf{#1.} \xspace}
\newcommand{\myquestion}[1]{\smallskip\noindent\textbf{#1?} \xspace}

\newcommand{\algcomment}[1]{\hfill {\color{blue} $\triangleright$ \emph{\small{#1}}}}

\newcommand{\myfullcomment}[1]{\STATE {\textcolor{gray}{\small\textit{\# #1}}}}

\newcommand{\mymethod}{\ensuremath{\mathsf{IMIA}}\xspace}

\newcommand{\mymethods}{\ensuremath{\mathsf{IMIA}}'s\xspace}

\newcommand{\mymethodgaussian}{\ensuremath{\mathsf{IMIA}}$_\text{Gaussian}$\xspace}

\newcommand{\mymethoddistill}{\ensuremath{\mathsf{IMIA}}$_\text{distill}$\xspace}

\newcommand{\mymethodshadow}{\ensuremath{\mathsf{IMIA}}$_\text{shadow}$\xspace}

\newcommand{\computationreduce}{5\%\xspace}

\begin{document}

\title{\Large \bf Imitative Membership Inference Attack}


\author{
{\rm Yuntao Du}\\
Purdue University
\and
{\rm Yuetian Chen}\\
Purdue University
\and
{\rm Hanshen Xiao}\\
Purdue University \& NVIDIA Research
\and
{\rm Bruno Ribeiro}\\
Purdue University
\and
{\rm Ninghui Li}\\
Purdue University
} 

\maketitle

A Membership Inference Attack (MIA) assesses how much a target machine learning model reveals about its training data by determining whether specific query instances were part of the training set.
State-of-the-art MIAs rely on training hundreds of shadow models that are independent of the target model, leading to significant computational overhead.
In this paper, we introduce Imitative Membership Inference Attack (\mymethod), which employs a novel imitative training technique to strategically construct a small number of target-informed imitative models that closely replicate the target model’s behavior for inference.
Extensive experimental results demonstrate that
\mymethod substantially outperforms existing MIAs in various attack settings while only requiring less than \computationreduce of the computational cost of state-of-the-art approaches.

\section{Introduction}
\label{sec:intro}

Over the past decade, machine learning (ML) has seen remarkable advances, with models such as neural networks increasingly being trained on sensitive datasets.
This trend has raised growing concerns about the privacy risks associated with these models.
Membership inference attacks (MIAs)~\cite{sp17miashokri} have been proposed to measure the degree to which a model leaks information by determining whether some instances were part of its training set. 
Closely related to Differential Privacy (DP)~\cite{dwork_dp,ccs13membership}, MIAs have become a widely adopted technique for empirical auditing of various trustworthy risks in ML models~\cite{arxiv20privacy_meter,tensorflow_mia,dsi,memorization,copyright}, and serve as key components for more sophisticated attacks~\cite{usenix21llm_extract,usenix23diffusion_extract,iclr25usage}.


MIAs in the black-box setting~\cite{csf18privacy, usenix21systematic, ccs22enhanced,sp22lira} typically exploit the model's behavioral discrepancy between its training instances (\ie members) and non-training instances (\ie non-members) for inference.
A widely-used strategy to explore the discrepancy is  \textit{shadow training}~\cite{sp17miashokri}, which involves training multiple shadow models on datasets drawn from the same distribution as the target model’s training set. 
For each query instance, the membership scores generated by the shadow models trained without the instance (shadow \textit{out} models) represent the \textit{out} distribution for the instance.  
In some settings, one also has shadow models that are trained with the instance as a member, and can use the scores generated by these shadow \textit{in} models to compute an \textit{in} distribution.  
Given a target model, one tries to tell which distribution the score from the target model fits better. 
MIAs~\cite{iclr23canary, ccs24rapid, ndss26cpmia, icml24rmia, sp22lira} utilizing shadow training have shown strong attack performance.


\begin{figure*}[t]
    \centering
    \includegraphics[width=0.99\linewidth]{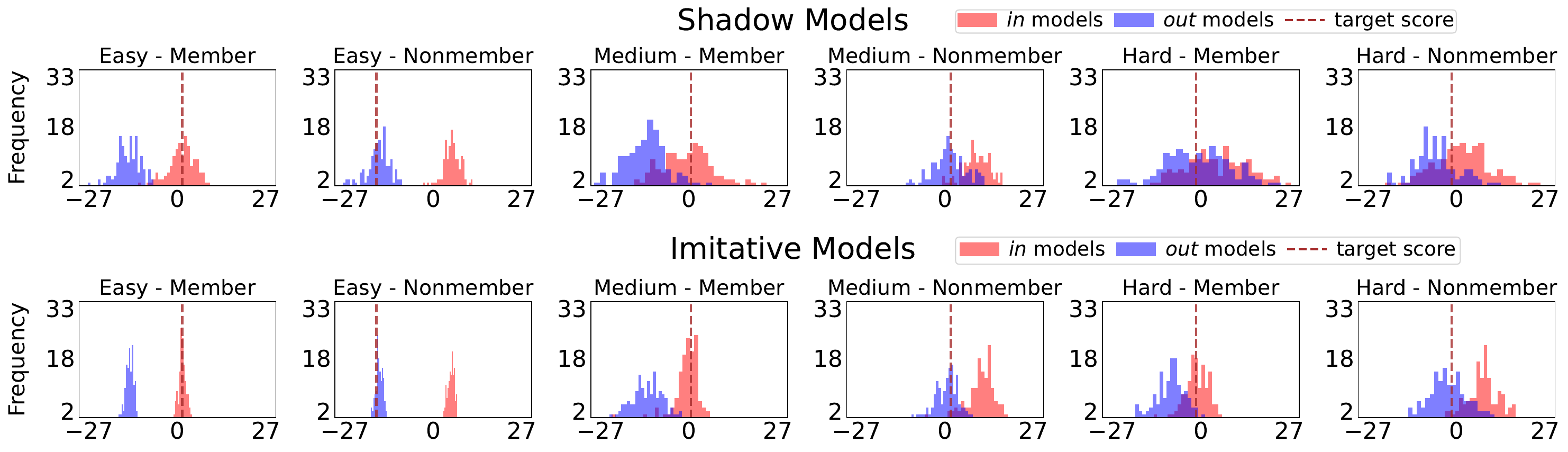}
    \caption{Distributions of membership scores (\ie scaled confidence scores, defined in~\Cref{sec:imia_framework}) for six CIFAR-100 instances with varying attack difficulty (easy, medium, hard). A larger overlap between \textit{in} (trained with the instance) and \textit{out} (trained without the instance) score distributions indicates greater difficulty in determining membership. The dashed vertical line represents each instance’s score on the target model. \textbf{Top row}: \textit{target-agnostic} shadow models show high predictive variance (with long tails and wide distributions) for both members and non-members, resulting in significant overlap that hampers reliable inference, especially for hard-to-attack instances.
    \textbf{Bottom row}: \textit{target-informed} imitative models exhibit more stable and well-separated distributions, enabling effective inference across all levels of difficulty. More examples are in~\Cref{fig:demo_model_stability_appendix}.
    Best viewed in color.}
    \label{fig:demo_model_stability}
\end{figure*}

A critical limitation of existing shadow-based MIAs lies in their \textbf{substantial computational overhead}.
Training each shadow model incurs non-trivial overhead, and state-of-the-art attacks like LiRA~\cite{sp22lira} and PMIA~\cite{ndss26cpmia} require training hundreds of shadow models to estimate the likelihood ratio for inference. 
This requirement imposes a substantial computational burden, which reduces the feasibility of using these state-of-the-art MIAs for practical privacy auditing~\cite{icml24rmia}.  
In addition, it also impedes research reproducibility within the privacy community (see~\Cref{sec:definition} for further discussion).

We observe that this inefficiency stems from the \textit{target-agnostic} design of shadow training.
That is, the current shadow training process fails to take advantage of knowing the target model under attack.  
As a result, the shadow models learn only the general patterns of members and non-members, rather than the specific behavioral discrepancies of the target. 
As shown in the top row of~\Cref{fig:demo_model_stability}, this target-agnostic approach causes shadow models to exhibit high predictive variance across instances with varying levels of attack vulnerability. 
Consequently, existing MIAs need to train a multitude of shadow models to capture this variability, resulting in substantial computational overhead and suboptimal performance.

To address this challenge, we introduce Imitative Membership Inference Attack (\mymethod), a novel approach that improves both attack efficiency and effectiveness.
At the core of \mymethod is \textit{imitative training}, a new shadow training technique that trains \textit{target-informed} imitative models to mimic the target model’s behavior. 
Specifically, we first train a set of imitative \textit{out} models by applying weighted logits (log of output confidence) matching to the target model's outputs, capturing the behavior of the target on non-member instances. 
We then leverage these models to continue training on specially selected ``pivot'' instances from the adversary’s dataset using standard cross-entropy loss, yielding a set of imitative \textit{in} models that capture the target model’s predictions on member instances.
This targeted approach enables us to pair each query instance with a set of \textit{in} and \textit{out} predictions estimated from the target model.
As illustrated in the bottom row of~\Cref{fig:demo_model_stability}, imitative models yield significantly more separable and stable estimations, enabling the adversary to better exploit behavioral discrepancies for more powerful inference.

We note that some attacks~\cite{ccs22enhanced,tifs25glira,ccs24seqmia} employ model distillation to construct shadow models from the target model. 
However, these methods directly apply the standard distillation technique~\cite{arxiv15distill} without tailoring it to capture the membership-related behaviors of the target model.
Furthermore, the resulting distillation models can only reflect behaviors on non-member instances, making it infeasible to exploit behavioral discrepancies for effective membership inference.
A more detailed discussion of the differences between our approach and distillation-based methods is provided in~\Cref{sec:discussion_imia}.

We compare the proposed attack with a broad range of MIAs on six benchmark datasets (4 image and 2 non-image datasets), under both non-adaptive and adaptive settings~\cite{ndss26cpmia} (also referred to as ``offline'' and ``online''~\cite{sp22lira}; see~\Cref{sec:definition} for details).
Experimental results show that \mymethod consistently outperforms all evaluated attacks (including distillation-based attacks) while using significantly fewer shadow models, with particularly strong gains in the increasingly advocated non-adaptive setting. 
For example, \mymethod surpasses the state-of-the-art non-adaptive attack (\ie PMIA~\cite{ndss26cpmia}) by over $\mathbf{9\times}$ in true positive rate at a zero false positive rate on Fashion-MNIST, while using only  \textbf{\computationreduce} of its computational resources. 
Similarly, \mymethod outperforms all adaptive attacks with far less computation. 
We also conduct comprehensive analyses to show the effectiveness of imitative training over shadow training, evaluate the impact of components in \mymethod, and assess the effectiveness of existing defenses against the proposed attack.
In summary, we make the following contributions:

\begin{itemize}
    \item We propose \mymethod, a new MIA that employs a novel target-informed imitative training technique to effectively capture the target model’s behavioral discrepancies for inference.
    \item We conduct extensive experiments across diverse datasets, models, and attack settings, showing that \mymethod consistently outperforms all MIAs with substantially less computation.
    \item We provide an in-depth analysis of \mymethods effectiveness, showing that its strength lies in imitative training and highlighting its advantages over shadow training.
\end{itemize}


\mypara{Organization}
The rest of this paper is organized as follows. 
~\Cref{sec:definition} introduces the MIA definitions and threat models. 
\Cref{sec:imia} describes the proposed attack, \mymethod, in detail.
\Cref{sec:experiments} presents the experimental evaluation. 
Related work is discussed in~\Cref{sec:related}, and the paper concludes in~\Cref{sec:conclusion}. 
\section{Problem Definition and Threat Models}
\label{sec:definition}

The goal of a membership inference attack (MIA)~\cite{sp17miashokri} is to determine whether some instances were included in the training data of a given model $f_\theta$.
In this paper, we consider the model to be a neural network classifier $f_\theta: \mathcal{X} \to \Delta^c$, where $f_\theta$ is a learned function that maps an input data sample $x \in \mathcal{X}$ to a probability distribution over $c$ classes, where $\Delta^c$ denotes the $c$-dimensional simplex.
Given a dataset $D$, we denote by $f_\theta \gets \mathcal{T}(D)$ the process of training a model $f_\theta$, parameterized by weights $\theta$, using a training algorithm $\mathcal{T}$ on dataset $D$.


In early literature~\cite{csf18privacy,arixv20revisiting}, membership inference is defined via a security game in which the adversary is asked to determine the membership of a single instance.
Recent work~\cite{ndss26cpmia} refined this definition to more precisely characterize the adversary’s capabilities, particularly whether the adversary is allowed to train shadow models using query instances.
Inspired by~\cite{ndss26cpmia}, we instantiate the membership inference security game to make it match the typical experimental setting of using half members and half non-members for evaluation, and define the following canonical security game:

\begin{definition}[Canonical Membership Inference Security Game]
\label{def:game} 
The following game is between a challenger and an adversary that both have access to a data distribution $\mathbb{D}$: 
\begin{enumerate}[leftmargin=*]
    \item The challenger samples a training dataset $D \sim \mathbb{D}$, trains a target model $f_\theta \gets \mathcal{T}(D)$ on the dataset $D$, and grants the adversary query access to the model $f_\theta$.
    \item The challenger randomly selects a member set $D_\mathrm{a} \subseteq D$ and samples a nonmember set $D_\mathrm{b}$ from $\mathbb{D}$, such that $D_\mathrm{b} \cap D =\emptyset$ and $|D_\mathrm{a}| =|D_\mathrm{b}|$. These two sets are combined to create a query set: $D_\mathrm{query} = D_\mathrm{a} \cup   D_\mathrm{b}$, which the challenger then sends to the adversary.
    \item The adversary responds with a set $D_\mathrm{g} \subseteq D_\mathrm{query}$, which represents that the adversary guesses that instances in $D_\mathrm{g}$ are used when training $f_\theta$, and instances in $D_\mathrm{query}  \setminus D_\mathrm{g}$ are not used when training $f_\theta$.
\end{enumerate}
\end{definition}

In this paper, we focus on membership inference attacks in black-box scenarios, where the adversary is granted oracle access to the target model $f_\theta$, but is not given its parameters.
Specifically, the adversary can query the model to obtain softmax output probabilities (using the default softmax temperature of 1) for any input instance.
State-of-the-art MIAs~\cite{sp22lira,iclr23canary,ccs24rapid,icml24rmia,ndss26cpmia} leverage shadow models~\cite{sp17miashokri} to analyze how the model's outputs change in response to discrepancies between its members versus non-members.
We assume that the adversary can construct a dataset by sampling from the data distribution $\mathbb{D}$, and use subsets of this dataset to train shadow models.
Depending on when the shadow models are trained, we consider the following two threat models.

\mypara{Non-Adaptive Setting}
In this setting, the adversary is only allowed to train shadow models \textit{before} the adversary learns the query set $D_\text{query}$ (\ie before step 2 in~\Cref{def:game}).  
That is, the adversary constructs an attack dataset $D_\text{adv}^\text{non-adapt}$ (from which the adversary will sample subsets to train shadow models) by sampling from $\mathbb{D}$, \ie $D_\text{adv}^\text{non-adapt} \sim \mathbb{D}$.
For most practical classification tasks, the sizes of $D_\text{adv}^\text{non-adapt}$ and $D_\text{query}$ are relatively small compared to the entire data distribution $\mathbb{D}$, meaning the probability of each instance belonging to both datasets is quite low.
As a result, for most query instances, the adversary can only observe the model’s behavior when the query point is not part of the target model’s training set, leading to shadow \textit{out} models available for inference.

This non-adaptive (offline) setting has attracted growing interest from recent studies~\cite{nips23quantile, ccs24seqmia, icml24rmia, ccs24rapid} because it models a realistic attack scenario: the adversary receives a sequence of membership queries and must respond without incurring the cost of retraining shadow models for each query.
\textbf{In this paper, we also follow these studies and mainly focus on the non-adaptive setting. }


\mypara{Adaptive Setting}
In this setting, the adversary is allowed to train shadow models \textit{after} receiving the query set $D_\text{query}$ (\ie after step 2 in~\Cref{def:game}). 
The adversary can therefore leverage $D_\text{query}$ for shadow training.
For each query instance, the adversary can train both shadow \textit{in} models (trained with the instance) and shadow \textit{out} models (trained without it), and exploit their behavioral differences for membership inference.

Recent studies~\cite{icml24rmia,nips23quantile,usenix25free} criticize this setting as unrealistic in practice, as it requires training new shadow models for each batch of query instances.
Nevertheless, we demonstrate in~\Cref{sec:imia_framework} that our proposed method can be adapted to the adaptive setting, achieving state-of-the-art performance while significantly reducing computational overhead.

\mypara{High Computational Cost in Both Settings}
State-of-the-art MIAs in both settings require training a large number of shadow models to achieve their best performance. 
For instance, LiRA~\cite{sp22lira} in the adaptive setting and PMIA~\cite{ndss26cpmia} in the non-adaptive setting both suggest training 256 shadow models.
The practical implication of this is severe: a single LiRA run on CIFAR-10 can take around 6 days on a modern A100 GPU using its official implementation.
Such high computational demands make it infeasible for practical privacy auditing~\cite{icml24rmia} and difficult to reproduce research results.
It has become common practice for subsequent studies~\cite{ccs24seqmia,iclr23canary,ccs24rapid} to use a computationally cheaper version of LiRA (\eg with 64 or 128 shadow models) as their strongest baseline for evaluation, acknowledging the original version is too costly to run.
However, since LiRA's performance degrades significantly with fewer shadow models in the adaptive setting (as shown in~\cite{icml24rmia} and evidenced in~\Cref{sec:exp_main}), this compromise leads to skewed evaluations against a weakened baseline.

\mypara{Assumptions}
In this paper, we adopt the standard (albeit sometimes implicit) assumptions used in the MIA literature~\cite{ndss26cpmia, sp22lira, iclr23canary, icml24rmia}.
Specifically, in the adaptive setting, the adversary is given the membership query set $D_\text{query}$, which is constructed using all training instances in $D$ as members and an equal number of non-member instances, \ie $D_\mathrm{a} = D$ and $|D_\mathrm{b}| = |D|$ (the equal-size assumption).
In the non-adaptive setting, we assume that the adversary can sample a dataset $D_\text{adv}^{\text{non-adapt}}$ from the same distribution as the training data, and that $D_\text{adv}^{\text{non-adapt}} \cap D_\text{query} =\emptyset$ (the disjoint assumption). 
We emphasize that the equal-size assumption and the disjoint assumption are used only to align with prior experimental setups; the effectiveness of our attack does not depend on them.  Discussion of attacks without these assumptions is provided in Appendix~\ref{appendix:non_assumpmtion}.


\section{Imitative Membership Inference Attack}
\label{sec:imia}

\mypara{Motivation of Imitative MIA}
Our work is motivated by a core limitation in most MIAs: the high predictive variance of membership signals from their target-agnostic shadow models.
These shadow models, trained by randomly sampling from the adversary’s dataset, only capture the general patterns of members and non-members across models, failing to account for the specific target model being attacked.
To address this, we introduce \textit{imitative training}, which leverages the target model's knowledge to train imitative models that explicitly mimic the target's behavior.
This design yields more informative membership signals and, ultimately, more effective attacks. 
The next subsection describes our approach in detail, and \Cref{sec:exp_imiative} demonstrate its effectiveness and efficiency.

\subsection{Attack Method}
\label{sec:imia_framework}


We begin by introducing the membership signal and the loss function used for imitative training. 
Next, we outline the imitative training process. 
Finally, we present the workflow of \mymethod, and detail how it can be applied to the adaptive setting.


\mypara{Scaled Confidence Score}
We follow~\cite{sp22lira} and define the membership signal of query instance $(x,y)$ on model $f$ as:
\begin{equation}
\label{equ:margin}
    \phi\left(f(x)_y\right)=\log \left(f(x)_y\right)-\log (\max _{y^{\prime} \neq y} f(x)_{y^{\prime}}),
\end{equation}
where $f(x)_y$ denotes the model's output probability for the true class $y$, and $\max_{y^\prime \neq y }f(x)_{y^\prime}$ is the highest probability among all incorrect classes.
This signal 
captures the model’s prediction confidence and is widely used in prior MIAs~\cite{sp22lira,iclr23canary,ndss26cpmia}, as it is a more effective indicator of membership than alternatives such as output loss.




\begin{algorithm}[t]
\caption{\textbf{Imitative Training.} We first train an imitative \textit{out} model by matching its output with the target model over $T_1$ epochs. We then continue training for $T_2$ epochs on pivot data to obtain an imitative \textit{in} model.}
\label{alg:imitative_training}
\begin{algorithmic}[1]
\REQUIRE Target model $f_\theta$, imitation dataset $D_\text{imitate}$, pivot dataset $D_\text{pivot}$, epochs $T_\text{warmup},T_1, T_2$,  learning rate $\eta$
\STATE Initialize a model $f_\psi$ with randomized parameters $\Psi$

\myfullcomment{train imitative out model}
\FOR{epoch $= 1, \dots, T_1$}
    \FOR{each batch $B$ in $D_\text{imitate}$}
        \IF{$\text{epoch} \le T_\text{warmup}$}
            \myfullcomment{warm-up with cross-entropy loss}
            \STATE $L \gets \frac{1}{|B|} \sum_{(x,y)\in B} \mathcal{L}_\text{ce}(x,y,f_\psi)$
        \ELSE
            \STATE $L \gets \frac{1}{|B|} \sum_{(x,y)\in B} \mathcal{L}_\text{imitate}(x,y,f_\psi,f_\theta)$ \algcomment{Eq. (\ref{equ:imitative_loss})}
        \ENDIF
        \STATE $\psi \gets \psi - \eta \nabla_{\psi} L$
    \ENDFOR
    \IF{$\text{epoch} = T_\text{warmup}$}
        \STATE $f \gets f_\psi$ \algcomment{warmup checkpoint}
    \ENDIF
\ENDFOR
\STATE $f_\text{out} \gets f_\psi$ \algcomment{save as imitative out model}

\myfullcomment{continue training to obtain imitative in model for pivots}
\STATE Initialize $f_\psi \gets f$
\FOR{epoch $= 1, \dots, T_2$}
    \FOR{each batch $B$ in $D_\text{pivot}$}
        \STATE $L_\text{ce} \gets \frac{1}{|B|} \sum_{(x,y)\in B} \mathcal{L}_\text{ce}(x,y,f_\psi)$ \algcomment{cross-entropy loss}
        \STATE $\psi \gets \psi - \eta \nabla_{\psi} L_\text{ce}$ 
    \ENDFOR
\ENDFOR
\STATE $f_\text{in}^\text{pivot} \gets f_\psi$ \algcomment{save as imitative in model}
\RETURN $f_\text{out},\ f_\text{in}^\text{pivot}$
\end{algorithmic}
\end{algorithm}

\mypara{Imitation Loss Function}
Instead of training shadow models that are independent of the target model, we train imitative models to explicitly imitate the behavior of the target model by aligning their outputs.
To achieve this, we introduce the \textit{imitation loss function} that minimizes the discrepancy between the output probabilities of the imitative model and the target model.
Formally, for a given instance $(x,y)$, let $f_\psi(x)_i$ and $f_\theta(x)_i$ be the softmax outputs of the imitative model $f_\psi$ and the target model $f_\theta$ on the $i$-th class, respectively. 
The imitation loss is defined as follows:
\begin{equation}
\label{equ:imitative_loss}
    \mathcal{L}_\text{imitate}(x,y, f_\psi, f_\theta) = \sum_{i=1}^c w_i(c) \left(\log(f_\psi(x)_i) - \log( f_\theta(x)_i)\right)^2,
\end{equation}
where $c$ is the total number of classes, and 
$w_i(c)$ is a class-specific weight that determines the importance of each class:
\begin{equation*}
\label{equ:weight_logits}
    w_i(c) = 
        \begin{cases}
            \frac{1+\sqrt{c}}{c+2\sqrt{c}}, & \text{if} \ i = y \ \text{or} \ i = \argmax_{y^\prime \neq y}f_\theta(x)_{y^\prime},\\
            
             \frac{1}{c+2 \sqrt{c}} &  \text{else}.
        \end{cases}
\end{equation*}
Here, the weighting scheme prioritizes the probability corresponding to the ground-truth class, $y$, and the most likely incorrect class, as predicted by the target model.
This design encourages imitative models to focus on the most indicative signals, which aligns with our attack signal in~\Cref{equ:margin}.

Our approach can be interpreted as a weighted version of \textit{logits matching} in model distillation~\cite{ijcai21distill}, with a weighting strategy specifically tailored to membership inference.
In~\Cref{sec:exp_ablation}, we compare it to the commonly used Kullback-Leibler (KL) divergence~\cite{arxiv15distill} and also explore other formulations of the imitation loss to demonstrate its effectiveness. 
A more detailed discussion of the connections to model distillation is provided in the next subsection.


\mypara{Imitative Training}
The imitative training procedure is outlined in~\Cref{alg:imitative_training}.
It requires query access to the target model and utilizes two datasets selected from the adversary's data to mimic the target model's behavior: an imitation dataset $D_\text{imitate}$ and a pivot dataset $D_\text{pivot}$ (to be detailed later).
The training proceeds in two stages.
In the first stage, we warm up the model using the standard cross-entropy loss for $T_\text{warmup}$ epochs, after which we switch to the proposed imitative loss and train for $T_1$ epochs to align the model’s outputs with those of the target model on $D_\text{imitate}$ (lines 3–17).
Since query instances are not in the adversary’s dataset in the non-adaptive setting, the resulting model serves as an imitative \textit{out} model, imitating how the target model behaves on non-member inputs.
In the second stage, we resume training from the warmup checkpoint for $T_2$ epochs on a pre-selected pivot dataset using the standard cross-entropy loss (lines 19–26).
This yields an imitative \textit{in} model for the pivot data, whose behaviors on these instances will serve as proxies to approximate how the target model behaves when a query is in the training set. 

\begin{figure}[t]
    \centering
    \includegraphics[width=0.47\textwidth]{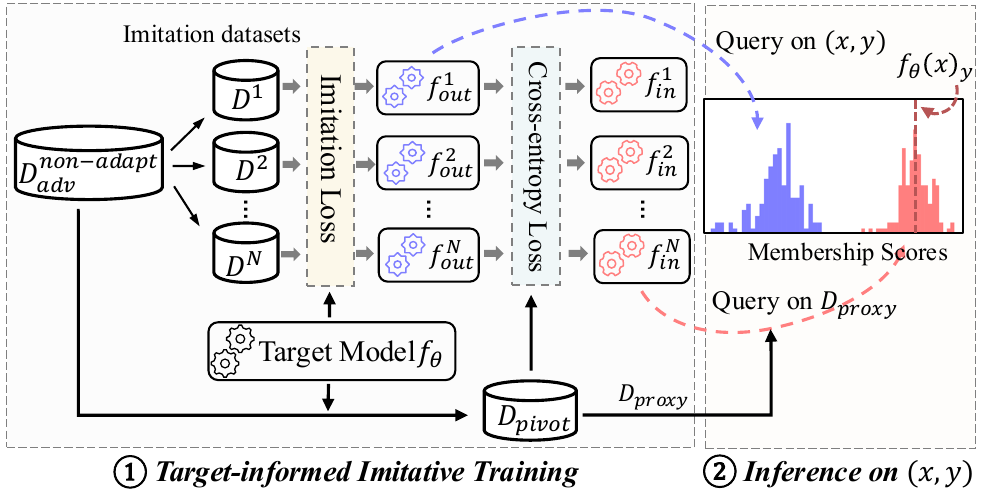}
    \caption{Demonstration of \mymethod in the non-adaptive setting. The adversary \ding{172} constructs imitative \textit{in} and \textit{out} models by mimicking the behaviors of the target model $f_\theta$, \ding{173} performs membership inference on query instance $(x,y)$ using trained imitative models and selected proxy data.}
    \label{fig:imia_framework}
\end{figure}

\begin{algorithm}[t!]
\caption{\textbf{Imitative Membership Inference Attack (\mymethod)} in the non-adaptive setting. The adversary first performs a one-time prepare phase, selecting pivot data and training $N$ imitative models via the proposed imitative training. During inference, the adversary queries each imitative \textit{out} model to obtain its \textit{out} confidence distribution for the query instance, then approximates its \textit{in} confidence distribution using proxies selected from pivot data.}
\label{alg:imia_nonadaptive}
\begin{algorithmic}[1]
\REQUIRE Target model $f_\theta$, adversary's dataset $D_\text{adv}^\text{non-adapt}$, query instance $(x,y)\in D_\text{query}$
\STATE \textcolor{gray}{\texttt{\# Prepare Phase: Imitative Training}}
\STATE $\mathcal{F}_{\text{out}} \gets \{\}, \ \mathcal{F}_{\text{in}}^{\text{pivot}} \gets \{\}$
\STATE $D_{\text{pivot}} \gets \texttt{SelectPivot}(f_\theta, D_\text{adv}^\text{non-adapt})$\algcomment{select pivots}

\FOR{$N$ \text{times}}
    \STATE $D_\text{imitate} \sim D_\text{adv}^\text{non-adapt}$ \algcomment{sample a dataset}
    \myfullcomment{train imitative models, see~\Cref{alg:imitative_training}}
    \STATE $f_\text{out}, f_{\text{in}}^{\text{pivot}} \gets \texttt{ImitativeTrain} (f_\theta, D_\text{imitate}, D_{\text{pivot}})$ 
    \STATE $\mathcal{F}_\text{out} \gets \mathcal{F}_\text{out} \cup \{f_\text{out}\}$ 
    \STATE $\mathcal{F}_{\text{in}}^{\text{pivot}} \gets \mathcal{F}_{\text{in}}^{\text{pivot}} \cup \{f_{\text{in}}^{\text{pivot}}\}$ 
\ENDFOR

\vspace{0.5em}
\STATE \textcolor{gray}{\texttt{\# Inference Phase: Query on $(x, y)$}}
\STATE $\mathcal{S}_{\text{out}} \gets \{\},\ \mathcal{S}_{\text{in}} \gets \{\}$
\myfullcomment{collect out confidence scores}
\FOR{each $f_\text{out} \in \mathcal{F}_{\text{out}}$}
    \STATE $\mathcal{S}_{\text{out}} \gets \mathcal{S}_{\text{out}} \cup \{\phi(f_\text{out}(x)_y)\}$
\ENDFOR
\myfullcomment{collect in confidence scores via proxies, see~\Cref{sec:imia_framework}}
\STATE $D_\text{proxy} \gets \texttt{FindProxy}(D_\text{pivot}, (x, y))$ \algcomment{find proxies}
\FOR{each $f_\text{in}^\text{pivot} \in \mathcal{F}_{\text{in}}^\text{pivot}$}
    \FOR{each $(u,v) \in D_\text{proxy}$}
        \STATE $\mathcal{S}_{\text{in}} \gets \mathcal{S}_{\text{in}} \cup \{\phi(f_\text{in}^\text{pivot}(u)_v)\}$
    \ENDFOR
\ENDFOR

\STATE $\bar{s}_\text{out} \gets \texttt{mean}(\mathcal{S}_{\text{out}})$
\STATE $\bar{s}_\text{in} \gets \texttt{mean}(\mathcal{S}_{\text{in}})$
\STATE $s_{\text{obs}} \gets \phi(f_\theta(x)_y)$ \algcomment{query target model}

\STATE \textbf{return} $\Lambda = (s_{\text{obs}} - \bar{s}_\text{out})^2 - (s_{\text{obs}} - \bar{s}_\text{in})^2 $

\end{algorithmic}
\end{algorithm}

\mypara{Attack Workflow}
The workflow of \mymethod is outlined in~\Cref{alg:imia_nonadaptive} and~\Cref{fig:imia_framework},
consisting of two phases: the prepare and inference phases.
In the prepare phase, we start by selecting a set of pivot data from the adversary's dataset $D_{\text{adv}}^{\text{non-adapt}}$, and then train $N$ imitative \textit{in} and imitative \textit{out} models via imitative training (lines 2-10).
During the inference phase, we first collect the \textit{out} distribution of scaled confidence scores for the query instance using the trained imitative out models (lines 13-16). 
Next, we select a set of proxy instances from the pivot dataset that are similar to the query instance and gather their \textit{in} distributions of scaled confidence scores via imitative \textit{in} models (lines 18-23). 
Finally, we estimate the means for both confidence distributions and compute a non-parametric score $\Lambda$ based on their distances to the query’s confidence score on the target model (lines 24–27).

The computation of the score $\Lambda$ can be interpreted as a Gaussian likelihood ratio test~\cite{sp22lira} with fixed variance, which allows us to frame the inference task as a hypothesis testing.
Note that the prepare phase can be done before accessing query instances and only needs to be performed once to answer any queries using the same set of imitative models. 
During the inference phase, no additional models are trained. 
The efficiency analysis of each phase is detailed in~\Cref{sec:exp_main}.


\mypara{Selecting Pivot Data}
Our attack requires a pivot dataset $D_\text{pivot}$ to train imitative \textit{in} models (line 3 in~\Cref{alg:imia_nonadaptive}).
An effective pivot dataset should (i) cover all classes of the data distribution, (ii) facilitate fast convergence of imitative models to improve efficiency.
To construct such a dataset, we adopt a simple yet effective heuristic: we query the target model $f_\theta$ with all instances from the adversary's dataset, compute their losses, and select the $k$ (set as 100 by default) instances with the lowest losses from each class to form $D_\text{pivot}$.
Our ablation study in~\Cref{sec:exp_ablation} shows \mymethod is robust to this heuristic.
While this simple approach works well in our experiments, more sophisticated pivot-selection strategies may further improve performance; we leave this as future work.

\mypara{Finding Proxy}
During membership inference, we retrieve a set of proxy samples from the pivot dataset (line 18 in~\Cref{alg:imia_nonadaptive}) to approximate the \textit{in} behavior of the query instance.
Prior work~\cite{ndss26cpmia} proposes several proxy-selection heuristics at global, class, and instance granularities.
We find that class-level selection consistently offers strong performance while remaining computationally efficient.
Therefore, we adopt this strategy: for a given query instance, we retrieve all samples from $D_\text{pivot}$ belonging to the same class and use their output distributions to approximate the query's \textit{in} behavior.

\mypara{\mymethod in the Adaptive Setting}
While \mymethod is primarily designed for the non-adaptive setting, it can also be applied to the adaptive attack setting.
As shown in~\Cref{alg:imia_adaptive}, for a query instance $(x,y)$, we train $N$ pairs of imitative \textit{in} models (trained with $(x,y)$) and imitative \textit{out} models (trained without $(x,y)$) on random subsets of the adversary's dataset $D_\text{adv}^\text{adapt}$ (lines 2–9).
This procedure directly learns the target model's behavior when the query instance is in its training set, thus eliminating the need for a pivot dataset and proxy selection.
The final score is computed by comparing the query's confidence score on the target model to the mean scores of the imitative \textit{in} and \textit{out} distributions (lines 11-14).
In addition, while the algorithm is described for a single query, the procedure can be easily parallelized across all query instances using the same set of imitative models.
Specifically, we follow~\cite {sp22lira} and construct the subsets such that each instance $(x,y) \in D_\text{query}$ appears in $N/2$ subsets, and train $N$ imitative models on these subsets. 
This ensures that the same $N$ imitative models are used to infer the membership for all instances in $D_\text{query}$.

\subsection{Discussion of \mymethod}
\label{sec:discussion_imia}

\mypara{Non-parametric Modeling}
State-of-the-art MIAs~\cite{sp22lira,ndss26cpmia} rely on \textit{parametric modeling} to compute final scores.
For instance, LiRA~\cite{sp22lira} estimates the likelihood ratio by fitting Gaussian distributions to scaled confidence scores.
However, accurately estimating the parameters of these distributions requires training a large number of shadow models, and the performance degrades significantly when computational resources are constrained~\cite{ndss26cpmia,icml24rmia}.
To address this, \mymethod adopts a non-parametric view and computes a score by measuring the squared distance between attack signals derived from imitative \textit{in} and imitative \textit{out} models. 
This design avoids the cost of parameter estimation, enabling \mymethod to outperform state-of-the-art MIAs while using less than \computationreduce of their computational cost.

It is worth noting that the core idea behind our approach (\ie imitative training) is not limited to low-resource settings. 
As demonstrated in~\Cref{sec:exp_main}, when computational resources are available (\eg training 256 models as in LiRA), adapting \mymethod to use LiRA's parametric modeling achieves even greater performance. 
This adaptability highlights the effectiveness and versatility of our imitative training framework across a broad spectrum of computational budgets.

\mypara{Connections with Model Distillation}
The imitative training procedure in \mymethod is conceptually related to model distillation, a technique used in several prior MIAs~\cite{ccs22enhanced,tifs25glira,ccs24seqmia,ccs22trajectory}. 
However, \mymethod differs from these works in several key aspects:
(i) \textit{Tailored loss function.} 
While prior attacks employ standard KL divergence for distillation, \mymethod introduces the imitation loss that is specifically designed to distill the structural logit information most indicative of membership.
(ii) \textit{Dual behavior modeling.} 
Previous MIAs that utilize model distillation only model the target's behavior on non-members. 
In contrast, our two-phase imitative training first creates an imitative \textit{out} model and then fine-tunes it on pivot data to create an imitative \textit{in} model.
This second phase explicitly encodes membership signals while preserving the target model's knowledge, allowing the adversary to take advantage of both in and out behaviors to mount a more powerful attack.
(iii) \textit{Significant performance gain.} 
In our experiments, we observe that prior distillation-based MIAs achieve subpar performance.
In contrast, \mymethod significantly outperforms all existing attacks while requiring substantially less computational overhead.

\begin{algorithm}[t]
\caption{\textbf{Imitative Membership Inference Attack (\mymethod)} in the adaptive setting. The adversary trains $N$ imitative \textit{in} and imitative \textit{out} models for the query instance.}
\label{alg:imia_adaptive}
\begin{algorithmic}[1]
\REQUIRE Target model $f_\theta$, adversary's dataset $D_\text{adv}^\text{adapt}$, query instance $(x,y)\in D_\text{query}$
\STATE $\mathcal{S}_{\text{out}} \gets \{\},\ \mathcal{S}_{\text{in}} \gets \{\}$
\FOR{$N$ \text{times}}
    \STATE $D_\text{tmp} \sim D_\text{adv}^\text{adapt}$ \algcomment{sample a dataset}
    \STATE $D_\text{out} \gets D_\text{tmp} \setminus \{(x,y)\}$
    \STATE $D_\text{in} \gets D_\text{tmp} \cup \{(x,y)\}$ 
    \myfullcomment{train imitative models, see~\Cref{alg:imitative_training}}
    \STATE $f_\text{out}, f_{\text{in}} \gets \texttt{ImitativeTrain} (f_\theta, D_\text{out}, D_\text{in})$ 
    \STATE $\mathcal{S}_{\text{out}} \gets \mathcal{S}_{\text{out}} \cup \{\phi(f_\text{out}(x)_y)\}$
    \STATE $\mathcal{S}_{\text{in}} \gets \mathcal{S}_{\text{in}} \cup \{\phi(f_\text{in}(x)_y)\}$
\ENDFOR

\STATE $\bar{s}_\text{out} \gets \texttt{mean}(\mathcal{S}_{\text{out}})$
\STATE $\bar{s}_\text{in} \gets \texttt{mean}(\mathcal{S}_{\text{in}})$
\STATE $s_{\text{obs}} \gets \phi(f_\theta(x)_y)$ \algcomment{query target model}

\STATE \textbf{return} $\Lambda = (s_{\text{obs}} - \bar{s}_\text{out})^2 - (s_{\text{obs}} - \bar{s}_\text{in})^2 $

\end{algorithmic}
\end{algorithm}




\section{Evaluation}
\label{sec:experiments}



\subsection{Experimental Setup}
\label{sec:exp_setup}

\mypara{Datasets}
We use four image datasets (\ie MNIST~\cite{mnist}, Fashion-MNIST~\cite{fmnist}, CIFAR-10~\cite{cifar10}, and CIFAR-100~\cite{cifar10}) for our main experiments. 
To demonstrate the generalizability of \mymethod, we also report attack results on two widely used non-image datasets (\ie Purchase and Texas~\cite{sp17miashokri}) in~\Cref{sec:exp_add}.
The dataset descriptions are provided in Appendix~\ref{appendix:data_description}.


\mypara{Network Architecture}
We consider four widely used neural network architectures for image classification: ResNet~\cite{cvpr16resnet}, VGG16~\cite{vgg}, DenseNet121~\cite{densenet}, and MobileNetV2~\cite{mobilenetv2}. 
To make these models compatible with evaluated datasets, we follow~\cite{ndss26cpmia} and apply several dataset-specific modifications: we adopt ResNet-18 for MNIST and FMNIST, use WideResNet-28 for CIFAR-10 and CIFAR-100, and adjust the input channels of these model architectures to accommodate both grayscale and RGB datasets.
We follow the training configurations used in prior work~\cite{sp22lira} to mitigate overfitting. 
Specifically, we use the SGD algorithm with a learning rate of $0.1$, momentum of $0.9$, and weight decay~\cite{krogh1991simple} set to $5 \times 10^{-4}$.
We also adopt a cosine learning rate schedule~\cite{loshchilov16sgdr} for optimization and apply data augmentation techniques~\cite{aaai20random} during training.
Nevertheless, in experiments, we find that some trained models still exhibit overfitting and yield low validation accuracy; in such cases, we rerun these models with different random seeds whenever validation accuracy falls below a threshold. The threshold, along with the accuracy of the target models, is reported in Appendix~\ref{appendix:acc_model}.

\mypara{Attack Baselines}
We compare our attacks against a broad range of state-of-the-art MIAs in our experiments:

\begin{itemize}
    \item \textit{LOSS}~\cite{csf18privacy} uses the loss of the query instance for inference.
    \item \textit{Entropy}~\cite{usenix21systematic} leverages a modified prediction entropy estimation for membership inference.
    \item \textit{Calibration}~\cite{iclr22calibrate} employs a technique called difficulty calibration, which adjusts the query instance's loss by calibrating it on shadow models.
    \item \textit{Attack-R}~\cite{ccs22enhanced} compares the loss of the query instance on the target model to its losses on shadow models. The final score is the proportion of shadow models for which the target model yields a lower loss.
    \item \textit{Attack-D}~\cite{ccs22enhanced} uses model distillation~\cite{arxiv15distill} to train shadow models and computes the same ratio scores as Attack-R.
    \item \textit{SeqMIA}~\cite{ccs24seqmia} leverages model distillation to replicate the learning process of the target model, extracting various membership metrics from its loss trajectory for attack.    
    \item \textit{LiRA}~\cite{sp22lira} adapts different strategies for two attack settings. In the adaptive setting, it trains both shadow \textit{in} and shadow \textit{out} models and applies a likelihood ratio test for inference. In the non-adaptive setting, it uses only shadow \textit{out} models to conduct a one-sided hypothesis test.
    \item \textit{Canary}~\cite{iclr23canary} improves upon LiRA by employing adversarial optimization to enhance inference performance.
    \item \textit{GLiRA}~\cite{tifs25glira} enhances LiRA with model distillation to obtain shadow \textit{out} models for a one-sided hypothesis test.
    \item \textit{RMIA}~\cite{icml24rmia} performs multiple pairwise likelihood ratio tests between the query instance and randomly selected population instances to compute the score.
    \item \textit{RAPID}~\cite{ccs24rapid} trains a neural network for membership inference by combining raw and calibrated loss values~\cite{iclr22calibrate}.
    \item \textit{PMIA}~\cite{ndss26cpmia} approximates the posterior odds test by using behaviors of the shadow model’s training set as proxies.
\end{itemize}

While all attacks can be applied in the non-adaptive setting, some methods are either not applicable in the adaptive setting (\ie GLiRA and PMIA) or exhibit the same performance in both settings (\ie LOSS, Entropy). 
Thus, we only evaluate them in the non-adaptive setting.
LiRA and Canary employ different strategies for two settings; we distinguish between their versions based on the experimental context.
Attack-D, SeqMIA, and GLiRA utilize distillation to train shadow models from the target model; we include them to demonstrate the effectiveness of imitative training over model distillation.

\mypara{Evaluation Procedures}
We divide each dataset into two disjoint subsets: the query set $D_\text{query}$ and the auxiliary set $D_\text{auxiliary}$. 
The target model is trained on a randomly sampled subset of $D_\text{query}$.
In the adaptive setting, we follow~\cite{sp22lira,iclr23canary} and provide the adversary with the same data as the query set to prepare their attack (\ie $D^\text{adapt}_\text{adv} = D_\text{query}$).
In the non-adaptive setting, consistent with~\cite{ccs24rapid,ndss26cpmia,icml24rmia}, the adversary's dataset is set to the auxiliary set (\ie $D^\text{nonadapt}_\text{adv} = D_\text{auxiliary}$), ensuring that the adversary cannot access queries when training shadow models.
Details about the data split are provided in~\Cref{tab:dataset_split}.

\mypara{Evaluation Metrics}
We use the following metrics:
\begin{itemize}
    \item \textit{TPR@0\%FPR.}
    This metric reflects the extent of privacy leakage under the strictest constraint, measuring the highest true positive rate where the attack makes no false positives.
    \item \textit{TPR@0.1\%FPR.} 
    This is a relaxed version of the previous metric, allowing more false-positive samples for evaluation. 
    \item \textit{Balanced Accuracy.} 
    This metric measures how often an attack correctly predicts membership (average case).
\end{itemize}


\begin{table*}[t]
\centering
\caption{Performance comparison of \textit{non-adaptive} attacks on ResNet across four image datasets (\ie MNIST, Fashion-MNIST, CIFAR-10, and CIFAR-100). 
\mymethod significantly outperforms the strongest baseline (underlined, PMIA~\cite{ndss26cpmia}) while requiring less than \textbf{\computationreduce} of its computational cost (see~\Cref{sec:exp_main} for efficiency comparison). 
The \%Imp. indicates the relative improvement of \mymethod compared to the strongest baseline. The best result is in bold.
}
\label{tab:imia_nonadapt_resnet}
\resizebox{0.98\textwidth}{!}{
\begin{tabular}{l|*{12}{c}}
\toprule
\multirow{2}{*}{\textbf{Method}} & \multicolumn{4}{c}{\textbf{TPR @ 0\% FPR}} & \multicolumn{4}{c}{\textbf{TPR @ 0.1\% FPR}} & \multicolumn{4}{c}{\textbf{Balanced Accuracy}} \\
\cmidrule(lr){
2-5}\cmidrule(lr){6-9}\cmidrule(lr){10-13}
&  MNIST & FMNIST & C-10 & C-100 & MNIST & FMNIST & C-10 & C-100 & MNIST & FMNIST & C-10 & C-100 \\
\midrule
LOSS & 0.00\% & 0.00\% & 0.00\% & 0.00\% & 0.01\% & 0.05\% & 0.00\% & 0.00\% & 52.87\% & 60.13\% & 59.97\% & 75.11\% \\
Entropy & 0.00\% & 0.00\% & 0.00\% & 0.00\% & 0.07\% & 0.05\% & 0.00\% & 0.00\% & 53.09\% & 60.06\% & 59.87\% & 74.90\% \\
Calibration & 0.18\% & 0.00\% & 0.05\% & 0.18\% & 0.47\% & 1.30\% & 0.74\% & 3.01\% & 52.21\% & 55.31\% & 57.74\% & 68.55\% \\
Attack-R & 0.00\% & 0.00\% & 0.00\% & 0.00\% & 0.00\% & 0.00\% & 0.00\% & 0.00\% & 52.58\% & 58.05\% & 60.02\% & 72.49\% \\
Attack-D & 0.00\% & 0.00\% & 0.00\% & 0.00\% & 0.00\% & 0.00\% & 0.00\% & 0.00\% & 52.58\% & 59.77\% & 59.83\% & 76.60\% \\
SeqMIA & 0.05\% & 0.00\% & 0.04\% & 0.05\% & 0.31\% & 1.30\% & 0.51\% & 3.21\% & 50.36\% & 55.34\% & 57.18\% & 69.07\%  \\
LiRA & 0.16\% & 0.16\% & 0.30\% & 0.36\% & 0.34\% & 1.00\% & 0.94\% & 2.71\% & 50.43\% & 52.80\% & 58.14\% & 67.41\% \\
Canary & 0.10\% & 0.12\% & 0.36\% & 0.43\% & 0.32\% & 0.84\% & 0.93\% & 2.72\% & 50.34\% & 52.70\% & 58.40\% & 67.33\% \\
GLiRA & 0.07\% & 0.07\% & 0.24\% & 0.11\% & 0.36\% & 0.92\% & 0.90\% & 3.88\% & 51.06\% & 55.57\% & 55.87\% & 73.87\% \\
RMIA & 0.17\% & 0.00\% & 0.03\% & 0.11\% & 0.49\% & 1.14\% & 0.83\% & 3.59\% & 53.20\% & 58.57\% & 60.04\% & 72.78\% \\
RAPID & 0.21\% & 0.00\% & 0.04\% & 0.25\% & 0.48\% & 1.30\% & 0.74\% & 3.09\% & 52.46\% & 55.38\% & 57.79\% & 68.63\% \\
PMIA & \underline{0.30\%} & \underline{0.25\%} & \underline{0.62\%} & \underline{0.89\%} & \underline{0.77\%} & \underline{3.51\%} & \underline{1.84\%} & \underline{5.01\%} & \underline{53.23\%} & \underline{60.51\%} & \underline{60.05\%} & \underline{77.64\%} \\
\midrule
\textbf{\mymethod} & \textbf{1.01\%} & \textbf{2.52\%} & \textbf{1.45\%} & \textbf{2.10\%} & \textbf{1.86\%} & \textbf{5.08\%} & \textbf{3.42\%} & \textbf{7.32\%} & \textbf{54.14\%} & \textbf{61.22\%} & \textbf{61.08\%} & \textbf{79.52\%} \\
\cellcolor[gray]{0.9}\%Imp.  & \cellcolor[gray]{0.9}236.67\% & \cellcolor[gray]{0.9}908.00\% & \cellcolor[gray]{0.9}133.87\% & \cellcolor[gray]{0.9}135.96\% & \cellcolor[gray]
{0.9}141.56\% & \cellcolor[gray]{0.9}44.73\% & \cellcolor[gray]{0.9}85.87\% & \cellcolor[gray]{0.9}46.11\% & \cellcolor[gray]{0.9}1.71\% & \cellcolor[gray]{0.9}1.17\% & \cellcolor[gray]{0.9}1.72\% & \cellcolor[gray]{0.9}2.42\%  \\
\bottomrule
\end{tabular}}
\end{table*}
\begin{table*}[t]
\centering
\caption{Performance comparison of \textit{adaptive} attacks on ResNet across four image datasets.
\mymethod outperforms the strongest baseline (underlined) while requiring less than \textbf{\computationreduce} of its computational cost. 
The best result is in bold.
}
\label{tab:imia_adapt_resnet}
\resizebox{0.98\textwidth}{!}{
\begin{tabular}{l|*{12}{c}}
\toprule
\multirow{2}{*}{\textbf{Method}} & \multicolumn{4}{c}{\textbf{TPR @ 0\% FPR}} & \multicolumn{4}{c}{\textbf{TPR @ 0.1\% FPR}} & \multicolumn{4}{c}{\textbf{Balanced Accuracy}} \\
\cmidrule(lr){
2-5}\cmidrule(lr){6-9}\cmidrule(lr){10-13}
&  MNIST & FMNIST & C-10 & C-100 & MNIST & FMNIST & C-10 & C-100 & MNIST & FMNIST & C-10 & C-100 \\
\midrule
Calibration & 0.16\% & 0.89\% & 0.19\% & 2.98\% & 0.59\% & 2.10\% & 0.78\% & 8.49\% & 52.26\% & 55.13\% & 57.39\% & 68.91\% \\
Attack-R & 0.00\% & 0.00\% & 0.00\% & 0.00\% & 0.00\% & 0.00\% & 0.00\% & 0.00\% & 52.95\% & 58.84\% & 60.71\% & 77.57\% \\
Attack-D & 0.00\% & 0.00\% & 0.00\% & 0.00\% & 0.00\% & 0.00\% & 0.00\% & 0.00\% & 52.59\% & 59.34\% & 59.62\% & 75.81\% \\
SeqMIA & 0.13\% & 0.76\% & 0.11\% & 1.63\% & 0.36\% & 1.82\% & 0.61\% & 5.84\% & 51.92\% & 52.17\% & 58.30\% & 69.15\%  \\
LiRA & \underline{0.80\%} & \underline{3.85}\% & 1.30\% & 5.41\% & \underline{2.01\%} & 6.03\% & 2.60\% & 18.03\% & \underline{54.01\%} & \underline{62.04\%} & \underline{61.03\%} & 81.39\% \\
Canary & 0.77\% & 3.72\% & \underline{1.31\%} & \underline{5.48\%} & 1.96\% & \underline{6.17\%} & \underline{2.63\%} & \underline{18.32\%} & 53.97\% & 61.98\% & 61.01\% & \underline{81.47\%} \\
RMIA & 0.56\% & 2.05\% & 0.21\% & 0.21\% & 0.91\% & 3.76\% & 0.64\% & 5.10\% & 53.86\% & 61.28\% & 61.08\% & 79.46\% \\
RAPID & 0.46\% & 1.96\% & 0.25\% & 0.42\% & 0.95\% & 3.51\% & 0.82\% & 5.26\% & 53.26\% & 61.42\% & 60.56\% & 79.73\% \\
\midrule
\textbf{\mymethod} & \textbf{1.33\%} & \textbf{4.62\%} & \textbf{2.33\%} & \textbf{8.52\%} & \textbf{2.35\%} & \textbf{7.10\%} & \textbf{3.61\%} & \textbf{19.82\%} & \textbf{54.25\%} & \textbf{62.43\%} & \textbf{61.30\%} & \textbf{82.94\%} \\
\cellcolor[gray]{0.9}\%Imp.  & \cellcolor[gray]{0.9}66.25\% & \cellcolor[gray]{0.9}20.00\% & \cellcolor[gray]{0.9}77.86\% & \cellcolor[gray]{0.9}55.47\% & \cellcolor[gray]
{0.9}16.92\% & \cellcolor[gray]{0.9}15.07\% & \cellcolor[gray]{0.9}37.26\% & \cellcolor[gray]{0.9}8.19\% & \cellcolor[gray]{0.9}0.44\% & \cellcolor[gray]{0.9}0.63\% & \cellcolor[gray]{0.9}0.44\% & \cellcolor[gray]{0.9}1.80\%  \\
\bottomrule
\end{tabular}}
\end{table*}

\mypara{Attack Setup}
For baselines, we train the recommended number of shadow models as specified in their papers. 
For instance, LiRA trains 256 shadow models to estimate the parametric likelihood ratio (see Appendix~\ref{appendix:baselines} for a complete list of shadow model counts).
In contrast, \mymethod uses a fixed set of $N=10$ imitative models to highlight its computational efficiency. 
For imitative training, we first warm up the models for $T_\text{warmup}=50$ epochs on MNIST and Fashion-MNIST, and $T_\text{warmup}=80$ epochs on CIFAR-10 and CIFAR-100.
After the warmup, we train the imitative \textit{out} models for $T_1=100$ epochs across all datasets, matching the number of epochs used in shadow training~\cite{sp22lira}.
For the imitative \textit{in} models, we continue training for $T_2 = 20$ using pivot data.
For CIFAR-10 and CIFAR-100, we follow~\cite{sp22lira,ndss26cpmia,iclr23canary} by querying each instance with $18$ random augmentations to compute the averaged membership score. 
Regarding hyperparameter configuration, we utilize the default settings from the original baseline implementations without further tuning across all experiments. 
We acknowledge this as a limitation; while these defaults may not be optimal for every dataset, we maintain this approach due to the prohibitive computational cost required for exhaustive tuning for every baseline.
The final results are reported as the average metrics over ten runs.

\subsection{Evaluation of \mymethod}
\label{sec:exp_main}

In this section, we evaluate \mymethod in terms of both attack performance and efficiency. 
Specifically, we benchmark \mymethod against state-of-the-art (SOTA) non-adaptive and adaptive MIAs, compare with attacks that use model distillation, and analyze its performance under varying computational budgets.

\mypara{Performance in the Non-Adaptive Setting}
We first evaluate \mymethod in the non-adaptive setting, and
the results on ResNet are shown in~\Cref{tab:imia_nonadapt_resnet}. 
\mymethod consistently outperforms all existing methods across the evaluated metrics. 
For instance, on MNIST, it achieves a true positive rate (TPR) of 1.01\% at zero false positive rate (FPR), which is at least three times as high as the best-performing baseline (\ie PMIA). 
The advantage is even more pronounced on Fashion-MNIST, with a 2.52\% TPR at the same FPR, over ten times higher than baselines.
These results represent a substantial advancement, as such datasets were previously considered hard to attack.
Although the improvement in balanced accuracy is relatively modest, this is consistent with prior studies~\cite{sp22lira,ccs22enhanced,icml24rmia}. 
Moreover, the consensus in the MIA community is that attacks should be evaluated using TPR at low FPR, where \mymethod shows a clear advantage. 
Similar trends are observed on other model architectures; results and Receiver Operating Characteristic (ROC) curves are provided in Appendix~\ref{appendix:imia_other_models}.

\mypara{Performance in the Adaptive Setting}
We also evaluate \mymethod with SOTA MIAs in the adaptive setting. 
The results for ResNet are shown in~\Cref{tab:imia_adapt_resnet}, while performance on other architectures can be found in Appenix~\ref{appendix:imia_other_models}. 
\mymethod consistently achieves the best attack performance across all datasets, surpassing strong attacks like LiRA and Canary. 
For instance, \mymethod achieves nearly double the performance of LiRA on CIFAR-10 at a TPR of 0\% FPR, with similar improvements observed across other datasets.


\mypara{Comparison with Distillation-based MIAs}
While the imitative training used in \mymethod is conceptually related to model distillation, it significantly outperforms distillation-based attacks (\ie Attack-D, SeqMIA, and GLiRA) in both adaptive and non-adaptive settings. 
For instance, Attack-D yields negligible TPR at low FPRs, and in the non-adaptive setting, \mymethod outperforms SeqMIA and GLiRA by at least a factor of five.
We attribute this performance gap to key limitations in existing distillation-based MIAs: 
(i) These attacks directly apply the standard model distillation process~\cite{arxiv15distill}, failing to capture the salient membership signals that are important for MIA.
(ii) They use distillation to create shadow \textit{out} models, which can only mimic the target's behavior on non-member instances.
In contrast, \mymethod explicitly caputer membership signals by training both imitative \textit{in} and \textit{out} models, leveraging their behavioral discrepancies for a more effective attack.

\begin{table}[t]
\centering
\caption{Runtime comparison of leading MIAs and \mymethod, measured in hours on a cluster of one A100 GPU. The computation reduction (\%) is calculated against the leading attacks in each setting (PMIA for non-adaptive, LiRA for adaptive).}
\label{tab:runtime}
\resizebox{0.49\textwidth}{!}{
\begin{tabular}{l|c|c|c|c|c|c}
\toprule
 & \textbf{MNIST} & \textbf{FMNIST} & \textbf{CIFAR-10} & \textbf{CIFAR-100} & \textbf{Purchase} & \textbf{Texas}\\
\midrule
\multicolumn{7}{c}{\textit{Non-Adaptive Setting}} \\
\midrule
LiRA & 14.30 & 14.41 & 142.32 & 284.62 & 2.92 & 3.22  \\ 
Canary & 22.46 & 23.34 & 182.22 & 336.67 & 10.59 & 4.10 \\ 
GLiRA & 8.51 & 8.65 & 72.62 & 144.39 & 1.53 & 1.68 \\ 
RMIA & 9.26 & 9.50 & 79.79 & 156.48 & 1.58 & 1.74 \\ 
Rapid & 11.21 & 11.35 & 88.42 & 163.35 & 1.62 & 1.77 \\ 
PMIA & 14.31 & 14.43 & 143.08 & 287.14 & 3.01 & 3.23 \\ 
\midrule
\mymethod & \textbf{0.64} & \textbf{0.65} & \textbf{6.72} & \textbf{13.43} & \textbf{0.12} & \textbf{0.13} \\ 
\cellcolor[gray]{0.9}\%Reduction & \cellcolor[gray]{0.9}95.53\% & \cellcolor[gray]{0.9}95.50\% & \cellcolor[gray]{0.9}95.30\% & \cellcolor[gray]{0.9}95.32\% & \cellcolor[gray]{0.9}96.01\% & \cellcolor[gray]{0.9}95.98\% \\
\midrule
\multicolumn{7}{c}{\textit{Adaptive Setting}} \\
\midrule
LiRA & 14.40 & 14.62 & 142.43 & 287.85 & 2.94 & 3.38 \\ 
Canary & 22.69 & 23.51 & 182.95 & 337.42 & 11.03 & 4.13 \\ 
RMIA & 10.45 & 10.98 & 80.02 & 154.23 & 1.59 & 1.75 \\ 
Rapid & 11.34 & 11.63 & 89.17 & 165.73 & 1.65 & 1.79 \\ 
\midrule
\mymethod & \textbf{0.72} & \textbf{0.73} & \textbf{6.96} & \textbf{14.08} & \textbf{0.14} & \textbf{0.16} \\ 
\cellcolor[gray]{0.9}\%Reduction & \cellcolor[gray]{0.9}95.00\% & \cellcolor[gray]{0.9}95.01\% & \cellcolor[gray]{0.9}95.11\% & \cellcolor[gray]{0.9}95.11\% & \cellcolor[gray]{0.9}95.24\% & \cellcolor[gray]{0.9}95.27\% \\
\bottomrule
\end{tabular}}
\end{table}



\begin{table}[t]
\centering
\caption{Runtime breakdown (in hours) of MIAs on MNIST.
The preparation cost refers to the time for training shadow (imitative) models, while the inference cost denotes the time for inferring the membership of query instances. GLiRA and PMIA are applicable only in the non-adaptive setting.}
\label{tab:runtime_breakdown}
\resizebox{0.49\textwidth}{!}{
\begin{tabular}{l|*{6}{c}|c}
\toprule
 & LiRA & Canary & GLiRA & RMIA & RAPID & PMIA & \mymethod \\
 \midrule
 \multicolumn{8}{c}{\textit{Non-Adaptive Setting}} \\
\midrule
Preparation & 14.22 & 8.57 & 8.43 & 8.35 & 10.59 & 14.22 & \textbf{0.62}   \\
Inference  & 0.08 & 13.89 &  0.08 & 0.91 
& 0.62 & 0.09 & \textbf{0.02}   \\
\midrule
\multicolumn{8}{c}{\textit{Adaptive Setting}} \\
\midrule
Preparation & 14.26 & 8.61 & - & 8.38 & 10.72 & - & \textbf{0.71}   \\
Inference  & 0.14 & 14.08 & - & 1.07 & 0.62 & - & \textbf{0.01}   \\
\bottomrule
\end{tabular}}
\end{table}

\mypara{Efficiency Analysis}
We measure the overall runtime of \mymethod in both adaptive and non-adaptive settings on a cluster with a single A100 GPU, comparing it to leading MIAs.
We do not include simple baselines (\eg LOSS attack~\cite{csf18privacy}) for comparison, as their performance is far inferior to \mymethod. 
As shown in \Cref{tab:runtime}, \mymethod requires substantially less computation. 
For instance, on CIFAR-10, the SOTA baselines (\ie PMIA and LiRA) require 6 days to perform their attack, whereas \mymethod completes in less than 7 hours, a reduction of more than 95\%. 
This efficiency is consistently observed across all datasets.

In~\Cref{tab:runtime_breakdown}, we also present a runtime breakdown on MNIST, separating the computation spent on preparation (\ie model training) and inference phases.
The results confirm that model training is the bottleneck for most MIAs, and \mymethod mitigates this by requiring far fewer models.
During inference, while most MIAs are relatively fast, \mymethod can even outperform them by computing a simple non-parametric score for inference.

\begin{figure}[t]
    \centering
    \subfigure[Non-adaptive setting.]
    {
    \includegraphics[width=0.46\linewidth]{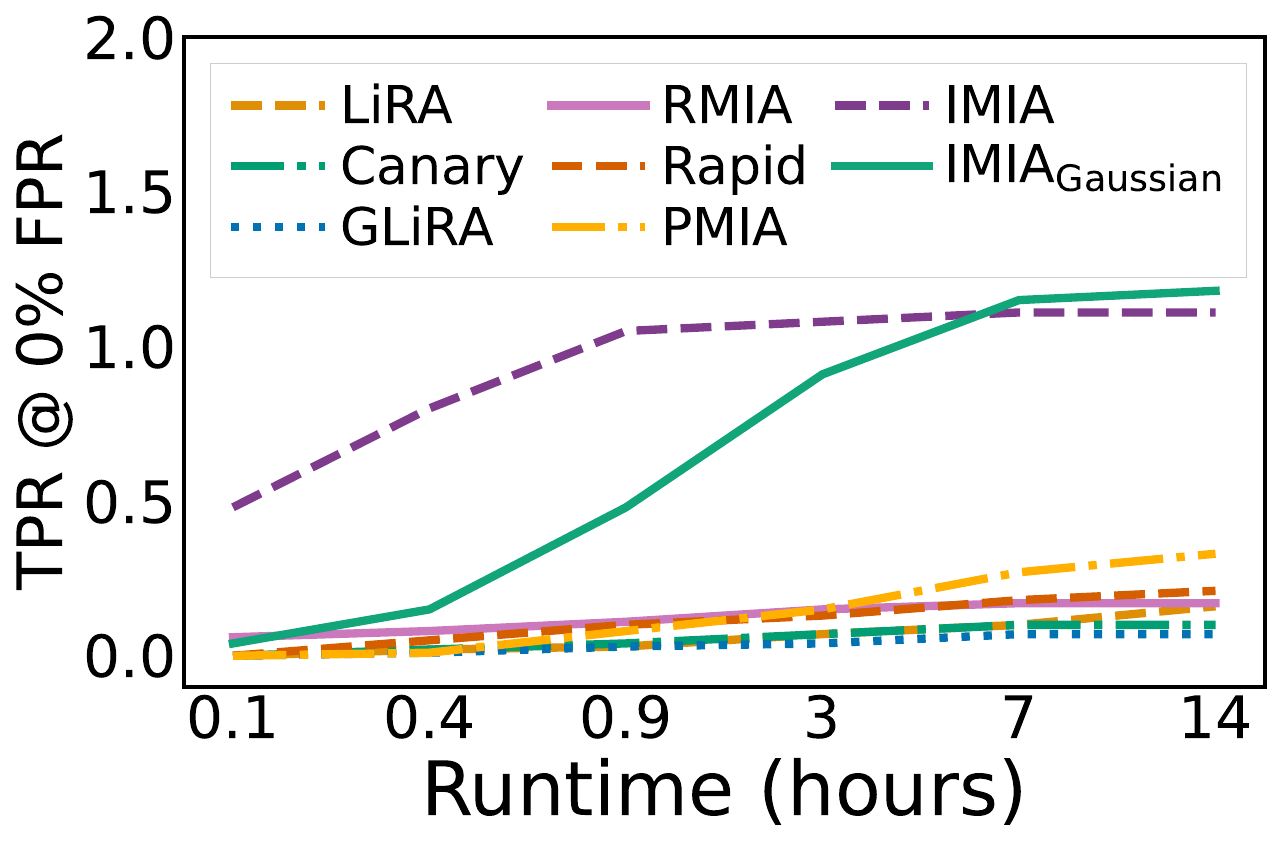}
    \label{fig:runtime_mnist_offline}
    }
    \subfigure[Adaptive setting.]
    {
    \includegraphics[width=0.46\linewidth]{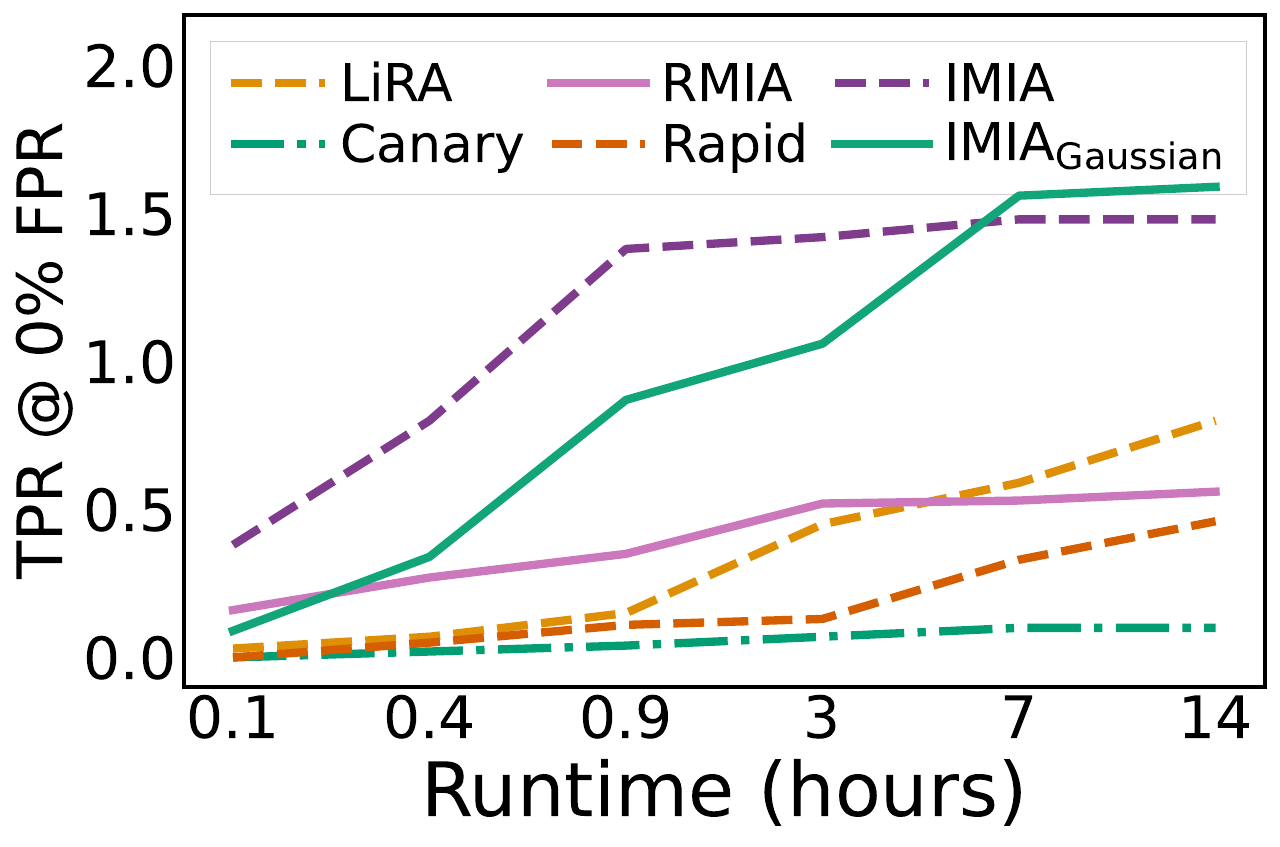}
    \label{fig:runtime_mnist_online}
    }
    \caption{Performance on MNIST under varying computational budgets. \mymethodgaussian employs a Gaussian likelihood ratio~\cite{sp22lira} on imitative models to compute the final scores.}
    \label{fig:runtime}
\end{figure}

\begin{figure*}[t]
    \centering
    \subfigure[Residuals of target-agnostic \textbf{shadow} models on \textit{members}.]
    {
    \includegraphics[width=0.22\linewidth]{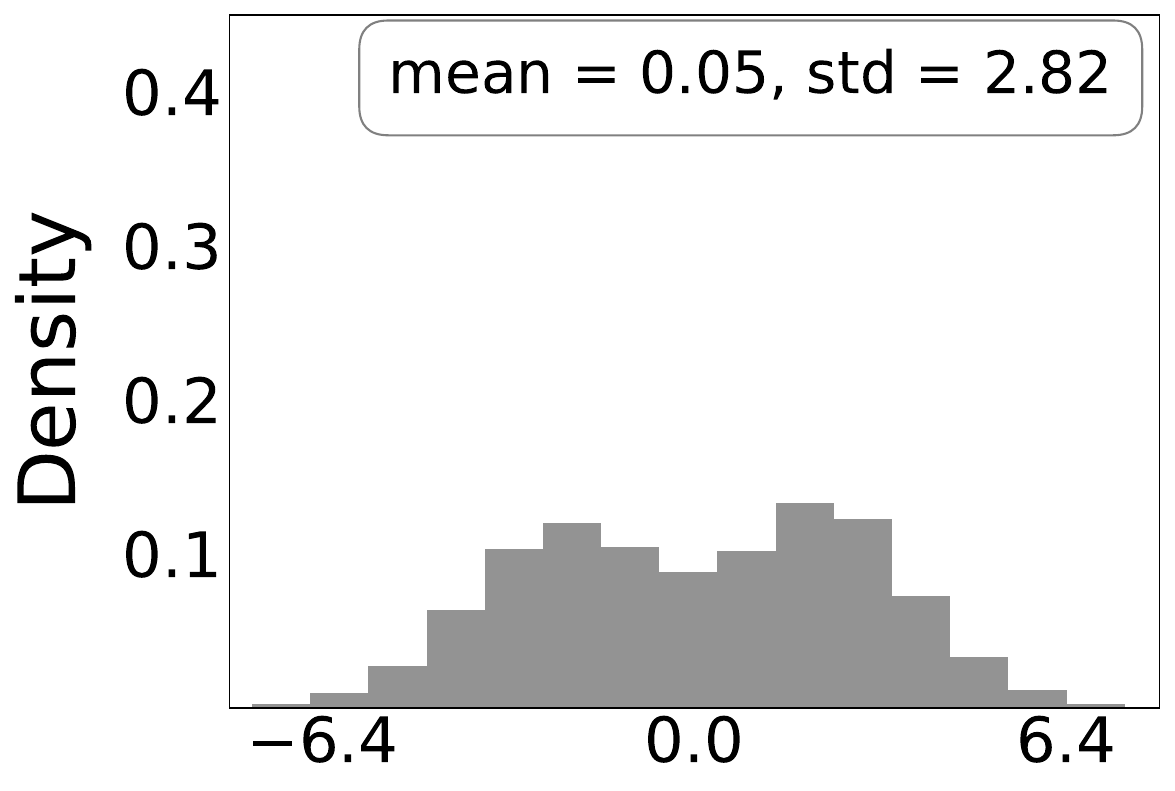}
    }
    \hspace{0.5mm}
    \subfigure[Residual of target-informed \textbf{imitative} models on \textit{members}.]
    {
    \includegraphics[width=0.22\linewidth]{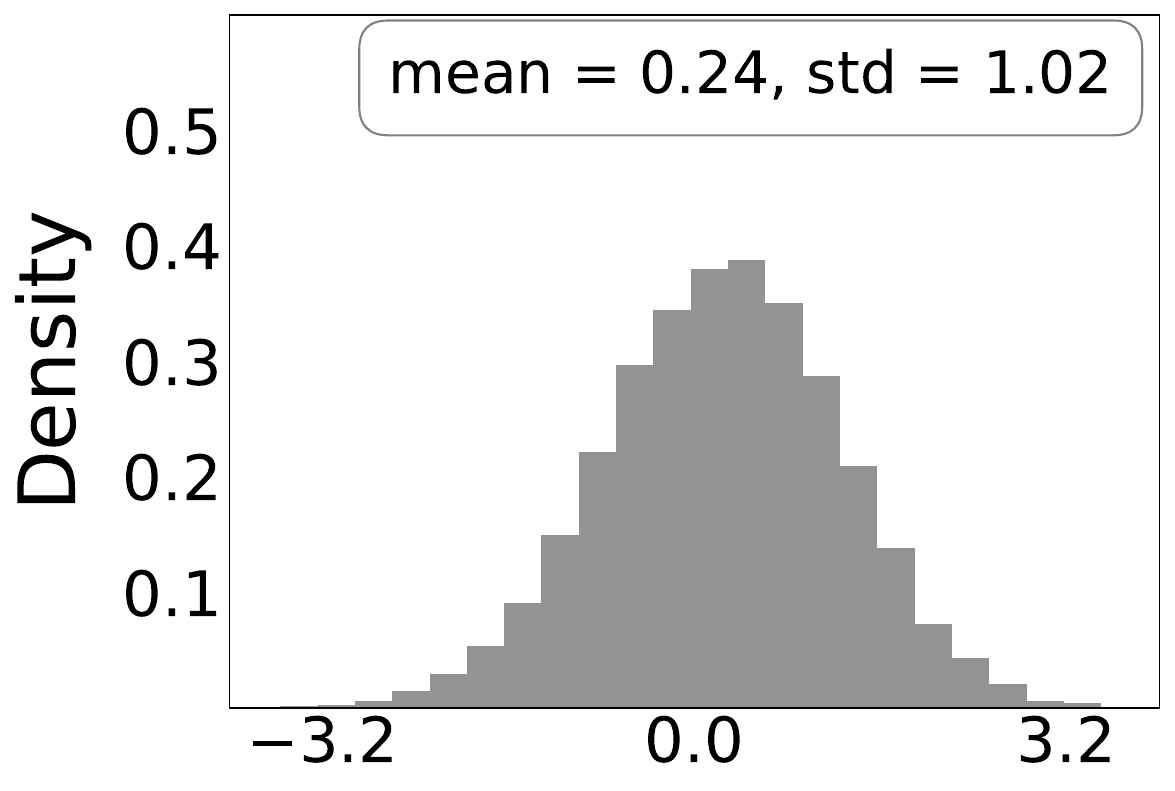}
    }
    \hspace{0.5mm}
    \subfigure[Residual of target-agnostic \textbf{shadow} models on \textit{non-members}.]
    {
    \includegraphics[width=0.22\linewidth]{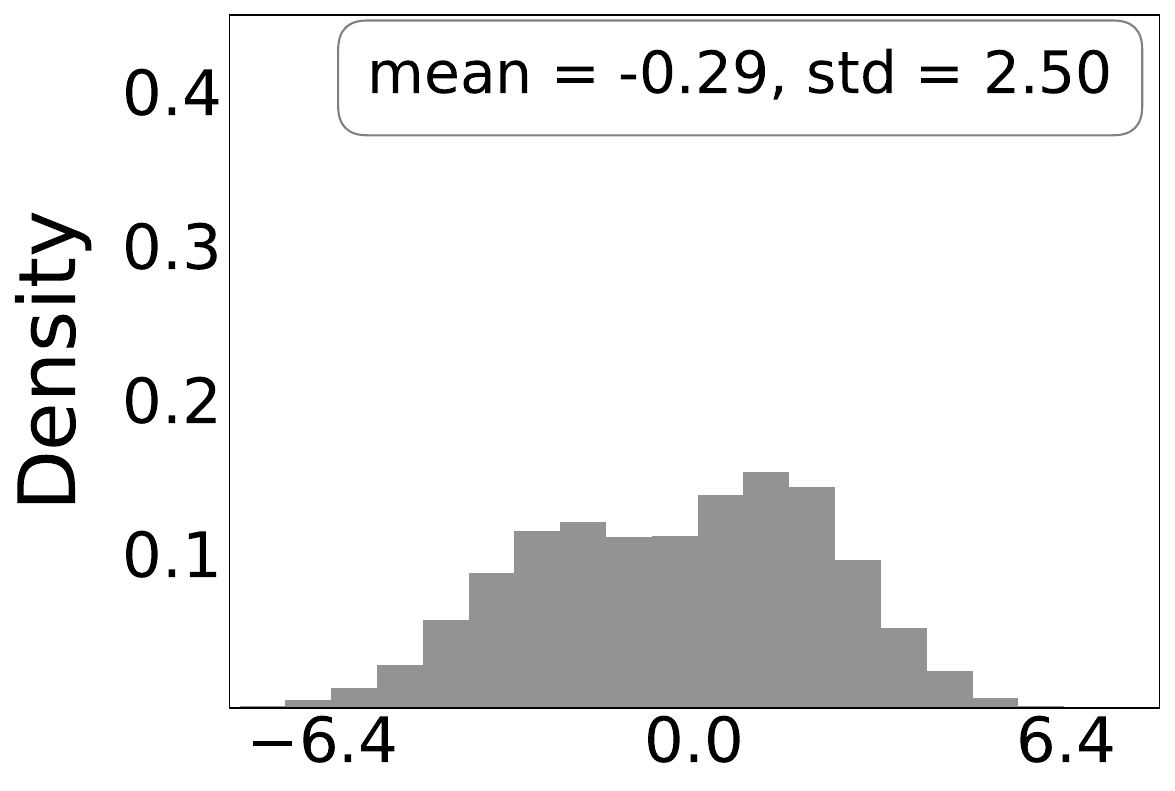}
    }
    \hspace{0.5mm}
    \subfigure[Residual of target-informed \textbf{imitative} models on \textit{non-members}.]
    {
    \includegraphics[width=0.22\linewidth]{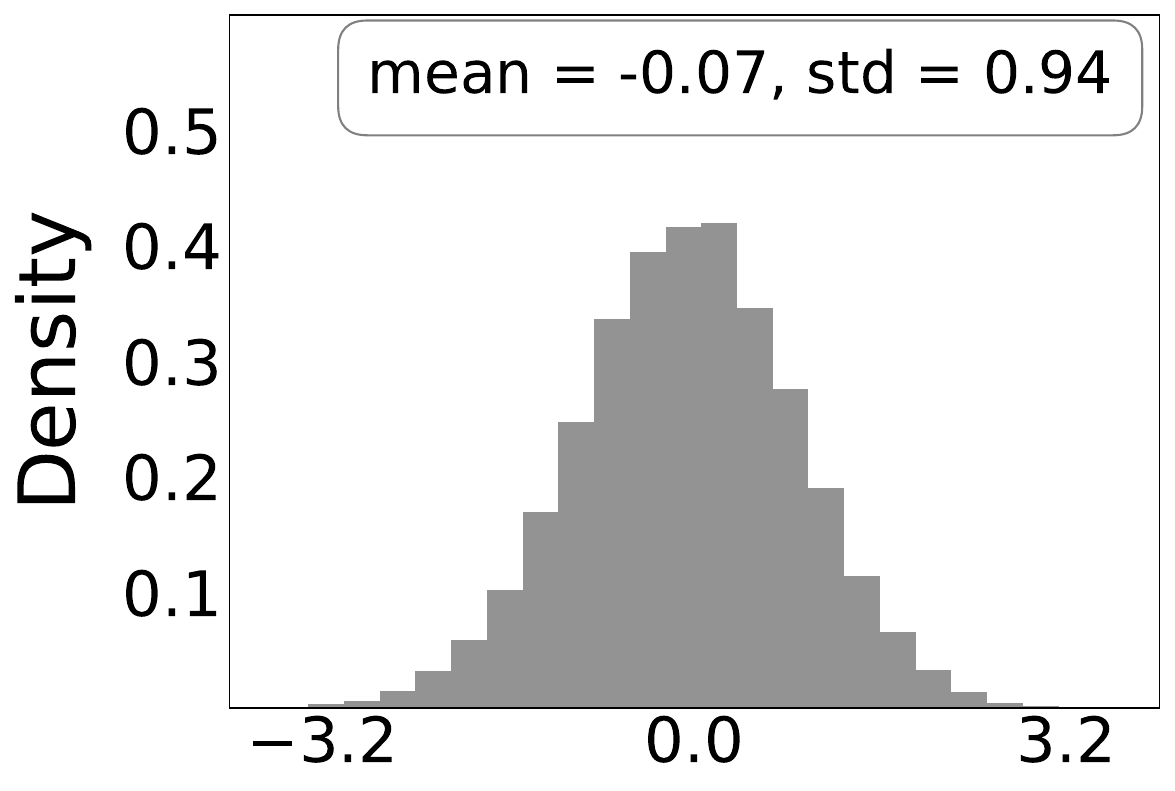}
    }
    \caption{Normalized residual distributions (defined in~\Cref{sec:exp_imiative}) of shadow vs.\ imitative models on CIFAR-100. The residuals of imitative models closely follow the standard normal distribution, indicating they better mimic the target model’s behavior.}
    \label{fig:z_score_hist_cifar100}
\end{figure*}

\mypara{Performance Across Computational Budgets}
To evaluate the trade-off between efficiency and attack effectiveness, we benchmark \mymethod against leading MIAs using an A100 GPU in both non-adaptive and adaptive settings. 
We also introduce \mymethodgaussian, a variant that leverages the same imitative training as \mymethod but employs parametric modeling for inference. 
Specifically, it follows LiRA~\cite{sp22lira} and fits the scaled confidence scores of imitative models as Gaussian distributions, and uses the likelihood ratio as the membership score. 

\Cref{fig:runtime} illustrates the performance under different computational budgets (equivalent to 1 to 256 shadow models).
\mymethod demonstrates superior performance across all settings and is particularly dominant in low-resource scenarios (\eg under 3 hours of training). 
In contrast, state-of-the-art MIAs typically require more computation to achieve their optimal performance. 
For example, LiRA in the adaptive setting shows substantial improvement when trained for 14 hours (\ie 256 shadow models), and evaluating it with fewer models would underestimate its efficacy.
Additionally, we observe that while \mymethodgaussian underperforms \mymethod when computational resources are limited, it surpasses \mymethod as more computation is available. 
This further underscores the versatility of the imitative training paradigm: imitative models can effectively mimic the target model's behavior, and increasing their number, in combination with parametric modeling, allows for better capture of subtle behavioral discrepancies between members and non-members, leading to more powerful attacks.

\begin{table}[t]
\centering
\caption{Performance stability on MNIST in the non-adaptive setting. All values are reported as percentages (\%).}
\label{tab:stability}
\resizebox{0.48\textwidth}{!}{
\begin{tabular}{l|c|c|c}
\toprule
& \textbf{TPR @ 0\%FPR} & \textbf{TPR @ 0.1\%FPR} & \textbf{Balanced Accuracy} \\
\midrule
LiRA & 0.16 $\pm$ 0.10 & 0.34 $\pm$ 0.15 & 50.43 $\pm$ 0.16  \\ 
Canary & 0.10 $\pm$ 0.08 & 0.32 $\pm$ 0.12 & 50.34 $\pm$ 0.15  \\ 
GLiRA & 0.07 $\pm$ 0.06 & 0.36 $\pm$ 0.08 & 51.06 $\pm$ 0.14  \\ 
RMIA & 0.17 $\pm$ 0.04 & 0.49 $\pm$ 0.11 & 53.20 $\pm$ 0.15  \\ 
RAPID & 0.21 $\pm$ 0.09 & 0.48 $\pm$ 0.14 & 52.46 $\pm$ 0.10  \\ 
PMIA  & 0.30 $\pm$ 0.11 & 0.77 $\pm$ 0.18  & 53.23 $\pm$ 0.17  \\ 
\midrule
\mymethod & \textbf{1.01 $\pm$ 0.04} & \textbf{1.86 $\pm$ 0.06} & \textbf{54.14 $\pm$ 0.07}  \\ 
\bottomrule
\end{tabular}}
\end{table}

\mypara{The Stability of Attacks}
We also investigate the performance stability of our attack. 
Specifically, we train 10 target models on randomly selected subsets and attack them using leading MIAs and \mymethod, then calculate the mean and standard deviation. 
The results on the non-adaptive setting on MNIST are in \Cref{tab:stability}. 
As observed, existing shadow-based MIAs (\eg LiRA and PMIA) exhibit high performance variability relative to their mean, whereas our target-informed imitative training approaches show better stability.
This demonstrates the reliability and robustness of our attack.

\subsection{Effectiveness of Imitative Training}
\label{sec:exp_imiative}


\myquestion{Does Imitative Training Better Mimic the Target Model}
We conduct a set of experiments comparing its ability to mimic the target model’s behavior against that of shadow training.
Specifically, we train 256 shadow models and imitative models in the adaptive setting.
We then collect the scaled confidence score distributions for both members and non-members by querying the shadow and imitative models. 
Prior studies~\cite{sp22lira,iclr23canary,ndss26cpmia} have empirically validated that these score distributions are well-approximated by a Gaussian distribution. 
Following this approach, we model each distribution as Gaussian and estimate its mean $\mu$ and standard deviation $\sigma$. 
Using these parameters, we compute the \textbf{normalized residual} of the target model’s true score $s$ as $r = \frac{s - \mu}{\sigma}$.
If the models' distributions truly replicate the target model's behavior, the residuals $r$ for both members and non-members should follow a standard normal distribution, \ie $r \sim N(0,1)$.

Figure~\ref{fig:z_score_hist_cifar100} presents the results for both members and non-members on CIFAR-100. 
Residuals from imitative models align much more closely with the standard normal distribution than those from shadow models, indicating that imitative models more faithfully reproduce the target model’s behavior.
In~\Cref{appendix:effectivness_imiative}, we present results for additional image and non-image datasets, along with quantile–quantile plots to assess the alignment with $N(0,1)$ and a likelihood-based analysis to quantify that imitative models better capture the target model's true behavior than shadow models.

\begin{table}[t]
\centering
\caption{Wasserstein distance between the \textit{in} and \textit{out} distributions of shadow and imitative models. A larger distance indicates that the models can better capture the target model's behavioral discrepancy between members and non-members.}
\label{tab:wasserstein}
\resizebox{0.47\textwidth}{!}{
\begin{tabular}{l|c|c|c|c|c|c}
\toprule
 & \textbf{MNIST} & \textbf{FMNIST} & \textbf{C-10} & \textbf{C-100} & \textbf{Purchase} & \textbf{Texas}\\
\midrule
Shadow models & 0.01 & 0.03 & 0.06 & 0.21 & 0.13 & 0.25  \\ 
Imitative models & \textbf{0.08} & \textbf{0.18} & \textbf{0.47} & \textbf{0.82} & \textbf{0.73} & \textbf{0.85} \\
\bottomrule
\end{tabular}}
\end{table}

\myquestion{Do Imitative Models Enhance Behavioral Discrepancy}
Effective imitative models should reliably replicate the behavioral discrepancy between members and non-members of the target model.
We quantify this discrepancy by computing the Wasserstein distance between the \textit{in} and \textit{out} distributions of shadow and imitative models, respectively.
The results in~\Cref{tab:wasserstein} show that the Wasserstein distance of imitative models is significantly larger than that of shadow models across datasets, which indicates that imitative models generate a much more separable signal between members and non-members.

\subsection{Ablation Study}
\label{sec:exp_ablation}


\begin{table}[t]
\centering
\caption{Impact of imitative training in \mymethod. We compare against two variants: \mymethoddistill, which replaces imitative training with standard model distillation, and \mymethodshadow, which uses target-agnostic shadow training.}
\label{tab:impact_imitative_training}
\resizebox{0.499\textwidth}{!}{
\begin{tabular}{l|cc|cc|cc}
\toprule
\multirow{2}{*}{} & \multicolumn{2}{c|}{\textbf{TPR @ 0\%FPR}} & \multicolumn{2}{c|}{\textbf{TPR @ 0.1\%FPR}} & \multicolumn{2}{c}{\textbf{Balanced Accuracy}} \\
 & MNIST & C-10 & MNIST & C-10 & MNIST & C-10 \\
\midrule
\multicolumn{7}{c}{\textit{Non-Adaptive Setting}} \\
\midrule
\mymethod & \textbf{1.01\%} & \textbf{1.45\%} & \textbf{1.86\%} & \textbf{3.42\%} & \textbf{54.14\%} & \textbf{61.08\%} \\ 
\mymethoddistill & 0.83\% & 0.96\% & 1.09\% & 2.46\% & 53.13\% & 60.08\% \\ 
\mymethodshadow & 0.28\% & 0.54\% & 0.69\% & 1.46\% & 53.20\% & 59.57\% \\
\midrule
\multicolumn{7}{c}{\textit{Adaptive Setting}} \\
\midrule
\mymethod & \textbf{1.33\%} & \textbf{2.33\%} & \textbf{2.35\%} & \textbf{3.61\%} & \textbf{54.25\%} & \textbf{61.30\%} \\ 
\mymethoddistill & 1.04\% & 1.69\% & 1.92\% & 2.78\% & 53.80\% & 60.47\% \\ 
\mymethodshadow & 0.63\% & 1.02\% & 1.24\% & 1.78\% & 53.06\% & 60.51\% \\
\bottomrule
\end{tabular}}
\end{table}
\begin{table}[t]
\centering
\caption{Impact of softmax temperature on \mymethod.}
\label{tab:imitative_temp}
\resizebox{0.49\textwidth}{!}{
\begin{tabular}{c|cc|cc|cc}
\toprule
\multirow{2}{*}{temperature} & \multicolumn{2}{c|}{\textbf{TPR @ 0\%FPR}} & \multicolumn{2}{c|}{\textbf{TPR @ 0.1\%FPR}} & \multicolumn{2}{c}{\textbf{Balanced Accuracy}} \\
 & MNIST & C-10 & MNIST & C-10 & MNIST & C-10 \\
 \midrule
0 & 0.72\% & 0.84\% & 1.43\% & 2.03\% & 54.01\% & 60.37\% \\ 
0.5 & 0.84\% & 1.12\% & 1.53\% & 2.50\% & 54.08\% & 60.42\% \\ 
1 & 1.01\% & \textbf{1.45\%} & \textbf{1.86\%} & \textbf{3.42\%} & 54.14\% & 61.08\% \\
2 & \textbf{1.04\%} & 1.43\% & 1.84\% & 3.41\% & \textbf{54.18\%} & 61.03\% \\ 
\bottomrule
\end{tabular}}
\end{table}

\mypara{Impact of Imitative Training}
We evaluate the effectiveness of imitative training by comparing \mymethod with two variants:
(i) \mymethoddistill, which replaces imitative training with standard model distillation using KL divergence as the loss function~\cite{arxiv15distill}; 
and (ii) \mymethodshadow, which substitutes imitative training with target-agnostic shadow training.
Table~\ref{tab:impact_imitative_training} reports results on MNIST and CIFAR-10. 
Both variants show substantial degradation across all metrics, especially in the low false-positive regime.
Using model distillation reduces average performance by over 20\%, while replacing our approach with shadow training cuts attack performance by at least half. 
These results demonstrate that effective imitation should capture the target’s membership signals, rather than its general behaviors across models.

We next examine how the softmax temperature of the target model's outputs affects attack performance. 
This softmax temperature controls the ``softness'' of the probability distributions used to train the imitative models. 
A temperature of zero means the imitative models are trained using only the target model’s hard labels, while higher temperatures generate smoother distributions that capture more nuanced details of the model's behavior.
As shown in~\Cref{tab:imitative_temp}, performance improves when shifting from hard labels to soft labels, with a standard temperature of one typically yielding the best results.

\begin{table}[t]
\centering
\caption{Impact of the pivot selection in \mymethod on MNIST. Selecting $k$ pivot instances per class with the lowest losses on the target model consistently outperforms random selection.}
\label{tab:pivot}
\resizebox{0.48\textwidth}{!}{
\begin{tabular}{c|c|c|c|c}
\toprule
Selection & $k$& \textbf{TPR @ 0\%FPR} & \textbf{TPR @ 0.1\%FPR} & \textbf{Balanced Accuracy} \\
\midrule
Loss & 50 & 1.00\% & 1.84\% & 51.13\%  \\ 
Random & 50 & 0.89\% & 1.77\% & 51.10\%  \\ 
Loss & 100 & 1.01\% & 1.86\% & 51.14\%  \\
Random & 100 & 0.91\% & 1.79\% & 51.12\%  \\ 
Loss & 1,000 & 1.02\% & 1.85\% & 51.14\%  \\  
Random & 1,000 & 0.92\% & 1.79\% & 51.11\%  \\ 
\bottomrule
\end{tabular}}
\end{table}

\mypara{Impact of Pivot Selection}
In the non-adaptive setting, the performance of \mymethod depends on the selection of a high-quality pivot dataset from $D^\text{non-adapt}_\text{adv}$ to train imitative \textit{in} models.
We investigate two selection strategies: 
(1) our proposed approach of choosing the $k$ instances per class with the lowest loss on the target model, and 
(2) a baseline that randomly selects $k$ instances per class.
We evaluate both strategies for different values of $k$, with the results presented in~\Cref{tab:pivot}.
As shown, the loss-based selection consistently yields better performance than random selection. 
Additionally, \mymethod is robust to the choice of $k$, with attack performance remaining stable as $k$ ranges from 50 to 1,000.

\begin{table}[t]
\centering
\caption{Different weighting strategies for the imitation loss. }
\label{tab:impact_imitative_weighting}
\resizebox{0.49\textwidth}{!}{
\begin{tabular}{l|cc|cc|cc}
\toprule
\multirow{2}{*}{Strategy} & \multicolumn{2}{c|}{\textbf{TPR @ 0\%FPR}} & \multicolumn{2}{c|}{\textbf{TPR @ 0.1\%FPR}} & \multicolumn{2}{c}{\textbf{Balanced Accuracy}} \\
 & MNIST & C-10 & MNIST & C-10 & MNIST & C-10 \\
\midrule
\multicolumn{7}{c}{\textit{Non-Adaptive Setting}} \\
\midrule
uniform & 0.85\% & 1.07\% & 1.44\% & 3.14\% & 53.96\% & 60.78\% \\ 
log & 0.96\% & 1.38\% & 1.79\% & 3.38\% & 54.12\% & 61.05\% \\ 
square root & \textbf{1.01\%} & \textbf{1.45\%} & \textbf{1.86\%} & \textbf{3.42\%} & \textbf{54.14\%} & \textbf{61.08\%} \\ 
linear & 0.92\% & 1.31\% & 1.75\% & 3.29\% & 54.11\% & 61.02\% \\ 
\midrule
\multicolumn{7}{c}{\textit{Adaptive Setting}} \\
\midrule
uniform & 1.01\% & 1.98\% & 2.04\% & 3.28\% & 54.16\% & 61.19\% \\ 
log & 1.29\% & \textbf{2.35\%} & 2.28\% & 3.60\% & 54.22\% & 61.27\% \\ 
square root & \textbf{1.33\%} & 2.33\% & \textbf{2.35\%} & \textbf{3.61\%} & \textbf{54.25\%} & \textbf{61.30\%} \\  
linear & 1.27\% & 2.30\% & 2.15\% & 3.48\% & 54.18\% & 61.23\% \\ 
\bottomrule
\end{tabular}}
\end{table}

\mypara{Impact of Different Weighting Strategies}
We evaluate how different weighting strategies in the imitation loss affect attack performance. 
We compare three strategies:
(i) \textit{uniform}, which assigns equal weight to all logits;
(ii) \textit{log}, which assigns weight $\log(c)$ to membership-related logits and 1 to others, where $c$ is the number of classes;
(iii) \textit{square root}, which assigns weight $\sqrt{c}$ to membership-related logits (used as the default);
(iv) \textit{linear}, which assign weigh $c$ to membership-related logits.
The results for both adaptive and non-adaptive settings are presented in~\Cref{tab:impact_imitative_weighting}.
As shown, ignoring membership signals for the imitative loss (\ie uniform strategy) leads to a noticeable performance drop. 
Assigning greater weight to these critical logits generally yields improved results. 
However, excessively large weights, as seen in the linear strategy, can cause imitative models to focus too narrowly, leading to unstable training and suboptimal performance.
Empirically, the square root strategy consistently yields the best results, suggesting it strikes an effective balance.

\begin{table}[t]
\centering
\caption{Impact of imitative training epochs $T_2$ in \mymethod. }
\label{tab:impact_imitative_epoch}
\resizebox{0.45\textwidth}{!}{
\begin{tabular}{c|cc|cc|cc}
\toprule
\multirow{2}{*}{$T_2$} & \multicolumn{2}{c|}{\textbf{TPR @ 0\%FPR}} & \multicolumn{2}{c|}{\textbf{TPR @ 0.1\%FPR}} & \multicolumn{2}{c}{\textbf{Balanced Accuracy}} \\
 & MNIST & C-10 & MNIST & C-10 & MNIST & C-10 \\
\midrule
\multicolumn{7}{c}{\textit{Non-Adaptive Setting}} \\
\midrule
10 & 0.97\% & 1.34\% & 1.72\% & 3.29\% & 54.12\% & 61.01\% \\ 
20 & \textbf{1.01\%} & \textbf{1.45\%} & \textbf{1.86\%} & \textbf{3.42\%} & 54.14\% & 61.08\% \\ 
50 & 0.99\% & 1.38\% & 1.75\% & 3.22\% & \textbf{51.18\%} & \textbf{61.13\%} \\
100 & 0.98\% & 1.42\% & 1.77\% & 3.34\% & 54.15\% & 61.07\% \\
\midrule
\multicolumn{7}{c}{\textit{Adaptive Setting}} \\
\midrule
10 & 1.02\% & 1.96\% & 1.94\% & 2.35\% & 54.25\% & 61.33\% \\
20 & 1.33\% & \textbf{2.33\%} & 2.35\% & 3.61\% & 54.25\% & 61.30\% \\  
50 & 1.34\% & 2.32\% & 2.37\% & \textbf{3.69\%} & \textbf{54.31\%} & \textbf{61.38\%} \\
100 & \textbf{1.35\%} & 2.30\% & \textbf{2.39\%} & 3.68\% & 54.30\% & 61.34\% \\
\bottomrule
\end{tabular}}
\end{table}

\mypara{Impact of Training Epochs}
The imitative training consists of two stages: an initial training phase of $T_1$ epochs for imitative \textit{out} models, followed by an additional $T_2$ epochs using pivot data to obtain the imitative \textit{in} models.
While $T_1=100$ is set to match the standard model training, our default of $T_2=20$ involves fewer epochs.
Here, we analyze the impact of varying $T_2$ on attack performance.
As shown in~\Cref{tab:impact_imitative_epoch}, fewer epochs (\eg $T_2=10$) lead to performance degradation in both adaptive and non-adaptive settings, as the model has not fully learned the behaviors of the target model. 
However, we observe that increasing the number of epochs does not necessarily result in improved performance, particularly in the non-adaptive setting. This may be because training with more epochs using cross-entropy loss causes the model to increasingly forget the target-informed behaviors, gradually reverting to the standard shadow training.
Therefore, we believe $T_2=20$ achieves a good balance, yielding strong attack performance while maintaining training efficiency.

\begin{table}[t]
\centering
\caption{Impact of different attack signals in \mymethod. }
\label{tab:impact_imitative_signals}
\resizebox{0.47\textwidth}{!}{
\begin{tabular}{l|cc|cc|cc}
\toprule
\multirow{2}{*}{Signal} & \multicolumn{2}{c|}{\textbf{TPR @ 0\%FPR}} & \multicolumn{2}{c|}{\textbf{TPR @ 0.1\%FPR}} & \multicolumn{2}{c}{\textbf{Balanced Accuracy}} \\
 & MNIST & C-10 & MNIST & C-10 & MNIST & C-10 \\
\midrule
\multicolumn{7}{c}{\textit{Non-Adaptive Setting}} \\
\midrule
loss & 0.41\% & 0.51\% & 0.75\% & 1.24\% & 53.99\% & 61.18\% \\ 
probability & 1.01\% & 1.45\% & 1.86\% & 3.42\% & 54.14\% & 61.08\% \\ 
pre-softmax & \textbf{1.12\%} & \textbf{1.57\%} & \textbf{1.89\%} & \textbf{3.45\%} & \textbf{54.22\%} & \textbf{61.35\%} \\ 
\midrule
\multicolumn{7}{c}{\textit{Adaptive Setting}} \\
\midrule
loss & 0.42\% & 0.48\% & 0.65\% & 1.27\% & 54.01\% & 61.22\% \\ 
probability & 1.33\% & 2.33\% & 2.35\% & 3.61\% & \textbf{54.25\%} & 61.30\% \\ 
pre-softmax & \textbf{1.35\%} & \textbf{2.35\%} & \textbf{2.48\%} & \textbf{3.70\%} & 54.20\% & \textbf{61.31\%} \\ 
\bottomrule
\end{tabular}}
\end{table}


\mypara{Impact of Different Membership Signals}
We analyze how the choice of membership signal affects the performance of \mymethod. 
The signal is used both to train imitative models and to compute final scores for inference. 
We compare three options:
(i) \textit{loss}, which uses the output loss of a query instance;
(ii) \textit{probability}, our default approach, which uses the softmax output probabilities;
(iii) \textit{pre-softmax}, which uses the outputs of the final layer before softmax. 
While all black-box MIAs assume access only to probabilities, using pre-softmax activation is also common in the implementation of many attacks~\cite{sp22lira, iclr23canary, icml24rmia, ndss26cpmia}.
Results for both adaptive and non-adaptive settings are shown in~\Cref{tab:impact_imitative_signals}.
As observed, using the prediction loss significantly degrades performance, and using pre-softmax activation instead of probabilities consistently yields better results. 
We also find that using pre-softmax activation to compute scaled confidence scores is numerically more stable and computationally simpler, as it avoids the logarithmic operation on probability values. 
These results are consistent with previous studies~\cite{sp22lira}, and we recommend using pre-softmax activation for our attack when available.


\mypara{Additional Ablation Studies}
We also evaluate the robustness of \mymethod under varying query budgets, scenarios where the adversary is unaware of the target model's architecture, and cases involving distribution mismatches between the adversary's data and the target's data. Due to space limitations, the results of these experiments are provided in Appendix~\ref{appendix:ablation}.

\subsection{Additional Investigations}
\label{sec:exp_add}

\begin{table}[t]
\centering
\caption{Performance comparison on non-image datasets under the \textit{non-adaptive setting}. \mymethod outperforms the strongest baseline (underlined) using less than \textbf{\computationreduce} of its computation.}
\label{tab:nonimage_nonadapt}
\resizebox{0.47\textwidth}{!}{
\begin{tabular}{l|cc|cc|cc}
\toprule
\multirow{2}{*}{\textbf{Method}} & \multicolumn{2}{c|}{\textbf{TPR @ 0\%FPR}} & \multicolumn{2}{c|}{\textbf{TPR @ 0.1\%FPR}} & \multicolumn{2}{c}{\textbf{Balanced Accuracy}} \\
 & Purchase & Texas & Purchase & Texas & Purchase & Texas \\
\midrule
LiRA & 0.01\% & 0.04\% & 0.06\% & 0.17\% & 62.12\% & 65.03\% \\ 
Canary & 0.01\% & 0.03\% & 0.04\% & 0.19\% & 62.32\% & 65.23\% \\ 
GLiRA & 0.02\% & 0.08\% & 0.35\% & 0.24\% & 66.73\% & 70.82\% \\ 
RMIA & 0.03\% & 0.12\% & 0.14\% & 0.46\% & 72.36\% & 80.52\% \\ 
RAPID & 0.03\% & 0.10\% & 0.17\% & 0.49\% & 73.72\% & 81.31\% \\
PMIA & \underline{0.05\%} & \underline{0.42\%} & \underline{2.28\%} & \underline{5.72\%} & \underline{78.38}\% & \underline{87.10\%} \\
\midrule
\mymethod & \textbf{0.51\%} & \textbf{0.82\%} & \textbf{9.54\%} & \textbf{10.61\%} & \textbf{82.45}\% & \textbf{89.90\%} \\
\cellcolor[gray]{0.9}\%Imp. & \cellcolor[gray]{0.9}920.00\% & \cellcolor[gray]{0.9}95.24\% & \cellcolor[gray]{0.9}318.42\% & \cellcolor[gray]{0.9}85.49\% & \cellcolor[gray]{0.9}5.19\% & \cellcolor[gray]{0.9}3.21\% \\
\bottomrule
\end{tabular}}
\end{table}

\begin{table}[t]
\centering
\caption{Performance comparsion on non-image datasets under the \textit{adaptive setting}. \mymethod outperforms the strongest baseline (underlined) using less than \textbf{\computationreduce} of its computation.}
\label{tab:nonimage_adapt}
\resizebox{0.47\textwidth}{!}{
\begin{tabular}{l|cc|cc|cc}
\toprule
\multirow{2}{*}{\textbf{Method}} & \multicolumn{2}{c|}{\textbf{TPR @ 0\%FPR}} & \multicolumn{2}{c|}{\textbf{TPR @ 0.1\%FPR}} & \multicolumn{2}{c}{\textbf{Balanced Accuracy}} \\
 & Purchase & Texas & Purchase & Texas & Purchase & Texas \\
\midrule
LiRA & 1.26\% & 10.28\% & 11.33\% & 24.24\% & 85.95\% & 90.10\% \\ 
Canary & \underline{1.28\%} & \underline{10.30\%} & \underline{11.43\%} & \underline{24.36\%} & \underline{85.99\%} & \underline{90.21\%} \\ 
RMIA & 0.41\% & 5.37\% & 4.73\% & 10.63\% & 84.62\% & 89.68\% \\ 
RAPID & 0.33\% & 9.64\% & 4.16\% & 15.73\% & 84.05\% & 90.05\% \\
\midrule
\mymethod & \textbf{1.98\%} & \textbf{10.98\%} & \textbf{16.57\%} & \textbf{24.88\%} & \textbf{86.71\%} & \textbf{90.83\%} \\
\midrule
\cellcolor[gray]{0.9}\%Imp. & \cellcolor[gray]{0.9}54.69\% & \cellcolor[gray]{0.9}6.60\% & \cellcolor[gray]{0.9}44.97\% & \cellcolor[gray]{0.9}2.13\% & \cellcolor[gray]{0.9}0.84\% & \cellcolor[gray]{0.9}0.69\% \\
\bottomrule
\end{tabular}}
\end{table}

\mypara{Attack on Non-Image Datasets}
We evaluate \mymethod on two widely-used non-image datasets, Purchase and Texas~\cite{sp17miashokri}, using a multilayer perceptron (MLP) as the target model and comparing it with leading MIAs.
Results in both non-adaptive and adaptive settings are presented in~\Cref{tab:nonimage_nonadapt} and~\Cref{tab:nonimage_adapt}, respectively.
As shown, \mymethod consistently outperforms existing attacks, particularly in terms of TPR at low FPR.
We notice that all evaluated MIAs perform worse on these non-image datasets than on image datasets like CIFAR-100, despite both having 100 classes.
We attribute this discrepancy to the smaller generalization gap in the models trained on non-image data.
For instance, our Purchase model shows a modest gap between its training and validation accuracies (97.7\% vs. 79.5\%), whereas the CIFAR-100 model exhibits a much larger gap indicative of significant overfitting (99.4\% vs. 68.9\%). 
Since MIAs typically exploit overfitting for attack, they are less effective against these better-generalized models.
Similar observations have been found in prior work~\cite{sp22lira,ndss26cpmia}.

\begin{table}[t]
\centering
\caption{Effectiveness of using DP-SGD against \mymethod on CIFAR-10 in the non-adaptive setting. $\sigma$ is the noise multiplier and $\varepsilon$ is the privacy budget for DPSGD.}
\label{tab:imia_dpsgd}
\resizebox{0.45\textwidth}{!}{
\begin{tabular}{ccccc}
\toprule
& \multicolumn{2}{c}{Clipping norm $C = 10$} & \multirow{2}{*}{\text{Model Acc}} & \multicolumn{1}{c}{\text{TPR @ 0.1 FPR\%}} \\
\cmidrule(lr){2-3}\cmidrule(lr){5-5}
& $\sigma$ & $\varepsilon$ & & \mymethod \\
\midrule
\text{No defense} & - & - & 90.41\% & 3.42\% \\
\midrule
 & 0 & $\infty$ & 81.46\% & 1.57\% \\
 & 0.2 & $>1000$ & 49.76\% & 0.32\% \\
\text{DP-SGD} & 0.5 & 31 & 37.55\% & 0.15\% \\
 & 1.0 & 4 & 32.10\% & 0.12\% \\
\bottomrule
\end{tabular}}
\end{table}

\begin{table}[t]
\centering
\caption{Performance of adaptive attacks against a ResNet model trained on CIFAR-10 using DP-SGD.}
\label{tab:pmia_dpsgd}
\resizebox{0.38\textwidth}{!}{
\begin{tabular}{ccccccc}
\toprule
 & \multicolumn{2}{c}{\text{TPR @ 0.1\% FPR }}  & \multicolumn{2}{c}{Balanced Accuracy} \\
 \cmidrule(lr){2-3}\cmidrule(lr){4-5}
 & $\sigma=0.2$ & $\sigma=0.5$ & $\sigma=0.2$ & $\sigma=0.5$ \\
\midrule
Calibration & 0.12\% & 0.11\% & 50.74\% & 50.07\% \\
Attack-R & 0.00\% & 0.00\% &  50.09\% & 50.06\% \\
Attack-D & 0.00\% & 0.00\% &  50.13\% & 50.60\% \\
SeqMIA & 0.07\% & 0.06\% &  50.05\% & 50.38\% \\
LiRA & 0.17\% & 0.11\% &  50.73\% & 50.22\% \\
Canary & 0.19\% & 0.11\% &  50.54\% & 50.28\% \\
RMIA & 0.15\% & 0.10\% &  50.42\% & 50.31\% \\
RAPID & 0.14\% & 0.11\% &  50.81\% & 50.76\% \\
 \midrule
\mymethod & \textbf{0.34\%} & \textbf{0.18\%}  & \textbf{51.05\%} & \textbf{50.82\%} \\
\bottomrule
\end{tabular}}
\end{table}

\mypara{Attack Against DP-SGD}
Differentially Private Stochastic Gradient Descent (DP-SGD)~\cite{ccs16dpsgd} is an effective defense mechanism against privacy attacks, including MIAs.
Following prior work\cite{sp22lira,ccs24rapid}, we evaluate the effectiveness of DP-SGD in defending against \mymethod.
Specifically, we fix the clipping norm $C$ to 10 and test \mymethod on a ResNet model trained on CIFAR-10 under the non-adaptive setting.
As shown in~\Cref{tab:imia_dpsgd}, even applying only gradient clipping (without noise addition) significantly reduces both model accuracy and the effectiveness of our attack.
We also assess the impact of DP-SGD in the adaptive setting in~\Cref{tab:pmia_dpsgd}, and observe consistent trends: as the privacy guarantee strengthens, all attacks experience reduced performance, and their performance gaps narrow.
Nevertheless, \mymethod consistently outperforms existing MIAs, especially in the low false-positive regime.

\section{Related Work}
\label{sec:related}

Neural networks have been known to be vulnerable to leaking sensitive information about their training datasets. 
Various attacks~\cite{usenix21llm_extract, ccs15inversion, ccs18property} have been proposed to quantify the extent of this data leakage and assess the associated risks.
In this paper, we focus on the membership inference attack (MIA)~\cite{sp17miashokri}, which aims to predict whether a specific data instance was included in a target model's training set. 
MIA has become a widely used tool to audit the data privacy of machine learning models~\cite{arxiv20privacy_meter, tensorflow_mia, sp21auditing, nips20auditing,ccs25synmeter}.
The first MIA against ML models was introduced by~\cite{sp17miashokri}, who also proposed the technique of shadow training.
MIAs and the shadow training have since been extended to various scenarios, including white-box~\cite{usenix20stolen, sp19comprehensive, icml19white}, black-box~\cite{long2018understanding, liu2022ml, usenix21systematic, ndss19ml-leak, arixv20revisiting, codaspy21mia}, label-only access~\cite{ccs21label, icml21label}, and knowledge distillation~\cite{nips23distill} settings. 
Existing black-box MIAs can be categorized into the following two groups.

\mypara{Non-Adaptive MIAs}
In this setting, the adversary trains shadow models (or avoids model training entirely) before accessing the membership queries.
Several techniques have been introduced, such as score functions~\cite{csf18privacy, usenix21systematic}, difficulty calibration~\cite{icml19white, iclr22calibrate, ccs24rapid}, loss trajectory~\cite{ccs24seqmia, ccs22trajectory}, model distillation~\cite{tifs25glira, ccs22enhanced}, quantile regression~\cite{nips23quantile}, and hypothesis testing~\cite{icml24rmia, ndss26cpmia}.
However, these attacks typically rely on target-agnostic shadow training and require training hundreds of shadow models to achieve good performance.

\mypara{Adaptive MIAs}
In the adaptive setting, the adversary can perform instance-by-instance behavioral analysis by training additional predictors~\cite{sp19comprehensive} or conducting parametric hypothesis testing~\cite{sp22lira, iclr23canary}.
A recent work~\cite{ndss26cpmia} goes beyond individual instance attacks by exploring the membership dependence among query instances.
However, these attacks suffer from high computational costs, and recent studies~\cite{icml24rmia, nips23quantile,usenix25free} argue that such adaptive attacks are impractical, as they require training new models for each batch of queries.

\mypara{Model Distillation in MIAs}
Model distillation~\cite{arxiv15distill,ijcai21distill} has been used in MIAs in two main approaches: 
(i) it is used to train more reliable shadow out models for inference, as explored in~\cite{ccs22enhanced,tifs25glira};
(ii) it serves as a tool to simulate the intermediate learning dynamics of the target model, and derive membership signals from this simulated learning process~\cite{ccs24seqmia,ccs22trajectory}.
However, these methods use standard distillation techniques and only model the target's behavior on non-members.
In contrast, we propose an imitative training approach tailored for MIAs, which mimics the target model's behavior for both members and non-members, significantly enhancing both attack efficacy and efficiency.

\mypara{Evaluation of MIAs}
Early studies~\cite{sp17miashokri, csf18privacy, ndss19ml-leak} evaluated membership inference as a binary classification task using metrics like balanced accuracy.
More recent research~\cite{sp22lira, eurosp20pragmatic} has emphasized evaluating attacks by true positive rate (TPR) at a very low false positive rate (FPR), and most MIAs perform poorly under this metric.
In line with the current consensus in the MIA literature, we also adopt TPR at low FPRs as the primary evaluation metric in this paper.

\section{Conclusion}
\label{sec:conclusion}

In this paper, we introduce \mymethod, a new MIA built upon a novel imitative training technique to train target-informed imitative models for inference.
Imitative training addresses the high variation of shadow models by strategically distilling membership signals from the target model. 
Extensive experiments in various attack settings demonstrate the effectiveness and efficiency of \mymethod.
One limitation of our paper is that it mainly focuses on classification models, leaving the generalization to generative models~\cite{arxiv17logan, icml23diffusion, arxiv24llmmia} unexplored.
Nonetheless, our work provides an efficient and practical attack, opening new directions for future research in MIAs.

\mypara{Acknowledgments}
This work was funded in part by the National Science Foundation (NSF) awards, CNS-2212160, CNS-2504819,  CNS-2247794, and CNS-2207204, Amazon Research Award, and CISCO Research Award.  Any opinions, findings, and conclusions or recommendations expressed in this material are those of the authors and do not necessarily reflect the views of the sponsors.

\section*{Ethical Considerations}

Our work introduces Imitative Membership Inference Attack (\mymethod), a more powerful and efficient approach to membership inference. 
We recognize our responsibility to carefully assess its ethical implications. 
Our analysis follows the stakeholder-based framework recommended by USENIX Security and is guided by the principles of The Menlo Report.

\mypara{Stakeholder-Based Analysis}
We identify several key stakeholders impacted by this research:

\begin{itemize}
    \item \textit{Machine Learning Practitioners.} 
    This is our primary audience. We provide ML practitioners with a more effective tool for privacy auditing, allowing them to identify vulnerabilities in their models and improve privacy protections. 
    \item \textit{Data Subjects.} 
    The publication of a more effective MIA could theoretically be used to infer personal information, leading to privacy violations. 
    However, our experiments exclusively use public, non-sensitive datasets, and we do not target any individuals or proprietary models. 
    Moreover, by publishing these attacks and their potential mitigations, as described below, our work helps strengthen the overall privacy and security of the deep learning ecosystem.
    \item \textit{Adversaries.} 
    Our work could potentially be used by adversaries to exploit privacy vulnerabilities. 
    In recognition of this, we discuss the mitigations below to minimize this risk.
\end{itemize}

\mypara{Mitigations}
We have taken the following concrete steps to mitigate potential harms:

\begin{itemize}
    \item \textit{Exclusive Use of Public, Non-Sensitive Data.} 
    Our research was conducted exclusively on public datasets (\eg CIFAR-10), which do not contain Personally Identifiable Information (PII). We did not use any private or sensitive data, nor did we target specific individuals or proprietary models, ensuring no one's privacy was violated during our research.
    \item \textit{Validation of Existing Defenses.}
    We evaluate that established defenses like DP-SGD remain an effective mitigation strategy against \mymethod. 
    By demonstrating that practical defenses exist, we lower the privacy risk introduced by our work and provide clear guidance for developers.
\end{itemize}

\mypara{Justification for Research and Publication}
The vulnerabilities that \mymethod exploits are fundamental to current deep learning practices.
It is highly likely that adversaries could independently discover similar techniques, and we believe it is crucial to proactively disclose this attack to the research community.
By publishing our findings, we enable defenders to address the vulnerabilities in their models before they are exploited in real-world scenarios.

We emphasize that the responsible use of such attack methods should be used responsibly to strengthen the security and privacy of ML systems, rather than maliciously exploiting them. 
We encourage the research community to use our findings in a manner that promotes ethical research and enhances privacy protections for users and data subjects.

\section*{Open Science}


To facilitate reproducibility, we have made our research artifacts publicly available.
The code repository is hosted at \url{https://github.com/zealscott/IMIA} and archived at \url{https://doi.org/10.5281/zenodo.17885393}.
The repository contains a detailed \texttt{README.md} with step-by-step instructions for environment setup and experiment execution. 
We also provide automated scripts for training both shadow and imitative models in adaptive and non-adaptive settings. 
The training process automatically handles the downloading of evaluated datasets. 
For attack evaluation, a unified interface, \texttt{run\_attack.py}, allows users to execute \mymethod alongside baselines seamlessly.
The codebase is modularized for extensibility: model architectures are located in the \texttt{models/} directory, attack implementations are in the \texttt{attacks/} directory, and hyperparameters are stored in the \texttt{config/} directory.






\bibliographystyle{plain}
\bibliography{ref}

\appendix

\section{Experimental Setups}

\subsection{Dataset Description}
\label{appendix:data_description}

\mypara{MNIST}
MNIST~\cite{mnist} is a benchmark dataset for evaluating handwritten digit classification algorithms. It contains 60,000 grayscale images of 28×28 pixels for training and 10,000 images for testing. The dataset consists of 10 classes, representing digits from 0 to 9.

\mypara{Fashion-MNIST}
Fashion-MNIST (FMNIST)~\cite{fmnist} is a dataset designed as a more challenging replacement for MNIST, containing 60,000 grayscale images of size 28×28 pixels for training and 10,000 images for testing. It consists of 10 classes representing different fashion items, such as t-shirts, trousers, and sneakers.

\mypara{CIFAR10} 
CIFAR-10~\cite{cifar10} is a benchmark dataset for general image classification tasks, containing 60,000 color images of size 32×32 pixels, equally distributed across 10 classes, including animals like cats, dogs, and birds, as well as vehicles like airplanes and trucks.

\mypara{CIFAR-100}
CIFAR-100~\cite{cifar10} is a dataset similar to CIFAR-10 but with a greater level of complexity, as it contains 100 classes instead of 10. It also includes 60,000 color images of size 32×32 pixels, with each class containing 600 images.

\mypara{Purchase}
Purchase~\cite{sp17miashokri} is a dataset of shopping records with 197,324 samples of 600 dimensions. Following
previous works~\cite{sp17miashokri,ccs24seqmia,sp22lira}, we cluster these data into 100 classes to train a Multi-Layer Perceptron (MLP) classifier.

\mypara{Texas}
This dataset is to predict a patient’s primary medical procedure using 6,170 binary features. Following~\cite{sp17miashokri}, the data comprises records from 67,330 patients, with the 100 most frequent procedures used as classification labels.

\begin{table}[t]
    \centering
    \caption{Prediction accuracy of used datasets.}
    \label{tab:dataset_acc}
    \resizebox{0.499\textwidth}{!}{
    \begin{tabular}{lcccccccccccc}
    \toprule
    \multirow{2}{*}{Model} & \multicolumn{2}{c}{MNIST} & \multicolumn{2}{c}{FMNIST} & \multicolumn{2}{c}{CIFAR-10} & \multicolumn{2}{c}{CIFAR-100} & \multicolumn{2}{c}{Purchase} & \multicolumn{2}{c}{Texas} \\
    \cmidrule(lr){2-3} \cmidrule(lr){4-5} \cmidrule(lr){6-7} \cmidrule(lr){8-9} \cmidrule(lr){10-11} \cmidrule(lr){12-13}
     & Train & Val & Train & Val & Train & Val & Train & Val & Train & Val & Train & Val \\
    \midrule
    ResNet & 100.0\% & 99.5\% & 100.0\% & 94.8\% & 99.7\% & 90.4\% & 99.4\% & 68.9\% & - & - & - & - \\
    VGG16 & 100.0\% & 99.7\% & 100.0\% & 95.6\% & 99.9\% & 91.7\% & 99.6\% & 72.3\% & - & - & - & - \\
    DenseNet121 & 100.0\% & 99.6\% & 100.0\% & 96.1\% & 99.9\% & 90.2\% & 99.9\% & 71.5\% & - & - & - & - \\
    MobileNetV2 & 100.0\% & 99.5\% & 100.0\% & 94.9\% & 99.9\% & 97.1\% & 100.0\% & 72.4\% & - & - & - & - \\
    MLP & - & - & - & - & - & - & - & - & 97.7\% & 79.5\% & 96.5\% & 78.7\% \\
    \bottomrule
    \end{tabular}}
\end{table}

\begin{table}[t]
    \centering
    \caption{Data partitioning scheme. The query dataset $D_\text{query}$ and the auxiliary dataset $D_\text{auxiliary}$  are strictly disjoint. The target model is trained on a randomly sampled subset from $D_\text{query}$. The reference dataset is used by some MIAs~\cite{ccs24rapid,ccs24seqmia} to train their attack models.}
    \label{tab:dataset_split}
    \resizebox{0.45\textwidth}{!}{
    \begin{tabular}{lccccc}
    \midrule
    \multirow{2}{*}{Dataset} & \multicolumn{2}{c}{$D_\text{query}$} & \multicolumn{3}{c}{$D_\text{auxiliary}$}  \\
    \cmidrule(lr){2-3} \cmidrule(lr){4-6} 
    & Train & Validation & Train & Validation & Reference  \\
    \midrule
    MNIST & 11,667  & 11,667 & 11,667  & 11,667 & 23,334  \\
    \midrule
    FMNIST & 11,667  & 11,667 & 11,667  & 11,667 & 23,334  \\
    \midrule
    CIFAR-10 & 10,000  & 10,000 & 10,000  & 10,000 & 20,000 \\
    \midrule
    CIFAR-100 & 10,000  & 10,000 & 10,000  & 10,000 & 20,000 \\
    \midrule
    Purchase & 32,887 & 32,887 & 32,887 & 32,887 & 65,774 \\ 
    \midrule
    Texas & 11,221 & 11,221 & 11,221 & 11,221 & 22,443 \\ 
    \midrule
    \end{tabular}}
\end{table}

\subsection{Accuracy of the Trained Target Model}
\label{appendix:acc_model}
We use validation accuracy thresholds to determine whether a trained model needs to be rerun. 
Specifically, we require a minimum validation accuracy of 0.8 for MNIST and FMNIST, 0.6 for CIFAR-10, and 0.4 for CIFAR-100. 
If a model fails to meet the corresponding threshold, we retrain it using a different random seed.
We report the training and validation accuracies of the target model in~\Cref{tab:dataset_acc}.

\subsection{Number of Shadow Models of Baselines}
\label{appendix:baselines}

A critical hyperparameter for shadow-based MIAs is the number of trained shadow models. 
For each baseline method, we adopted the number of shadow models recommended in the original paper, if available. 
When a number was not specified, we select the maximum number of shadow models until the attack performance saturates to ensure its optimal attack performance.
Specifically, we use the following configurations: 10 shadow models for Calibration~\cite{iclr22calibrate} and SeqMIA~\cite{ccs24seqmia}; 128 shadow models for Canary~\cite{iclr23canary}, GLiRA~\cite{tifs25glira}, RMIA~\cite{icml24rmia}, RAPID~\cite{ccs24rapid}; and 256 shadow models for Attack-R~\cite{ccs22enhanced}, Attack-D~\cite{ccs22enhanced}, LiRA~\cite{sp22lira}, and PMIA~\cite{ndss26cpmia}.
The number of shadow models is consistent across all datasets. 
\section{Additional Experimental Results}

\begin{figure}[t]
    \centering
    \subfigure[Non-Adaptive setting]
    {
    \includegraphics[width=0.46\linewidth]{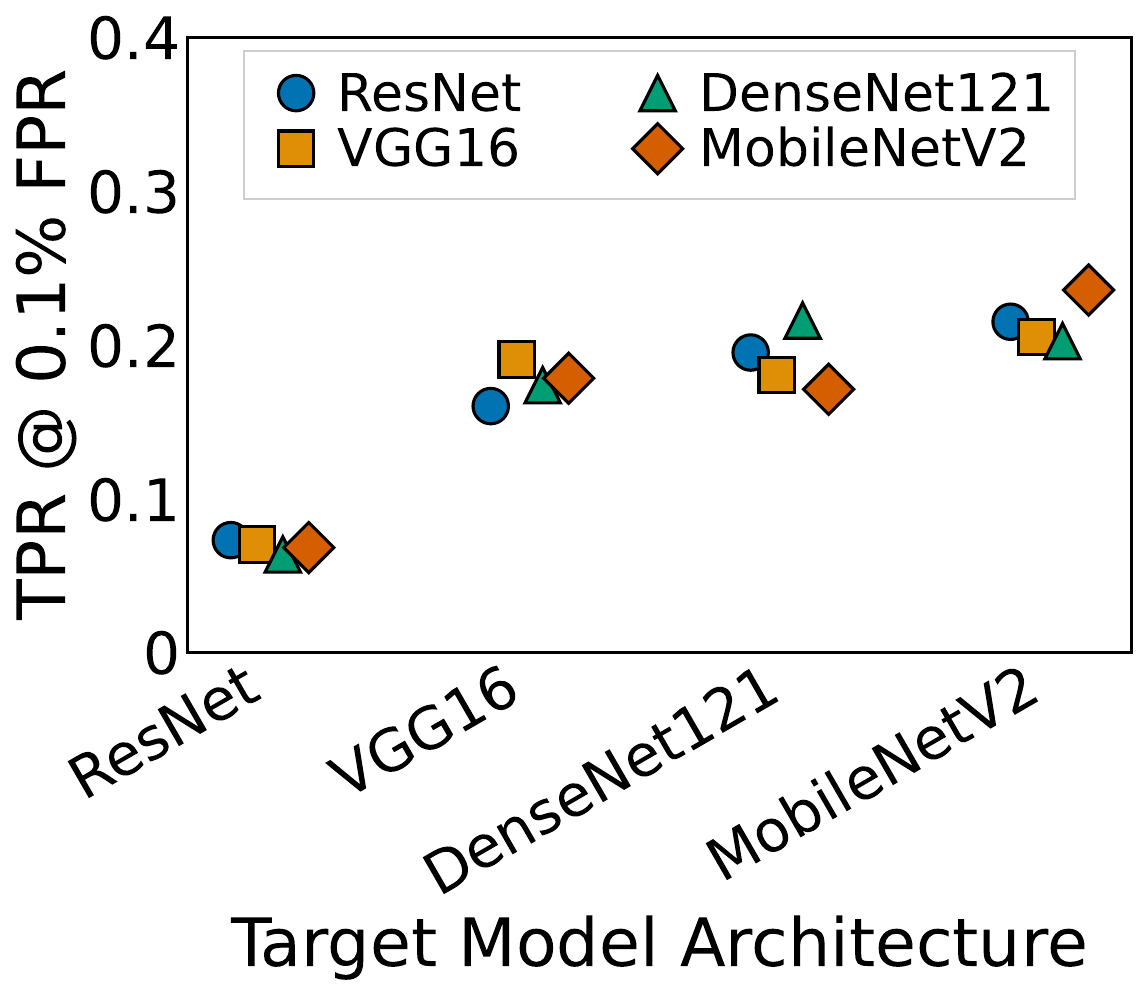}
    \label{fig:nonadaptive_arch}
    }
    \subfigure[Adaptive setting]
    {
    \includegraphics[width=0.46\linewidth]{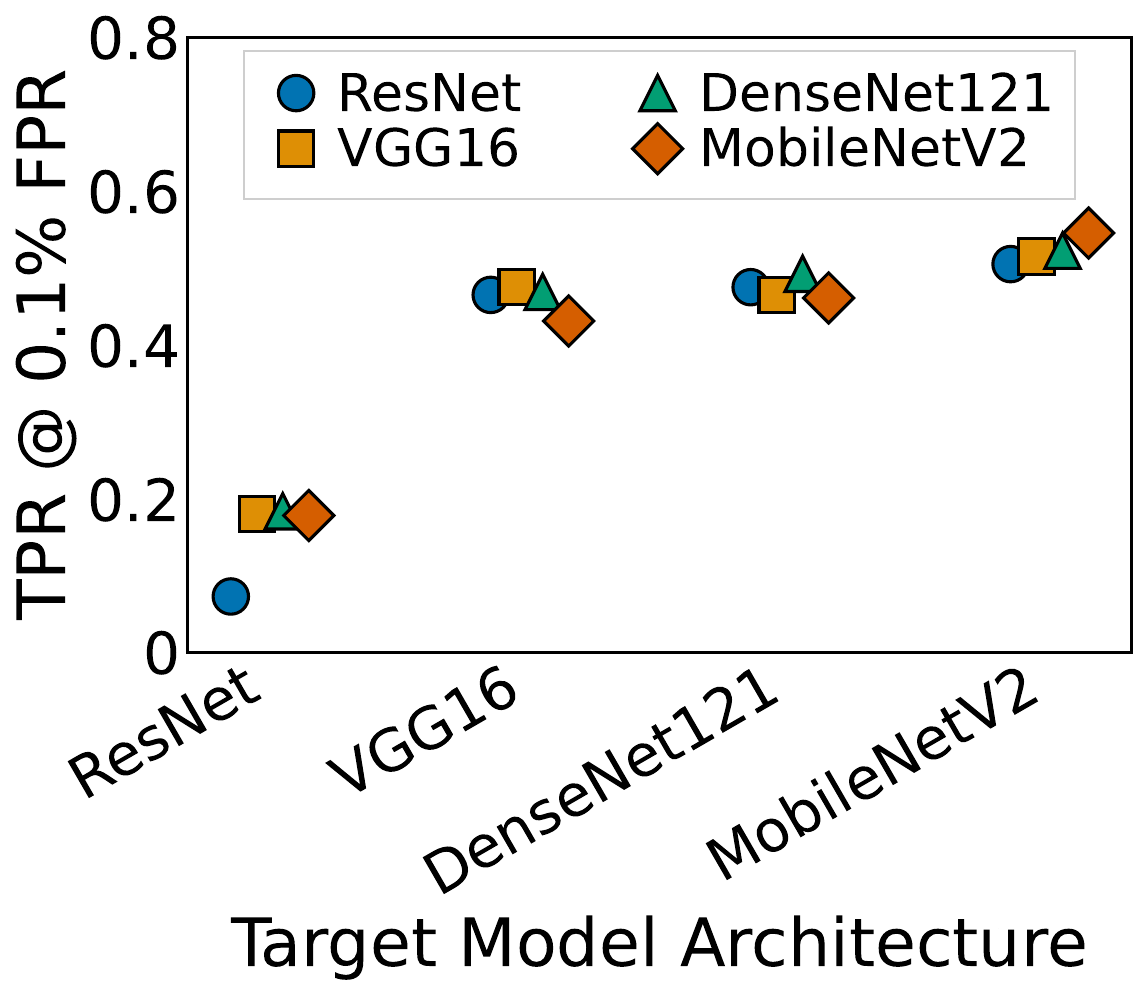}
    \label{fig:adaptive_arch}
    }
    \caption{The impact of architecture differences between the target model and the imitative models trained on CIFAR-100.}
    \label{fig:impact_arch}
\end{figure}

\begin{figure}[t]
    \centering
    \subfigure[Balanced Accuracy]
    {
    \includegraphics[width=0.46\linewidth]{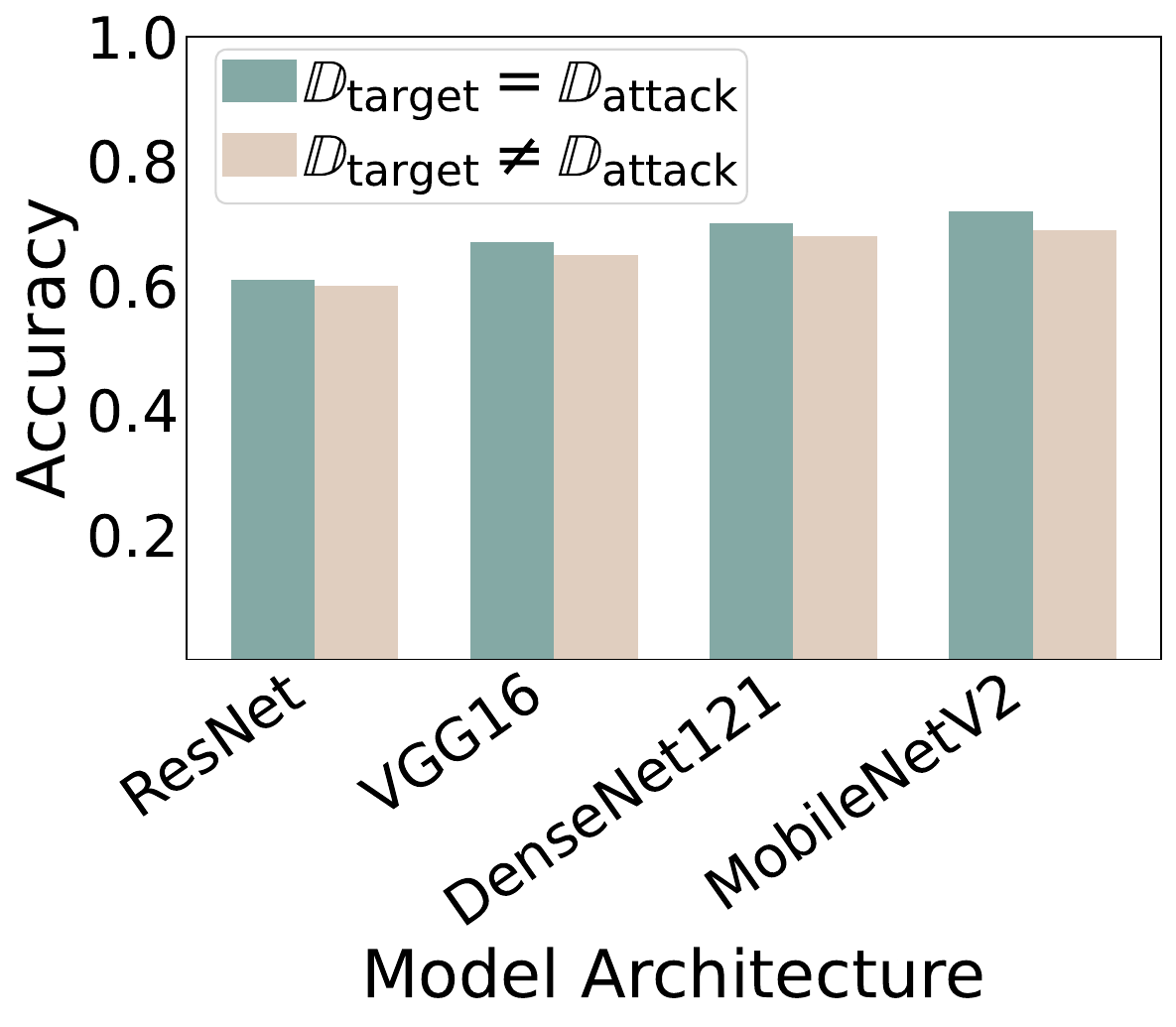}
    \label{fig:shift_accuracy}
    }
    \subfigure[TPR @ 0.1\% FPR]
    {
    \includegraphics[width=0.46\linewidth]{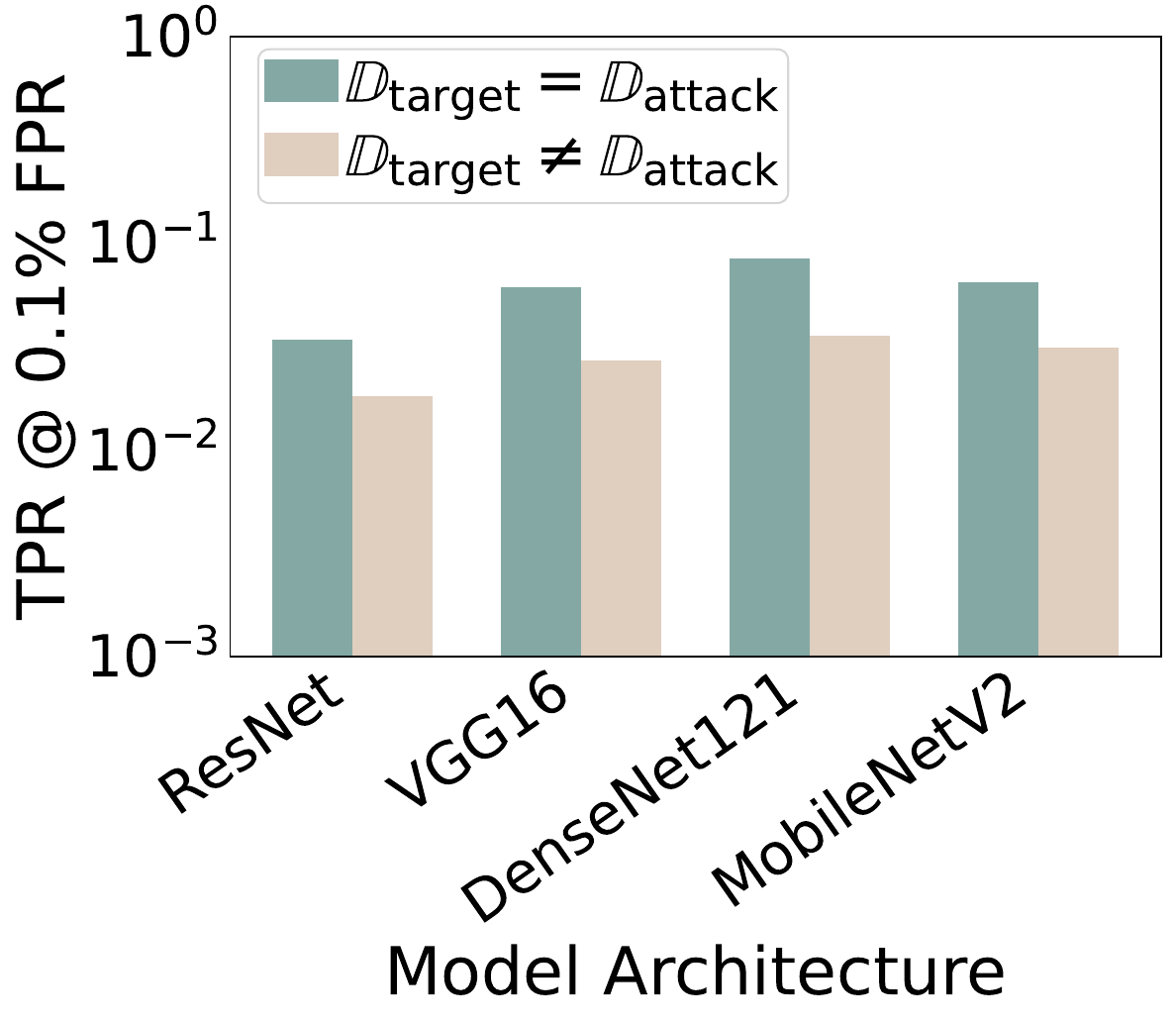}
    \label{fig:shift_tpr}
    }
    \caption{The impact of distribution shift between the target model training dataset and the attacker's dataset.}
    \label{fig:impact_shift}
\end{figure}

\begin{figure*}[t]
    \centering
    \includegraphics[width=0.99\linewidth]{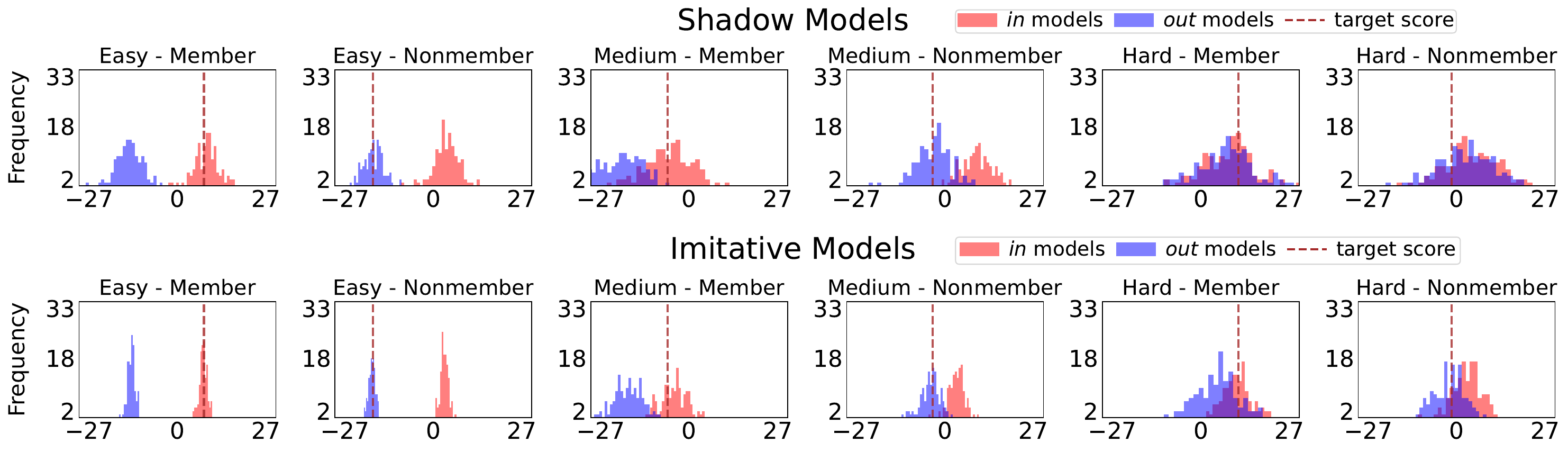}
    \caption{Distributions of scaled confidence scores for six additional CIFAR-100 instances with varying attack difficulty (easy, medium, hard). The dashed vertical line represents each instance’s score on the target model. \textbf{Top row}: \textit{target-agnostic} shadow models show high predictive variance (with long tails and wide distributions) for both members and non-members, resulting in significant overlap that hampers reliable inference, especially for hard-to-attack instances.
    \textbf{Bottom row}: \textit{target-informed} imitative models exhibit more stable and well-separated distributions, enabling effective inference across all levels of difficulty.}
    \label{fig:demo_model_stability_appendix}
\end{figure*}

\begin{figure*}[t]
    \centering
    \subfigure[Residuals of target-agnostic \textbf{shadow} models on \textit{members}.]
    {
    \includegraphics[width=0.22\linewidth]{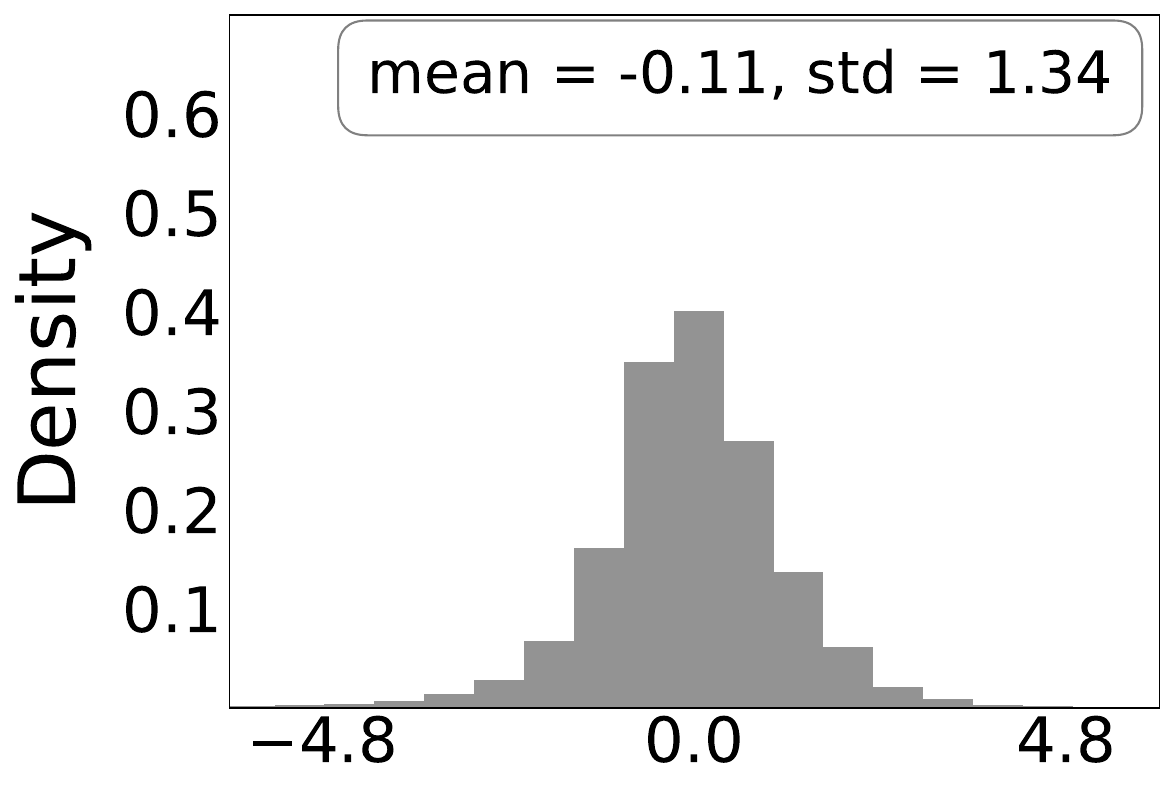}
    }
    \hspace{0.5mm}
    \subfigure[Residuals of target-informed \textbf{imitative} models on \textit{members}.]
    {
    \includegraphics[width=0.22\linewidth]{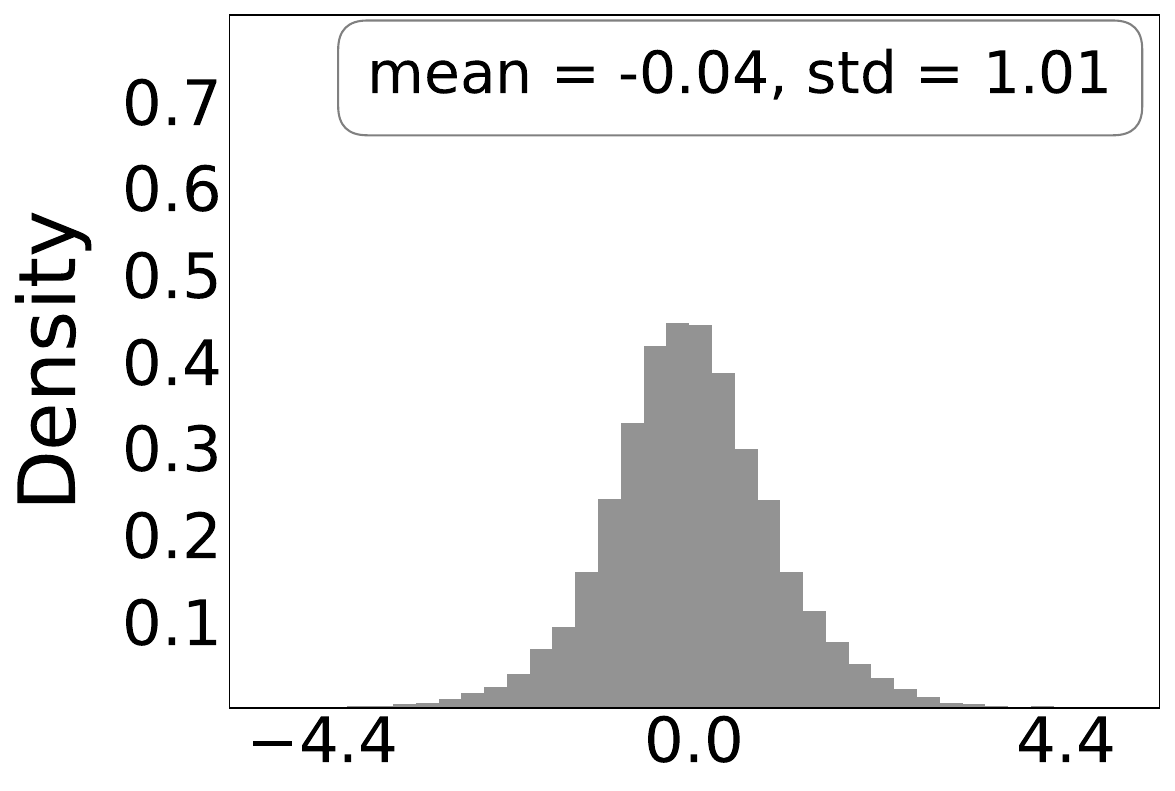}
    }
    \hspace{0.5mm}
    \subfigure[Residuals of target-agnostic \textbf{shadow} models on \textit{non-members}.]
    {
    \includegraphics[width=0.22\linewidth]{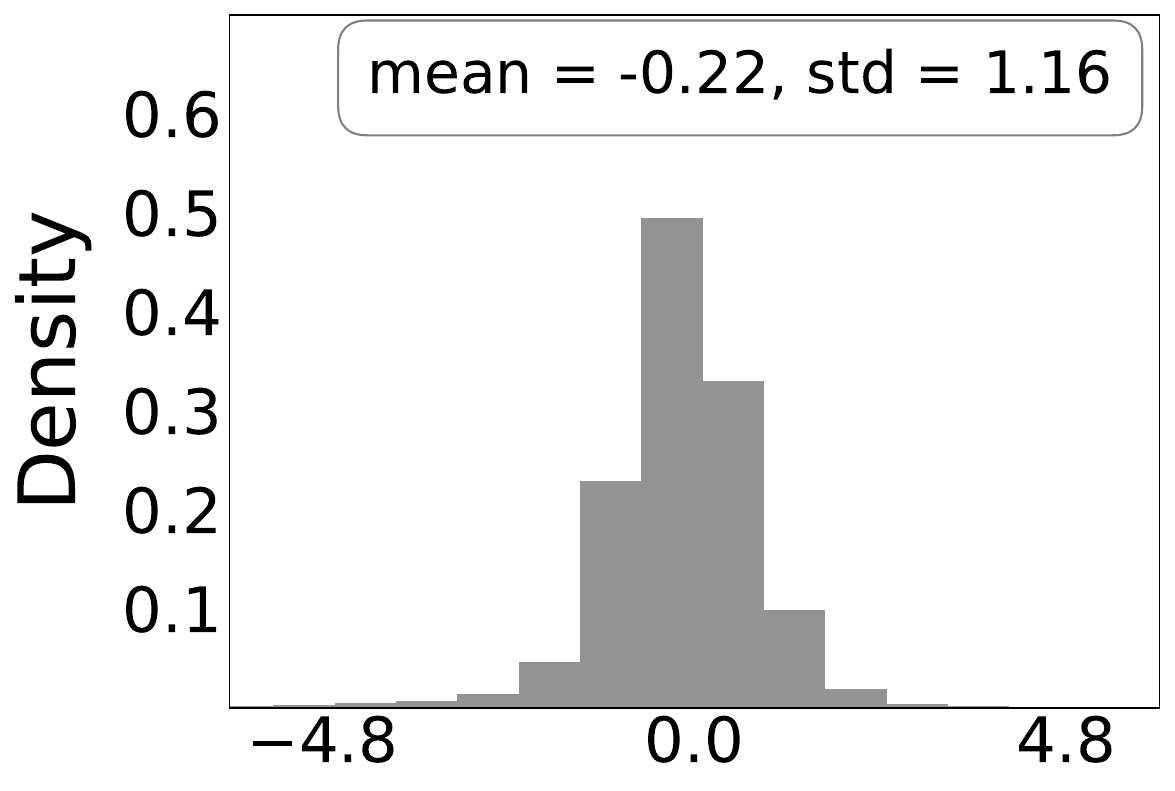}
    }
    \hspace{0.5mm}
    \subfigure[Residuals of target-informed \textbf{imitative} models on \textit{non-members}.]
    {
    \includegraphics[width=0.22\linewidth]{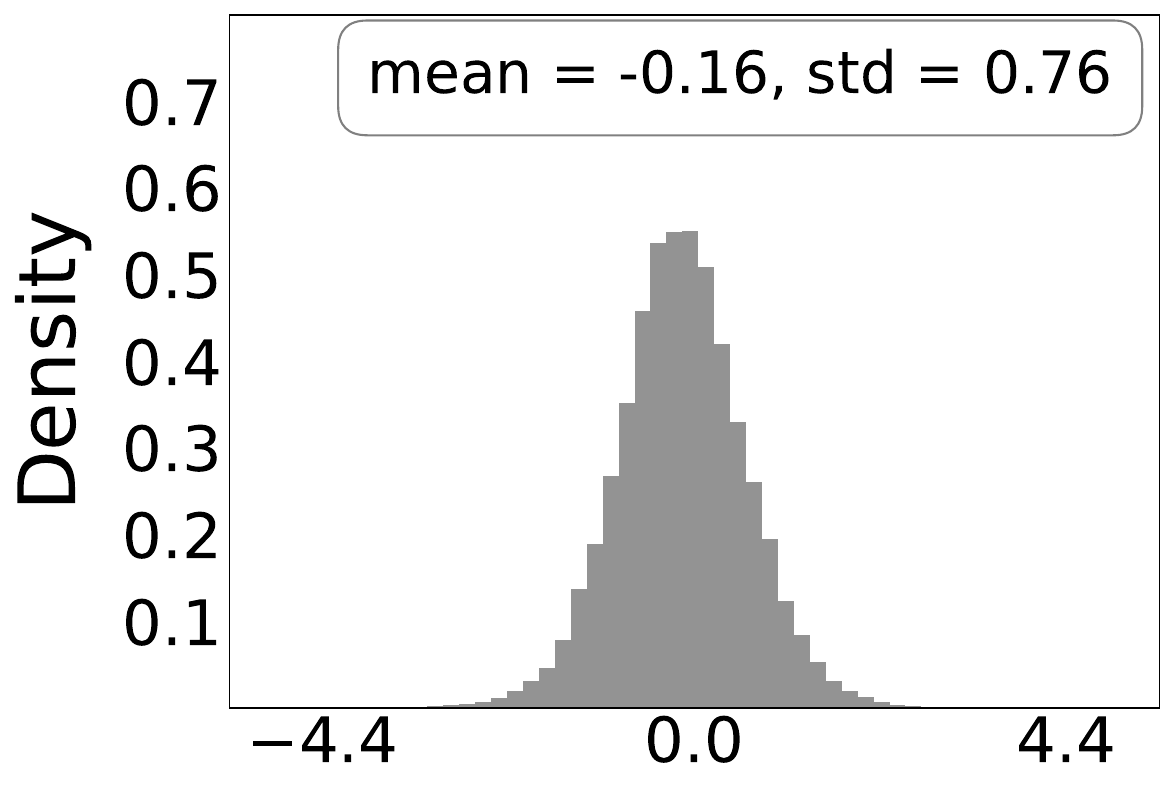}
    }
    \caption{Normalized residual distributions  of shadow models vs.\ imitative models on MNIST.}
    \label{fig:z_score_hist_mnist}
\end{figure*}

\begin{figure*}[t]
    \centering
    \subfigure[Residuals of target-agnostic \textbf{shadow} models on \textit{members}.]
    {
    \includegraphics[width=0.22\linewidth]{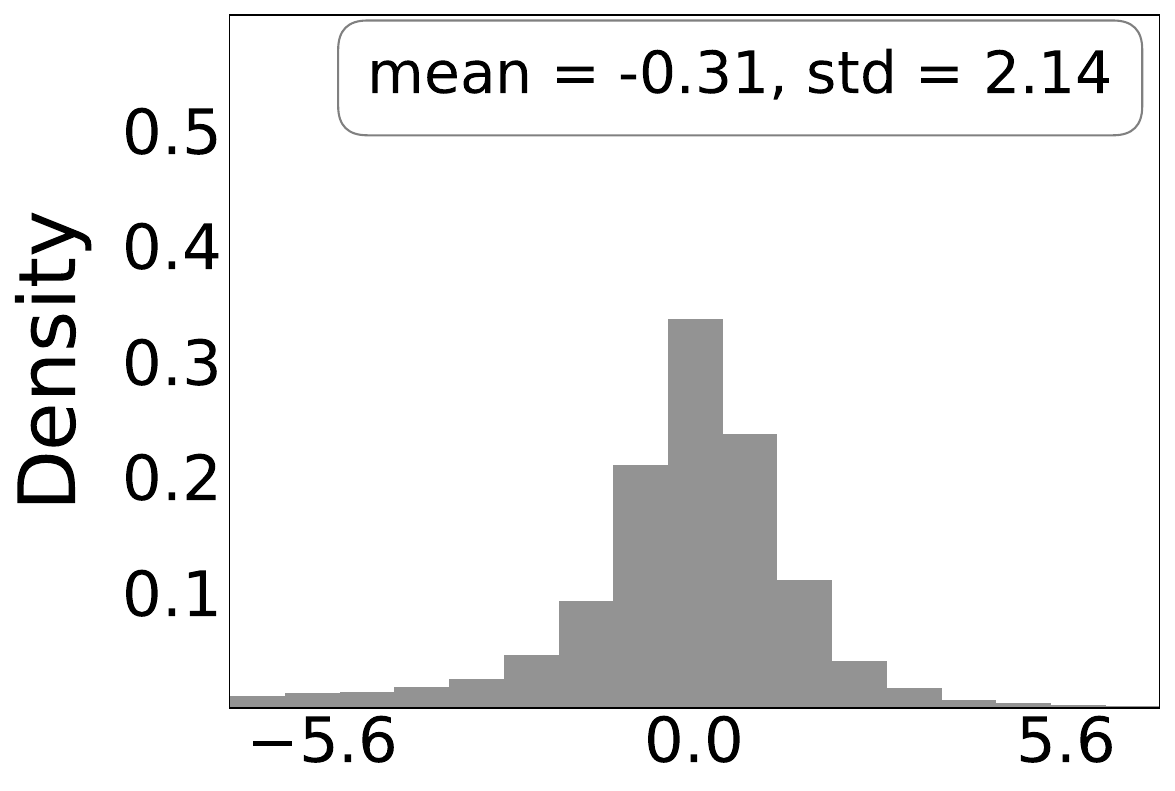}
    }
    \hspace{0.5mm}
    \subfigure[Residuals of target-informed \textbf{imitative} models on \textit{members}.]
    {
    \includegraphics[width=0.22\linewidth]{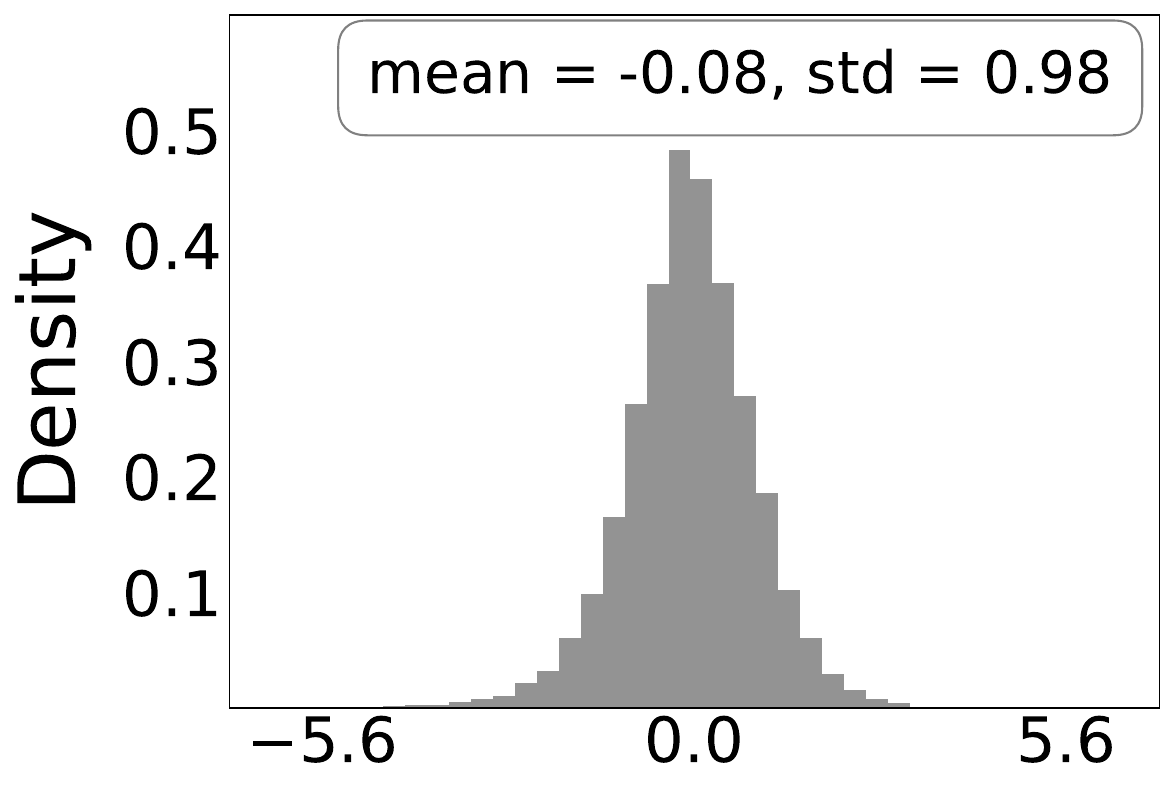}
    }
    \hspace{0.5mm}
    \subfigure[Residuals of target-agnostic \textbf{shadow} models on \textit{non-members}.]
    {
    \includegraphics[width=0.22\linewidth]{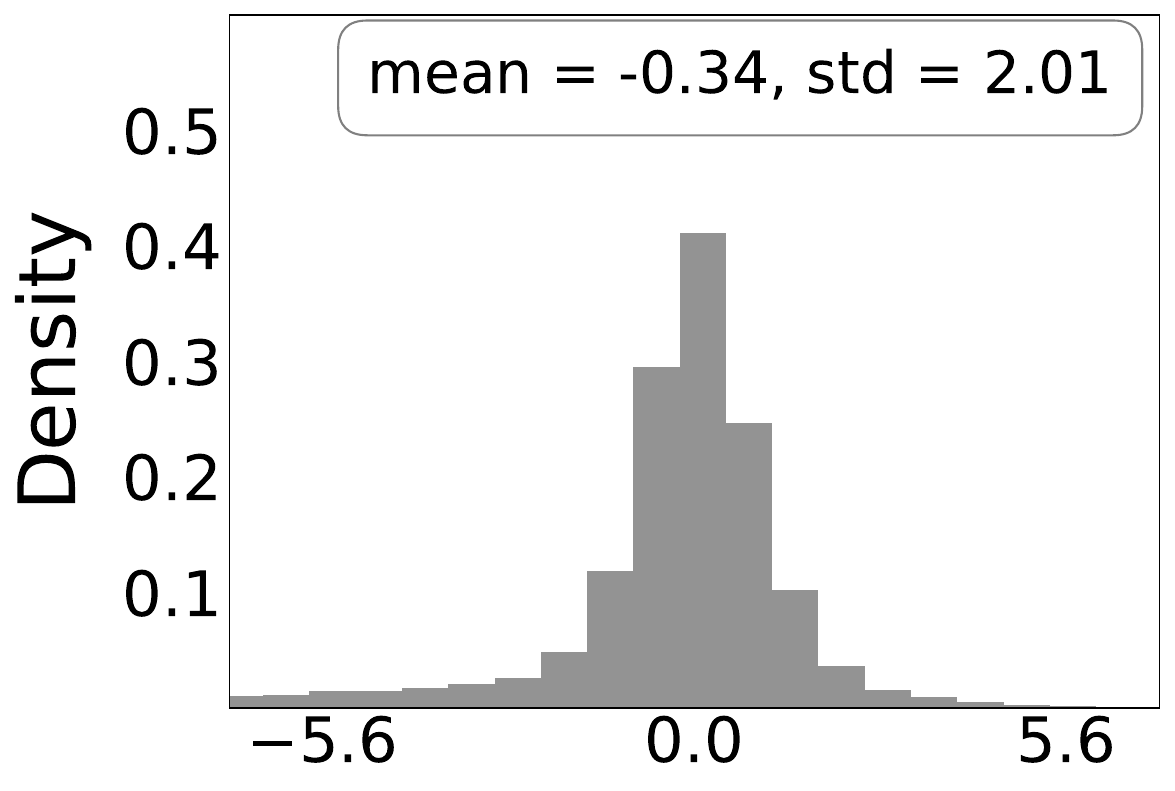}
    }
    \hspace{0.5mm}
    \subfigure[Residuals of target-informed \textbf{imitative} models on \textit{non-members}.]
    {
    \includegraphics[width=0.22\linewidth]{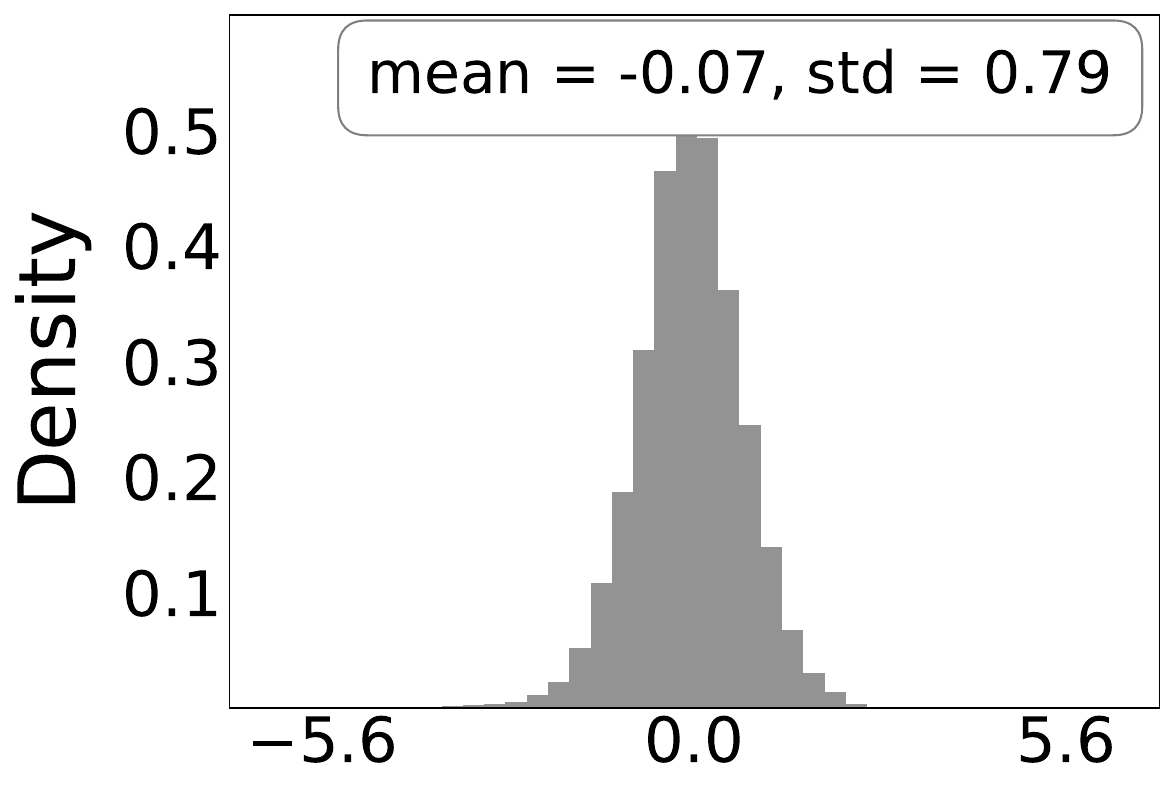}
    }
    \caption{Normalized residual distributions  of shadow models vs.\ imitative models on FMNIST.}
    \label{fig:z_score_hist_fmnist}
\end{figure*}

\begin{figure*}[t]
    \centering
    \subfigure[Residuals of target-agnostic \textbf{shadow} models on \textit{members}.]
    {
    \includegraphics[width=0.22\linewidth]{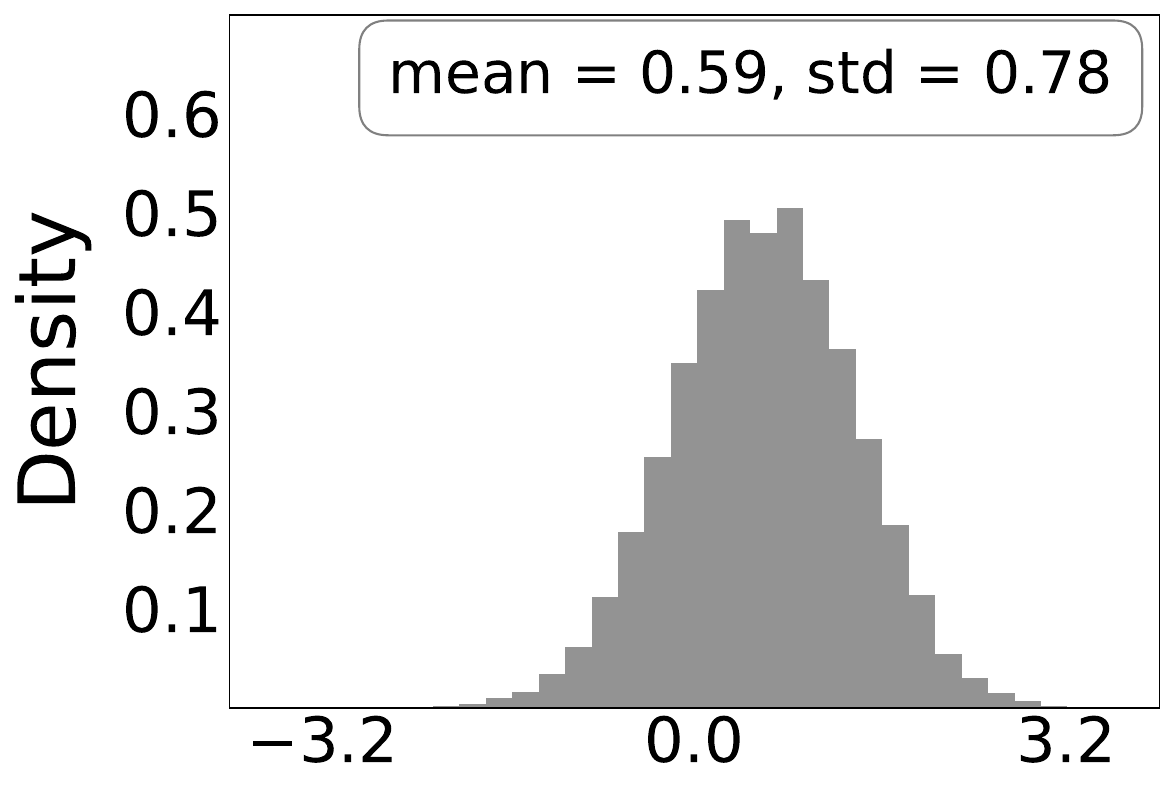}
    }
    \hspace{0.5mm}
    \subfigure[Residuals of target-informed \textbf{imitative} models on \textit{members}.]
    {
    \includegraphics[width=0.22\linewidth]{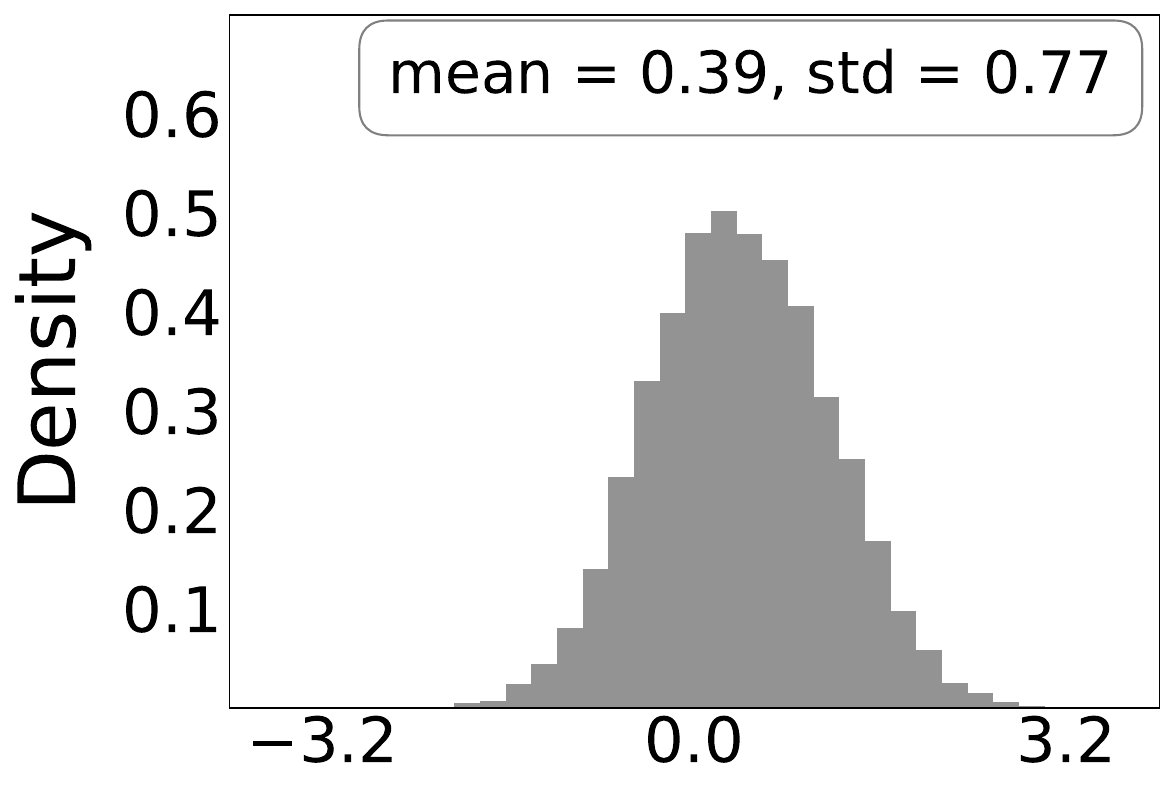}
    }
    \hspace{0.5mm}
    \subfigure[Residuals of target-agnostic \textbf{shadow} models on \textit{non-members}.]
    {
    \includegraphics[width=0.22\linewidth]{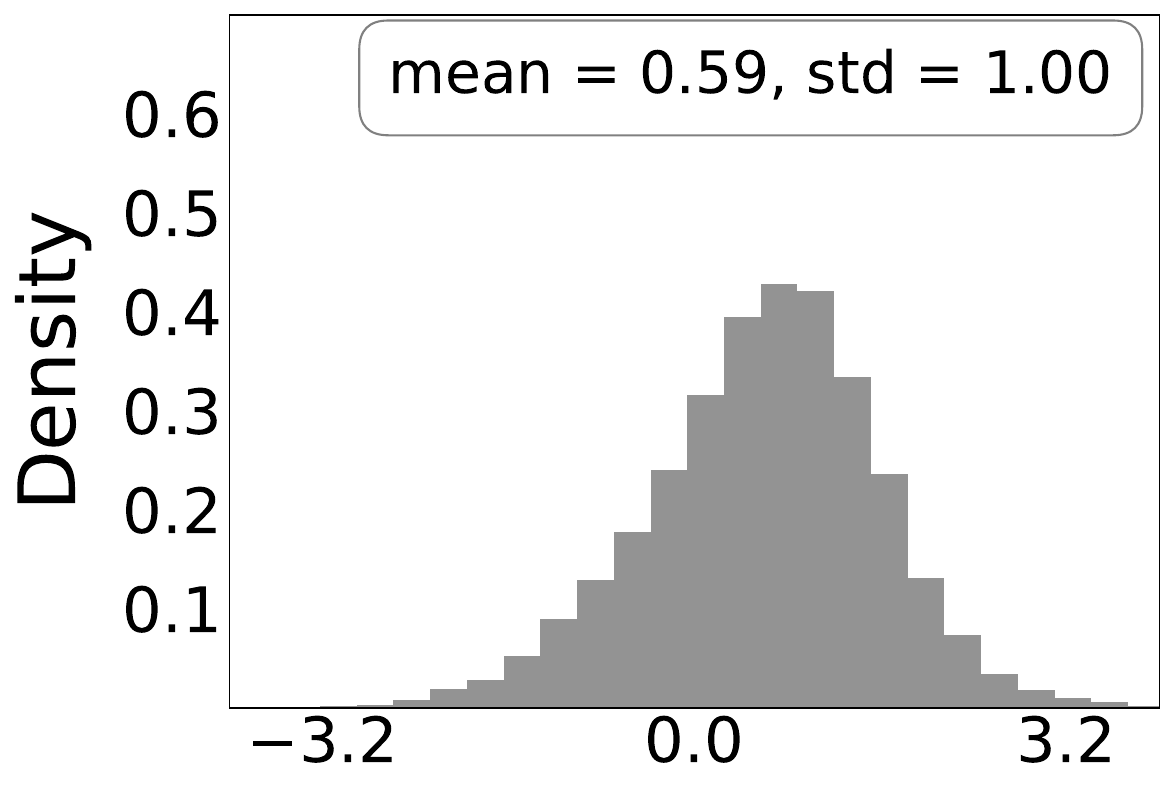}
    }
    \hspace{0.5mm}
    \subfigure[Residuals of target-informed \textbf{imitative} models on \textit{non-members}.]
    {
    \includegraphics[width=0.22\linewidth]{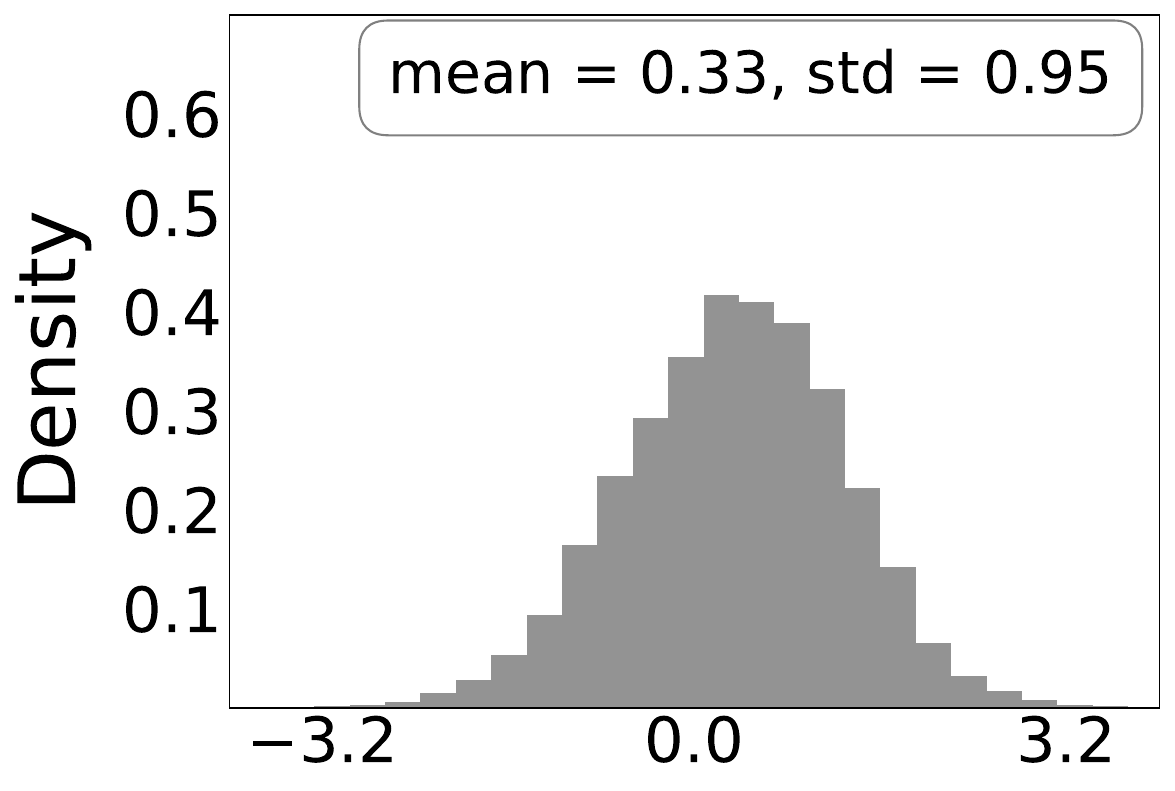}
    }
    \caption{Normalized residual distributions  of shadow models vs.\ imitative models on CIFAR-10.}
    \label{fig:z_score_hist_cifar10}
\end{figure*}

\begin{figure*}[t]
    \centering
    \subfigure[Residuals of target-agnostic \textbf{shadow} models on \textit{members}.]
    {
    \includegraphics[width=0.22\linewidth]{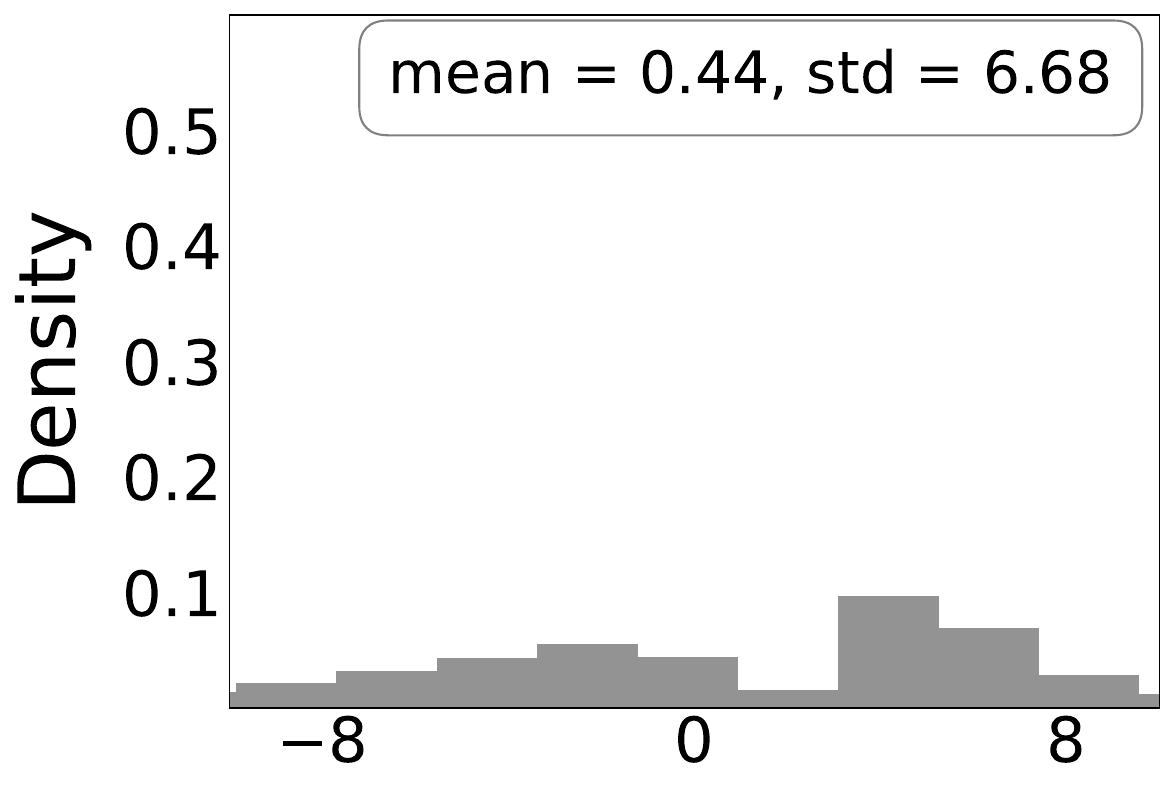}
    }
    \hspace{0.5mm}
    \subfigure[Residuals of target-informed \textbf{imitative} models on \textit{members}.]
    {
    \includegraphics[width=0.22\linewidth]{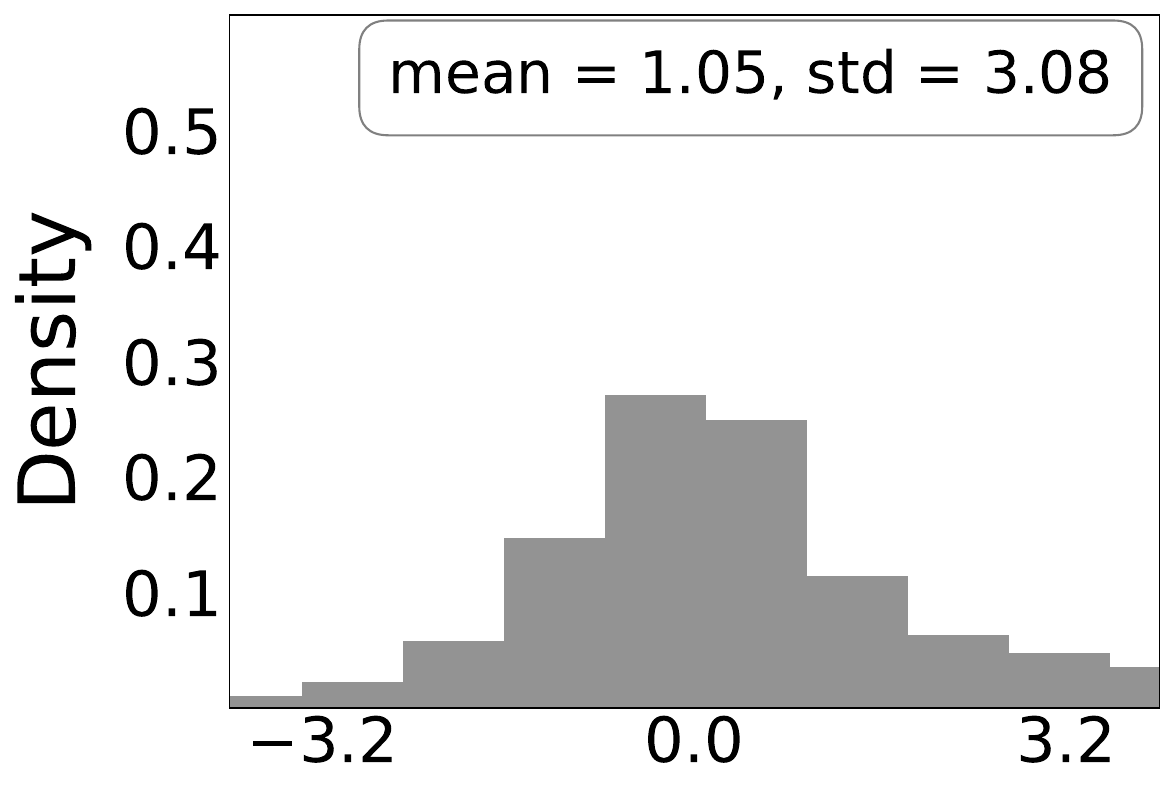}
    }
    \hspace{0.5mm}
    \subfigure[Residuals of target-agnostic \textbf{shadow} models on \textit{non-members}.]
    {
    \includegraphics[width=0.22\linewidth]{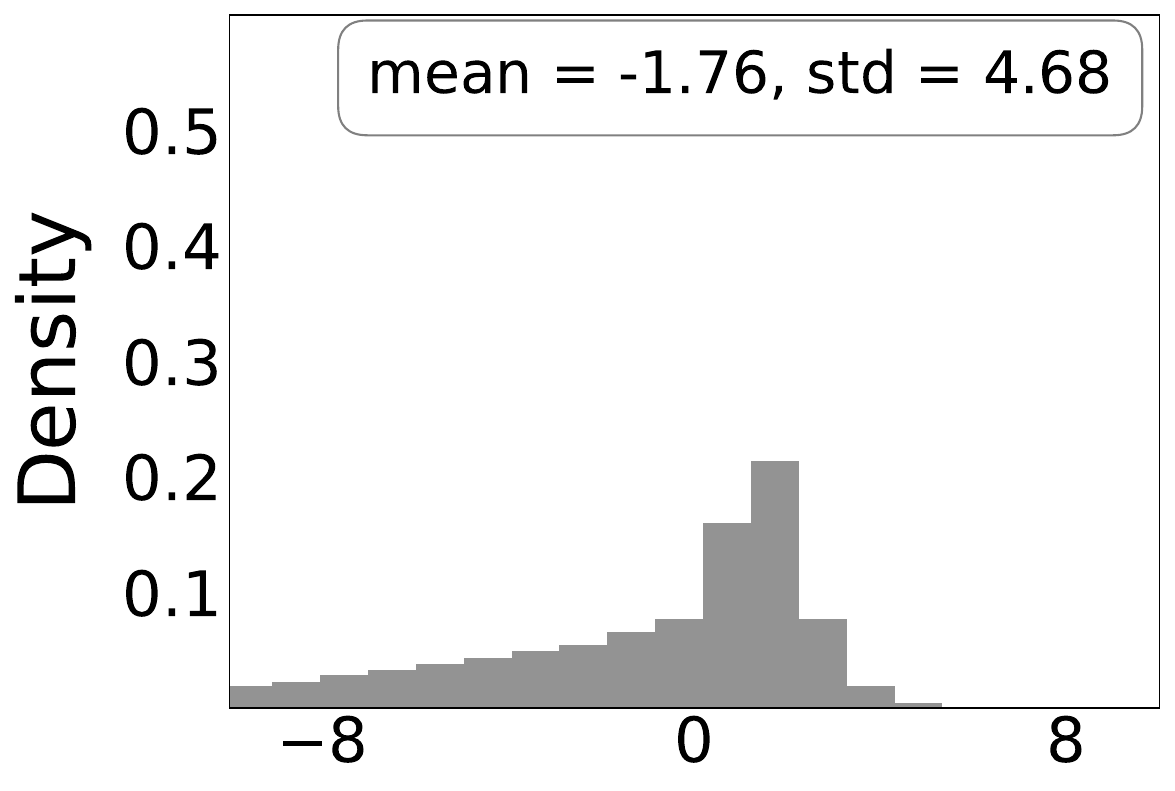}
    }
    \hspace{0.5mm}
    \subfigure[Residuals of target-informed \textbf{imitative} models on \textit{non-members}.]
    {
    \includegraphics[width=0.22\linewidth]{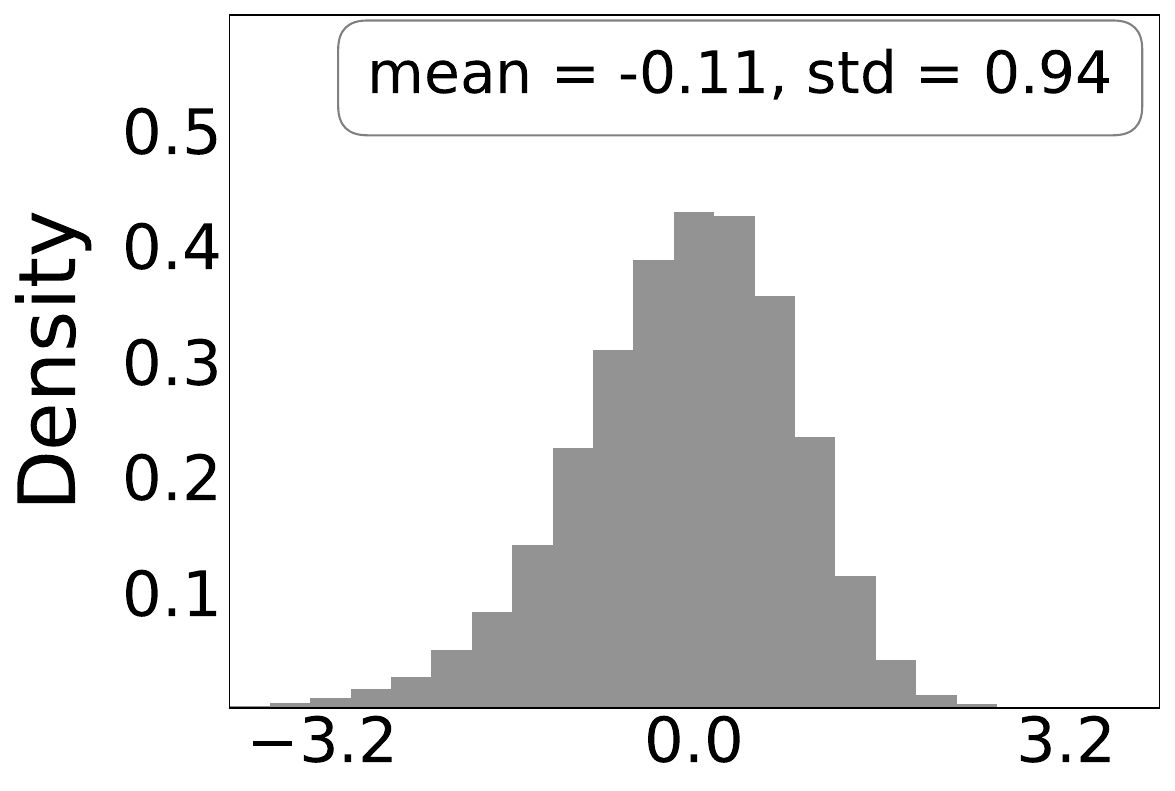}
    }
    \caption{Normalized residual distributions  of shadow models vs.\ imitative models on Purchase.}
    \label{fig:z_score_hist_purchase}
\end{figure*}

\begin{figure*}[t]
    \centering
    \subfigure[Residuals of target-agnostic \textbf{shadow} models on \textit{members}.]
    {
    \includegraphics[width=0.22\linewidth]{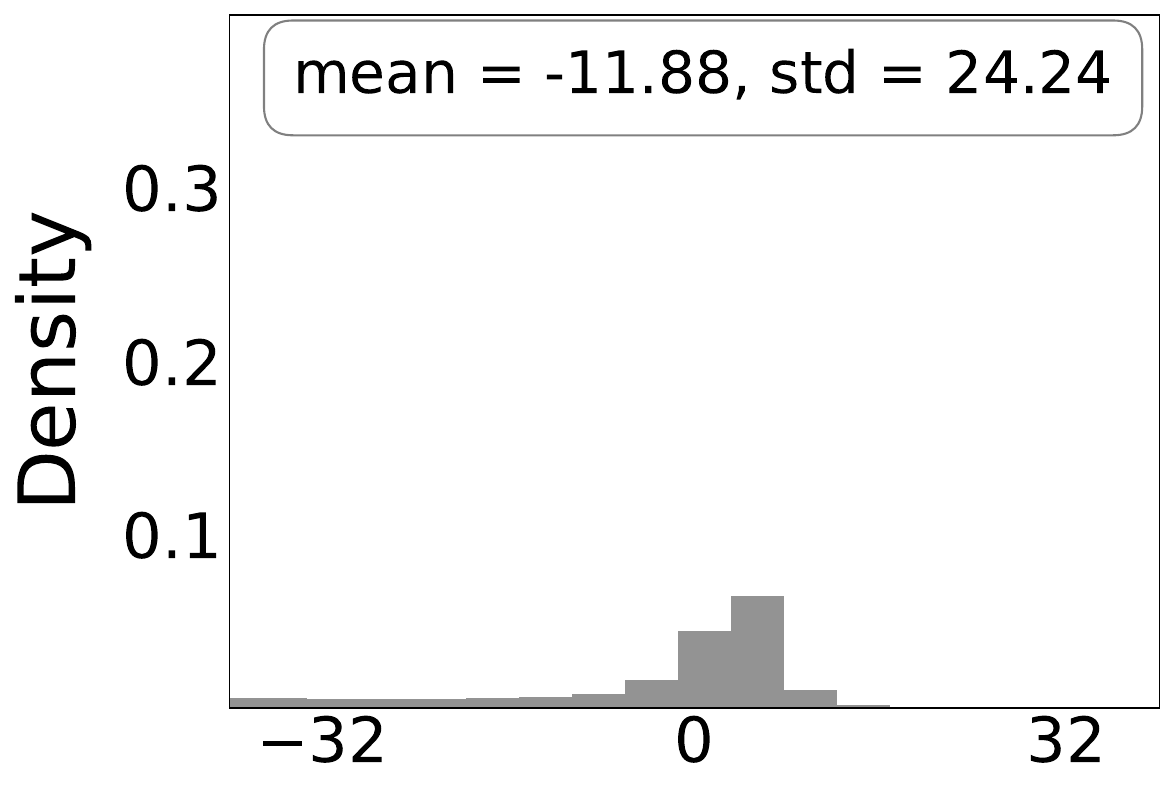}
    }
    \hspace{0.5mm}
    \subfigure[Residuals of target-informed \textbf{imitative} models on \textit{members}.]
    {
    \includegraphics[width=0.22\linewidth]{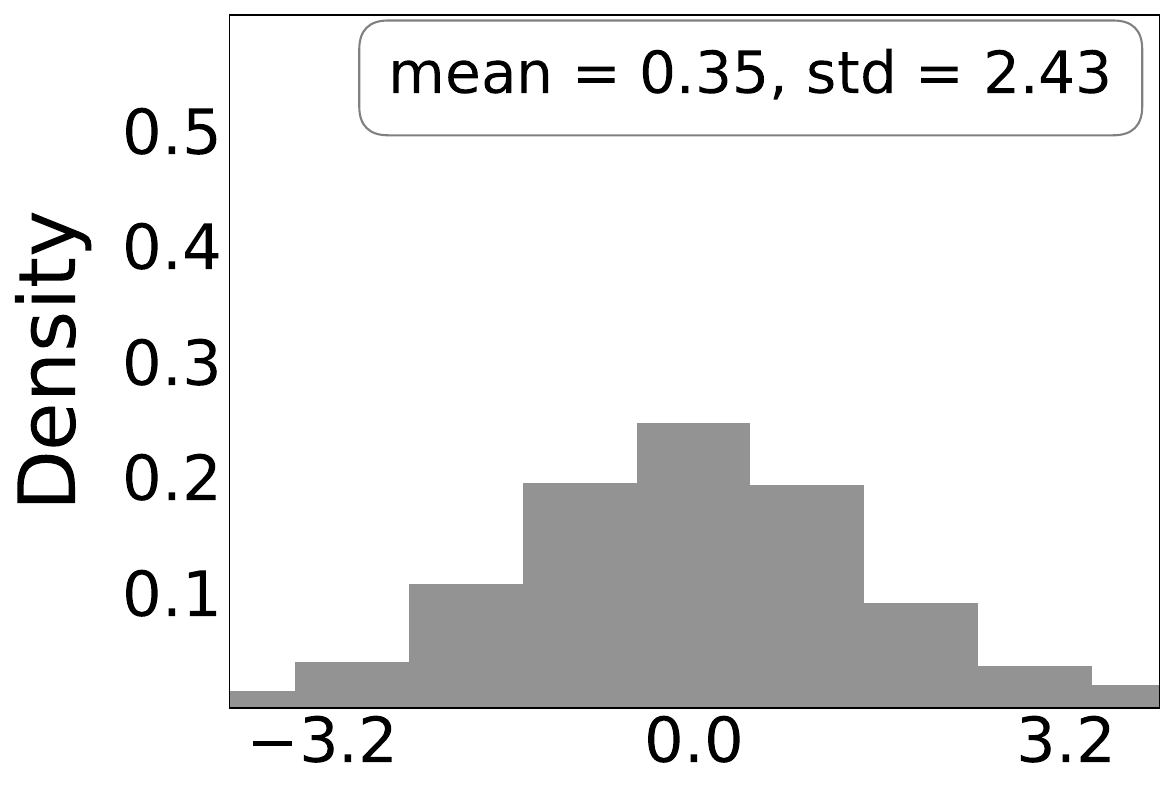}
    }
    \hspace{0.5mm}
    \subfigure[Residuals of target-agnostic \textbf{shadow} models on \textit{non-members}.]
    {
    \includegraphics[width=0.22\linewidth]{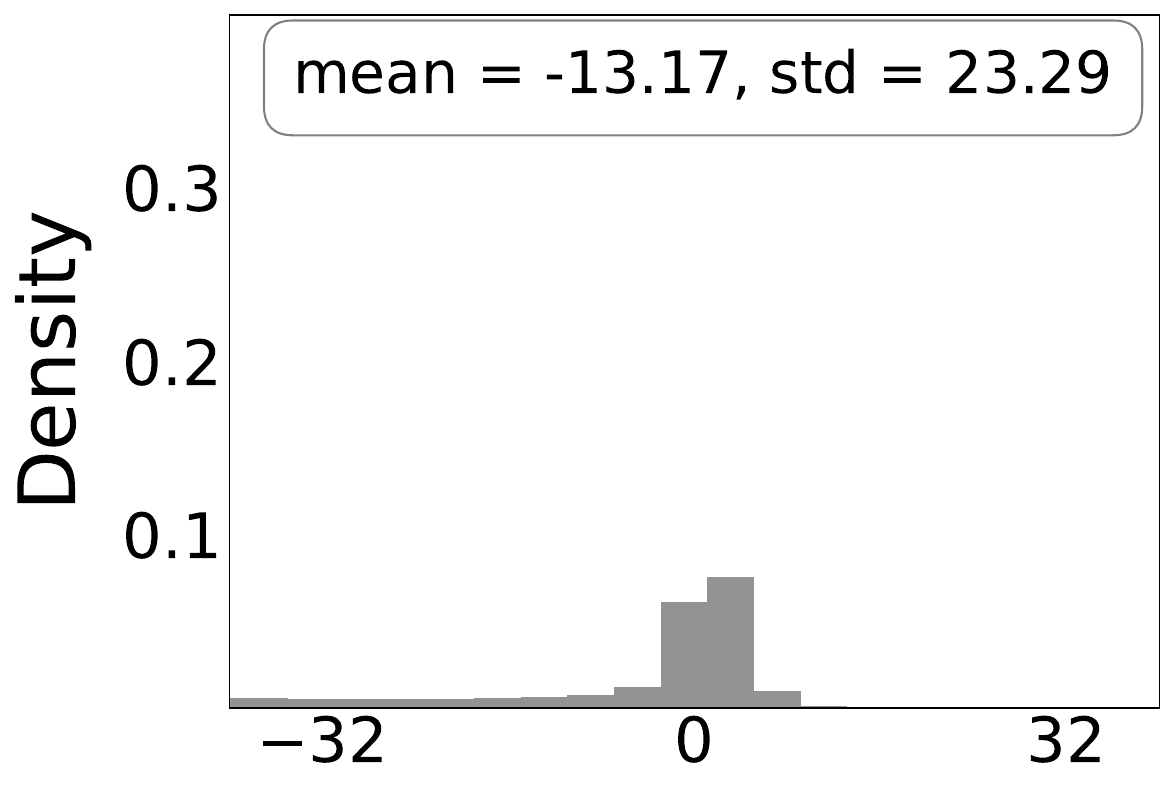}
    }
    \hspace{0.5mm}
    \subfigure[Residuals of target-informed \textbf{imitative} models on \textit{non-members}.]
    {
    \includegraphics[width=0.22\linewidth]{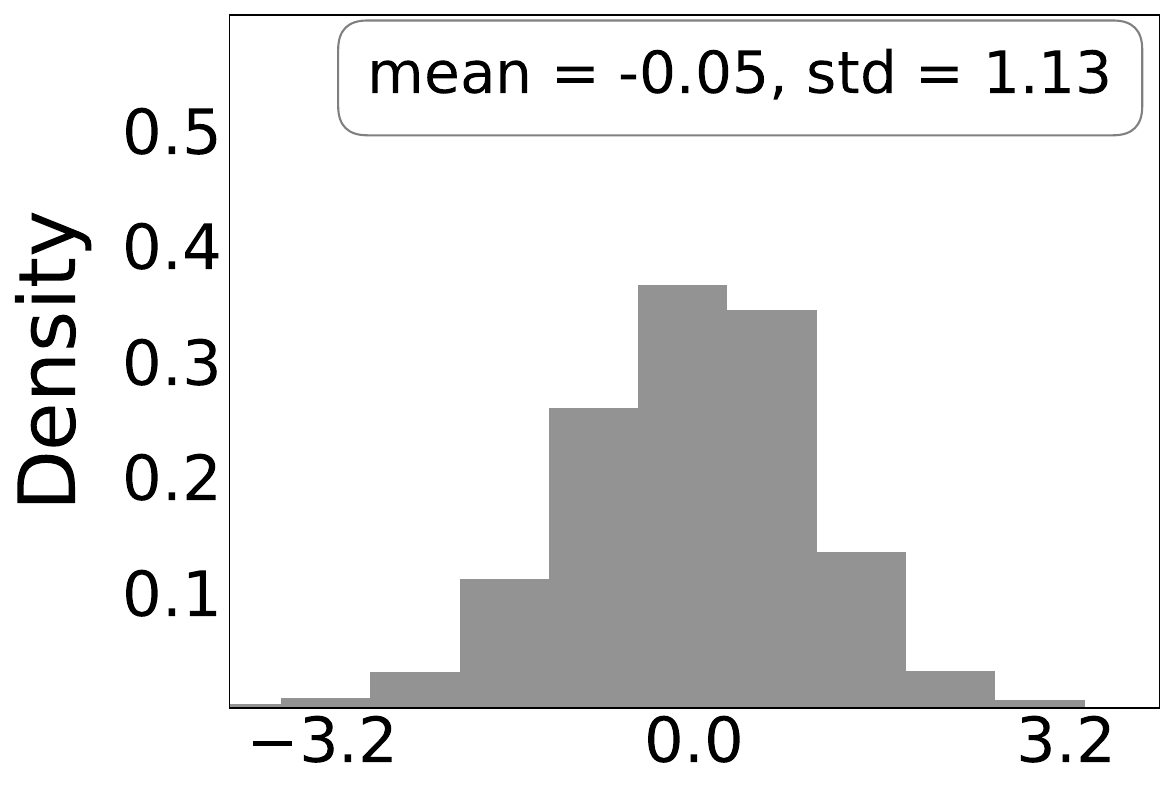}
    }
    \caption{Normalized residual distributions  of shadow models vs.\ imitative models on Texas.}
    \label{fig:z_score_hist_texas}
\end{figure*}

\begin{figure*}[t]
    \centering
    \subfigure[Q-Q plot of \textbf{shadow} models on \textit{members}.]
    {
    \includegraphics[width=0.22\linewidth]{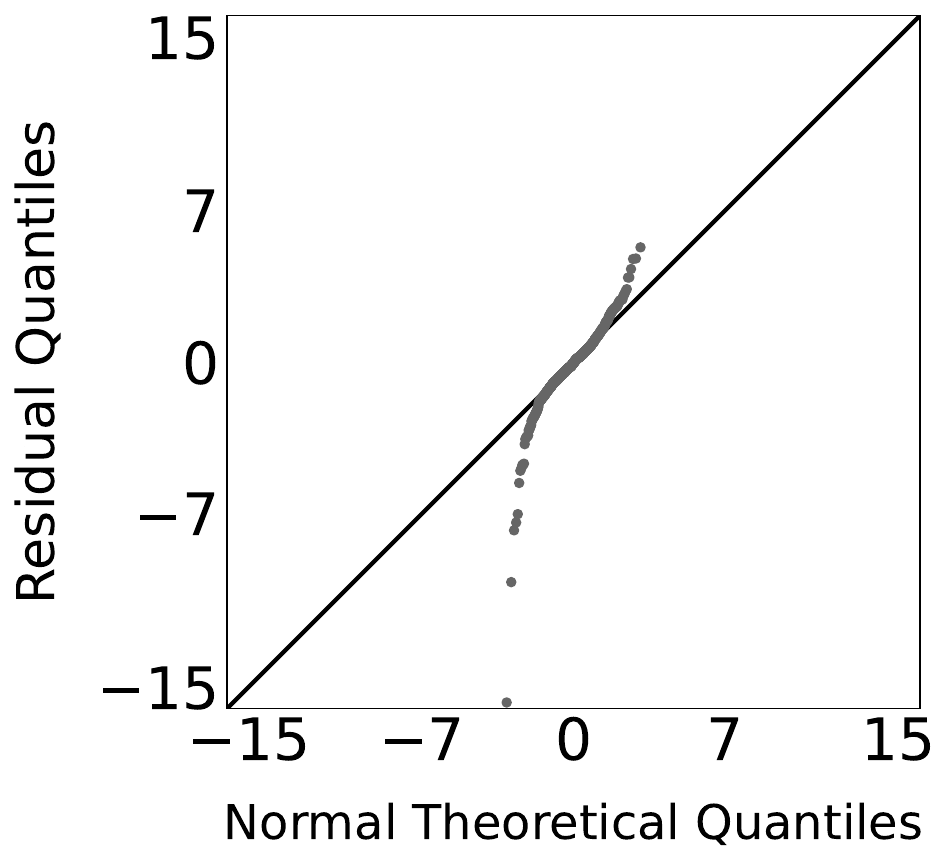}
    }
    \hspace{0.5mm}
    \subfigure[Q-Q plot of \textbf{imitative} models on \textit{members}.]
    {
    \includegraphics[width=0.22\linewidth]{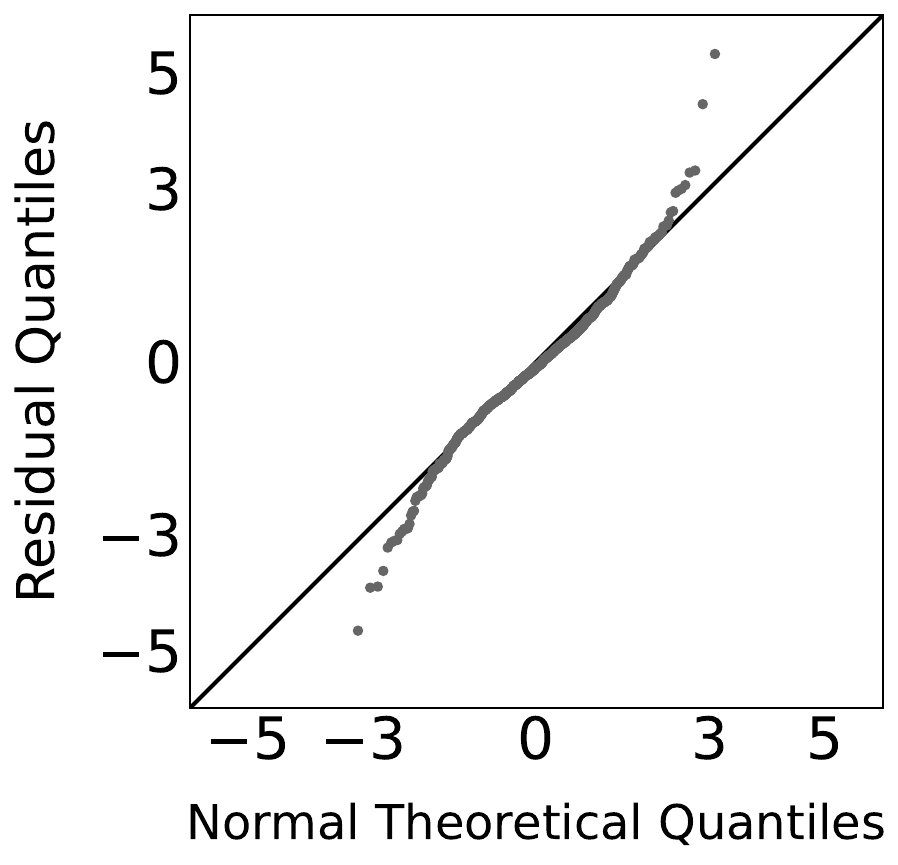}
    }
    \hspace{0.5mm}
    \subfigure[Q-Q plot of \textbf{shadow} models on \textit{non-members}.]
    {
    \includegraphics[width=0.22\linewidth]{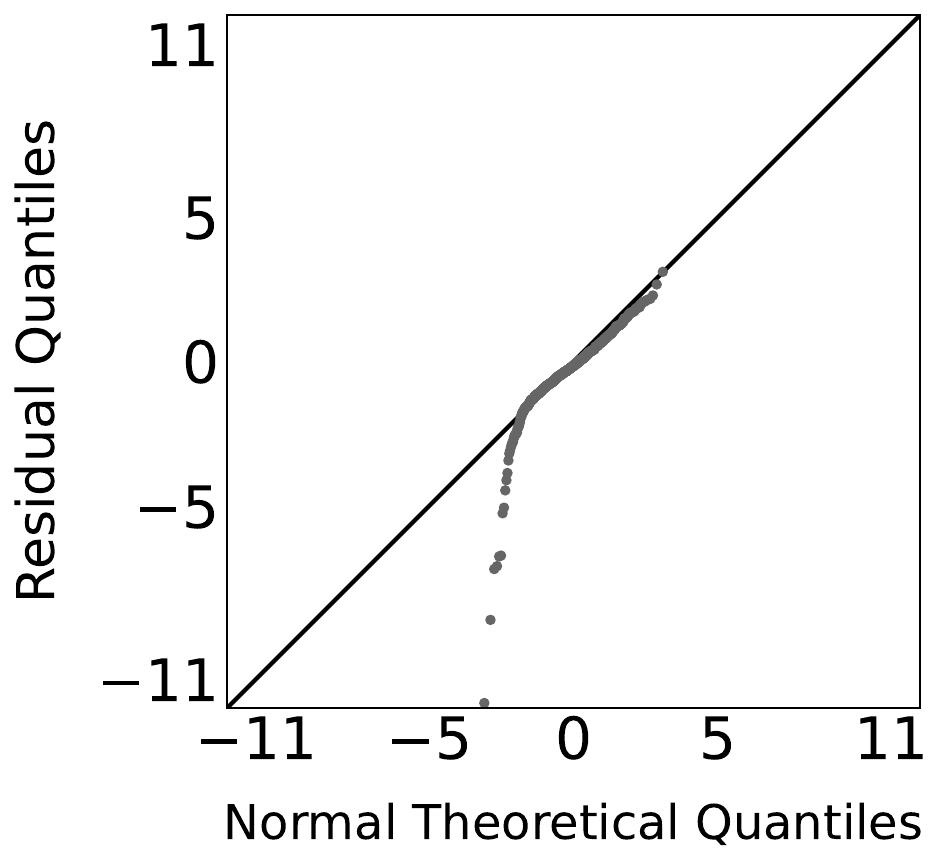}
    }
    \hspace{0.5mm}
    \subfigure[Q-Q plot of \textbf{imitative} models on \textit{non-members}.]
    {
    \includegraphics[width=0.22\linewidth]{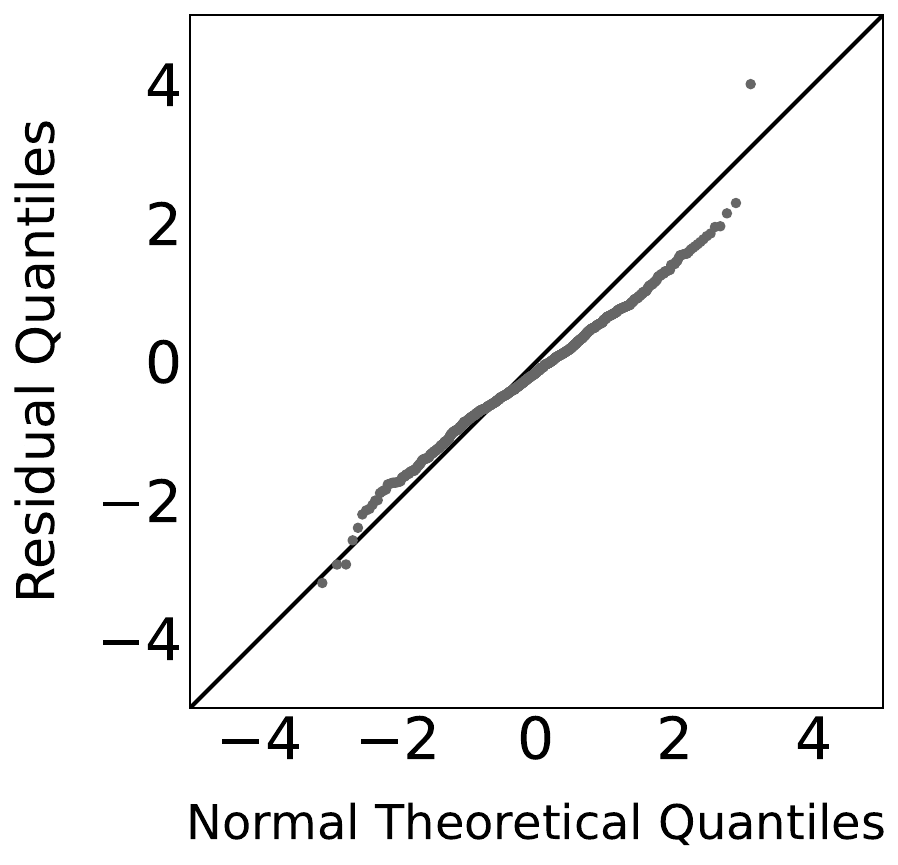}
    }
    \caption{Quantile-Quantile (Q-Q) plots of shadow models vs.\ imitative models on MNIST. The Q-Q plots compare the normalized residual distributions of shadow and imitative models against a standard normal distribution. The imitative models' residuals align closely with the theoretical normal distribution (diagonal), demonstrating a near-perfect distributional fit for both members and non-members.}
    \label{fig:z_score_qq_mnist}
\end{figure*}

\begin{figure*}[t]
    \centering
    \subfigure[Q-Q plot of \textbf{shadow} models on \textit{members}.]
    {
    \includegraphics[width=0.22\linewidth]{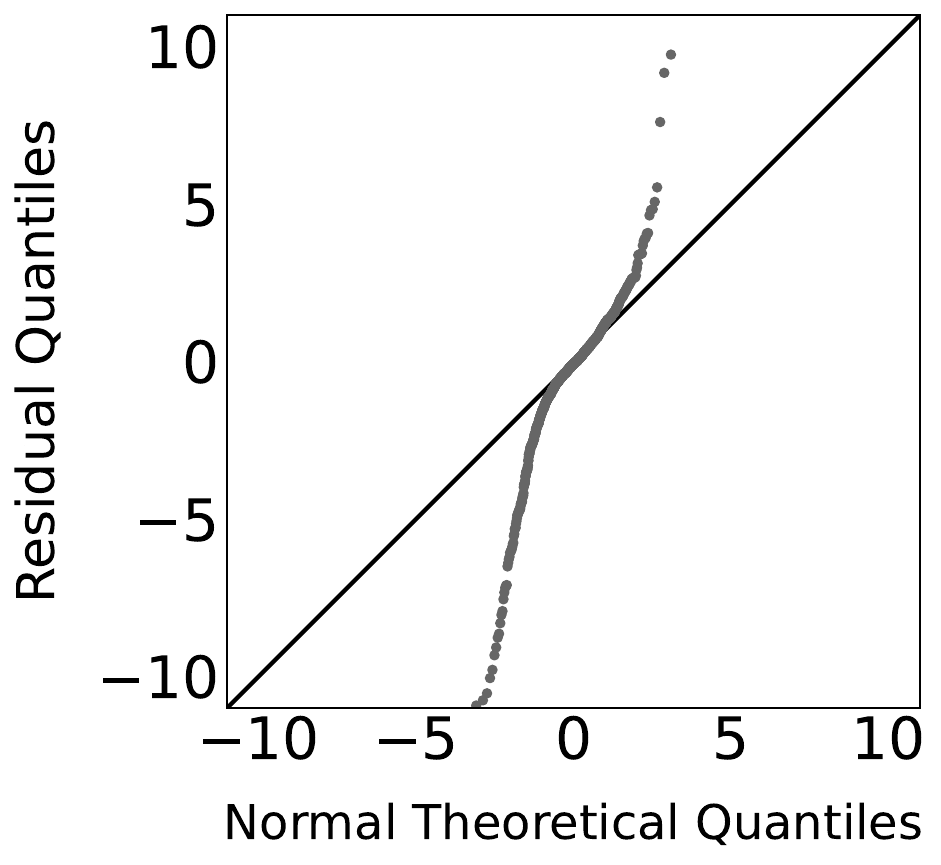}
    }
    \hspace{0.5mm}
    \subfigure[Q-Q plot of \textbf{imitative} models on \textit{members}.]
    {
    \includegraphics[width=0.22\linewidth]{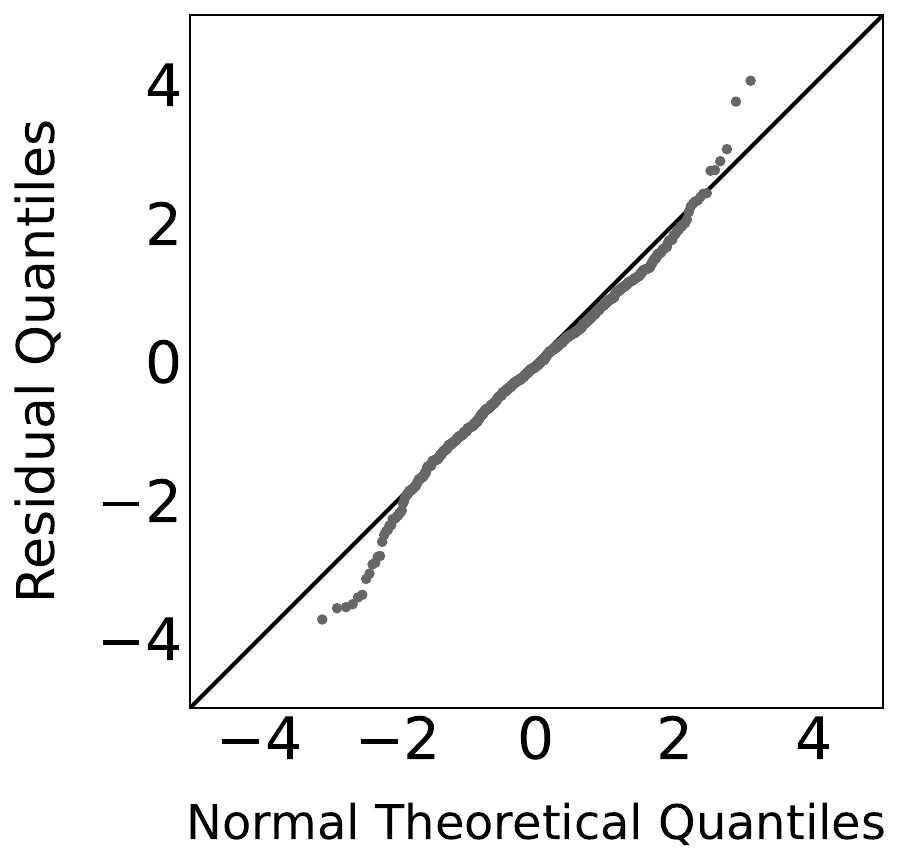}
    }
    \hspace{0.5mm}
    \subfigure[Q-Q plot of \textbf{shadow} models on \textit{non-members}.]
    {
    \includegraphics[width=0.22\linewidth]{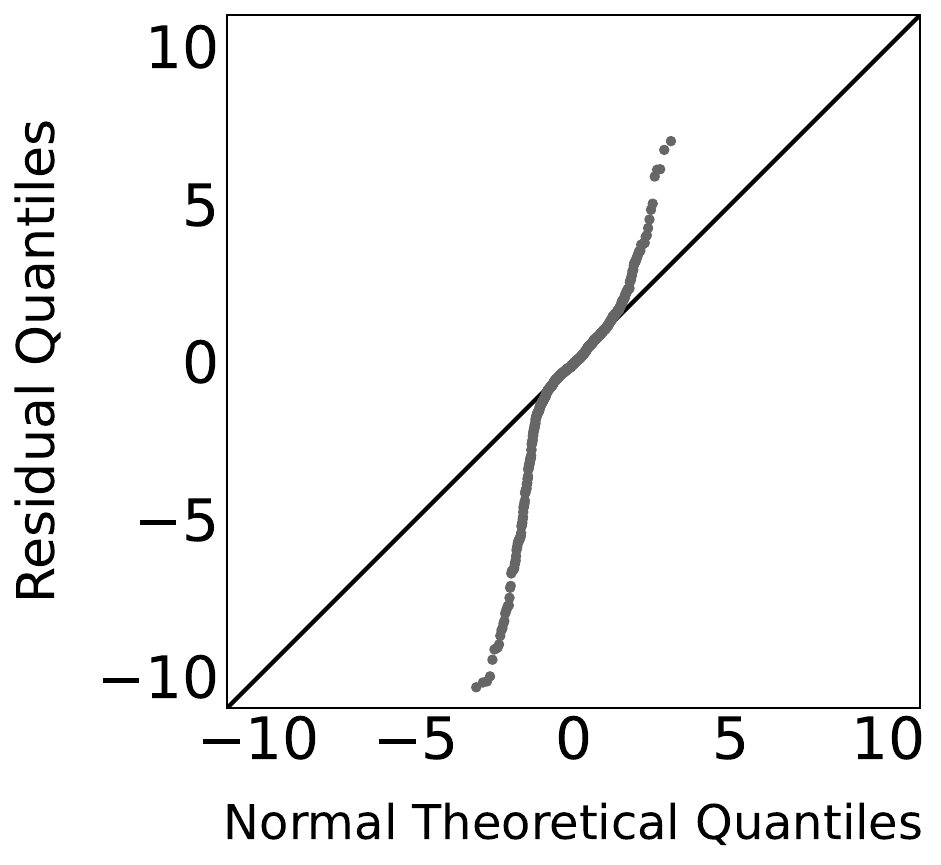}
    }
    \hspace{0.5mm}
    \subfigure[Q-Q plot of \textbf{imitative} models on \textit{non-members}.]
    {
    \includegraphics[width=0.22\linewidth]{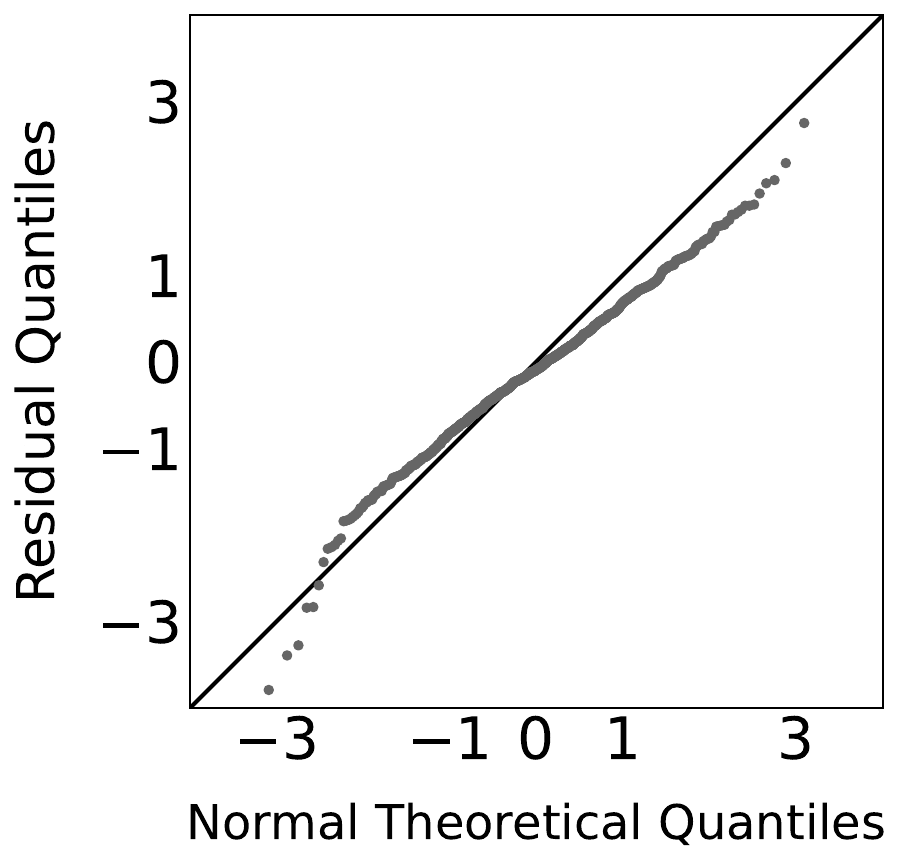}
    }
    \caption{Quantile-Quantile (Q-Q) plots of shadow models vs.\ imitative models on FMNIST.}
    \label{fig:z_score_qq_fmnist}
\end{figure*}

\begin{figure*}[t]
    \centering
    \subfigure[Q-Q plot of \textbf{shadow} models on \textit{members}.]
    {
    \includegraphics[width=0.22\linewidth]{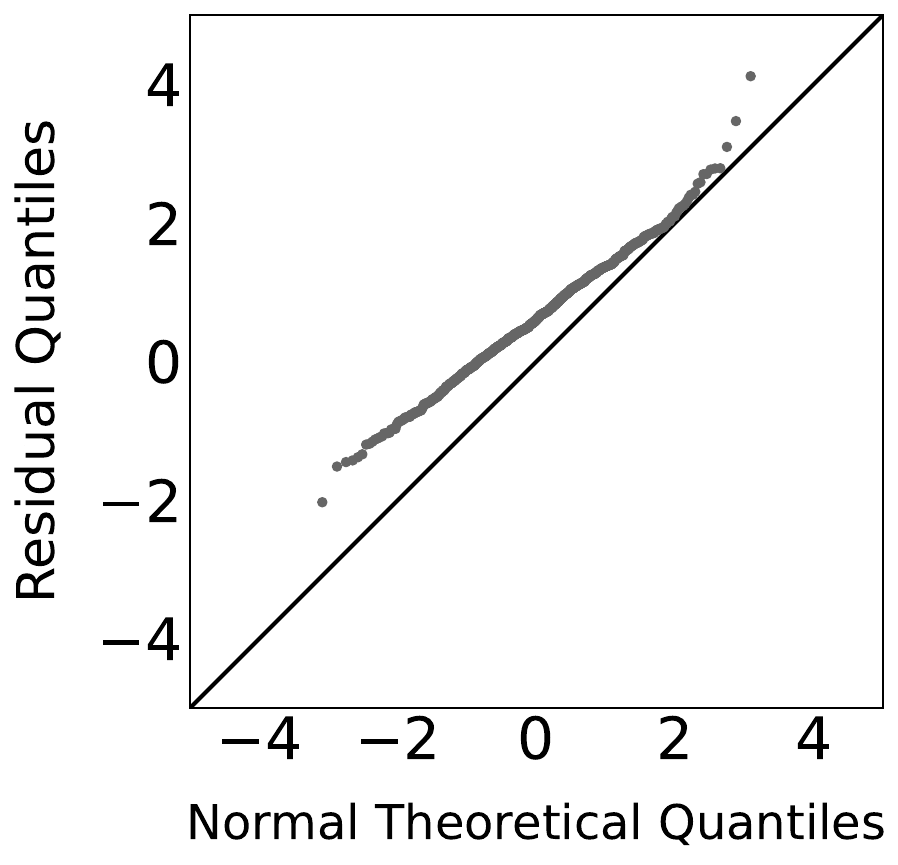}
    }
    \hspace{0.5mm}
    \subfigure[Q-Q plot of \textbf{imitative} models on \textit{members}.]
    {
    \includegraphics[width=0.22\linewidth]{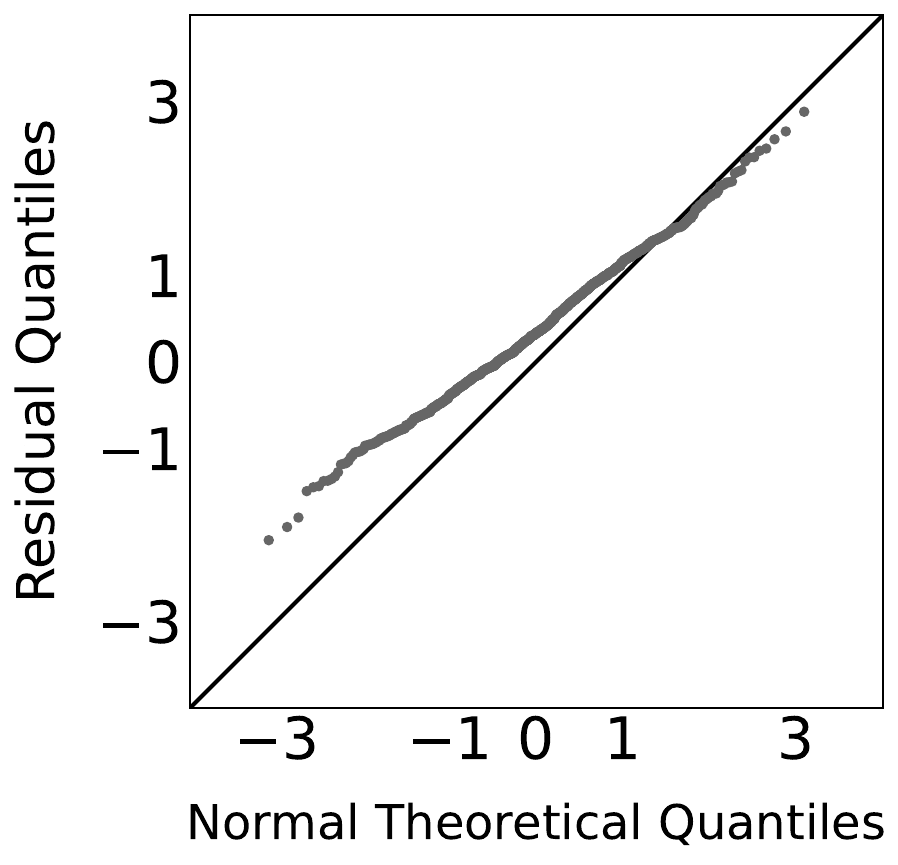}
    }
    \hspace{0.5mm}
    \subfigure[Q-Q plot of \textbf{shadow} models on \textit{non-members}.]
    {
    \includegraphics[width=0.22\linewidth]{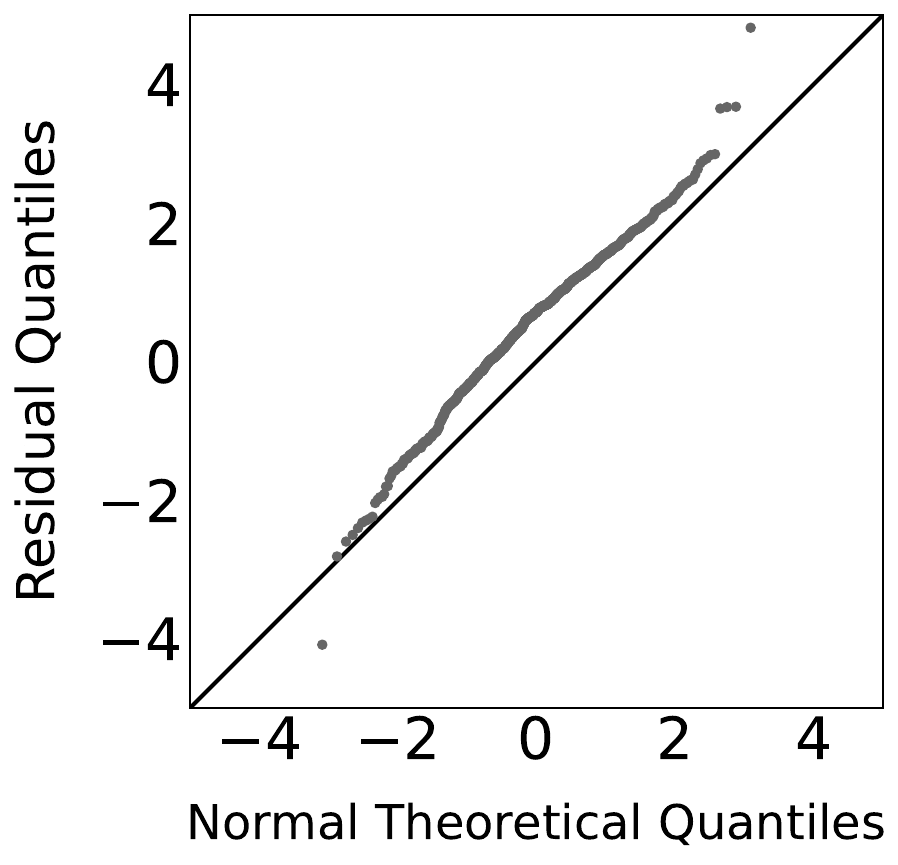}
    }
    \hspace{0.5mm}
    \subfigure[Q-Q plot of \textbf{imitative} models on \textit{non-members}.]
    {
    \includegraphics[width=0.22\linewidth]{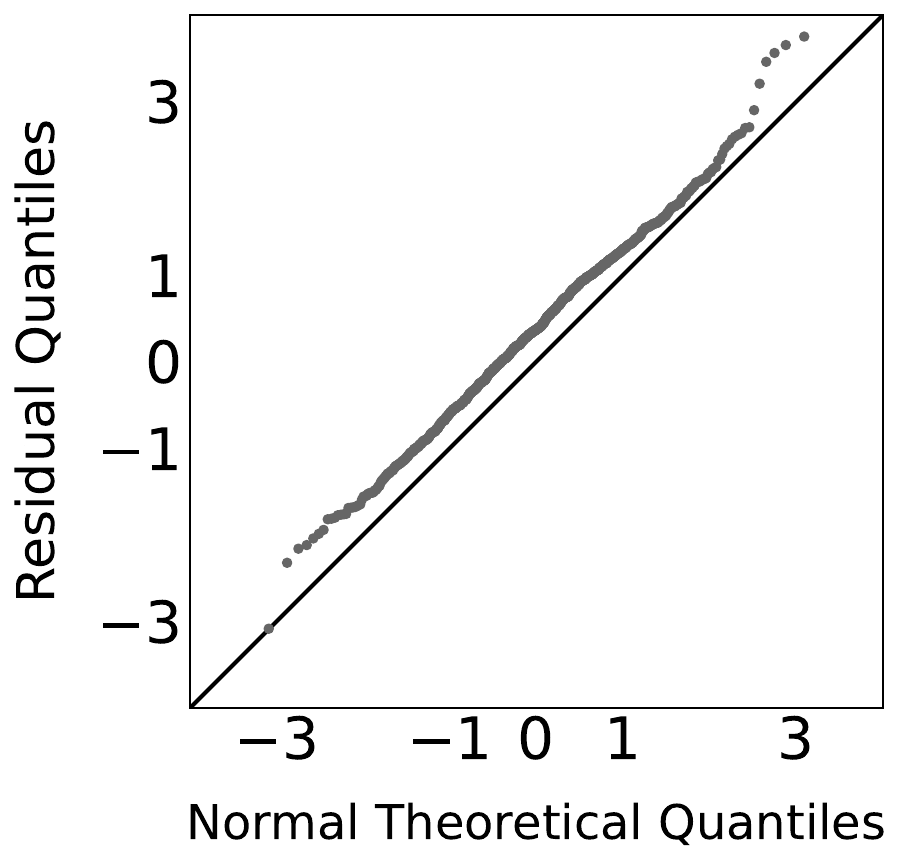}
    }
    \caption{Quantile-Quantile (Q-Q) plots of shadow models vs.\ imitative models on CIFAR-10. }
    \label{fig:z_score_qq_cifar10}
\end{figure*}

\begin{figure*}[t]
    \centering
    \subfigure[Q-Q plot of \textbf{shadow} models on \textit{members}.]
    {
    \includegraphics[width=0.22\linewidth]{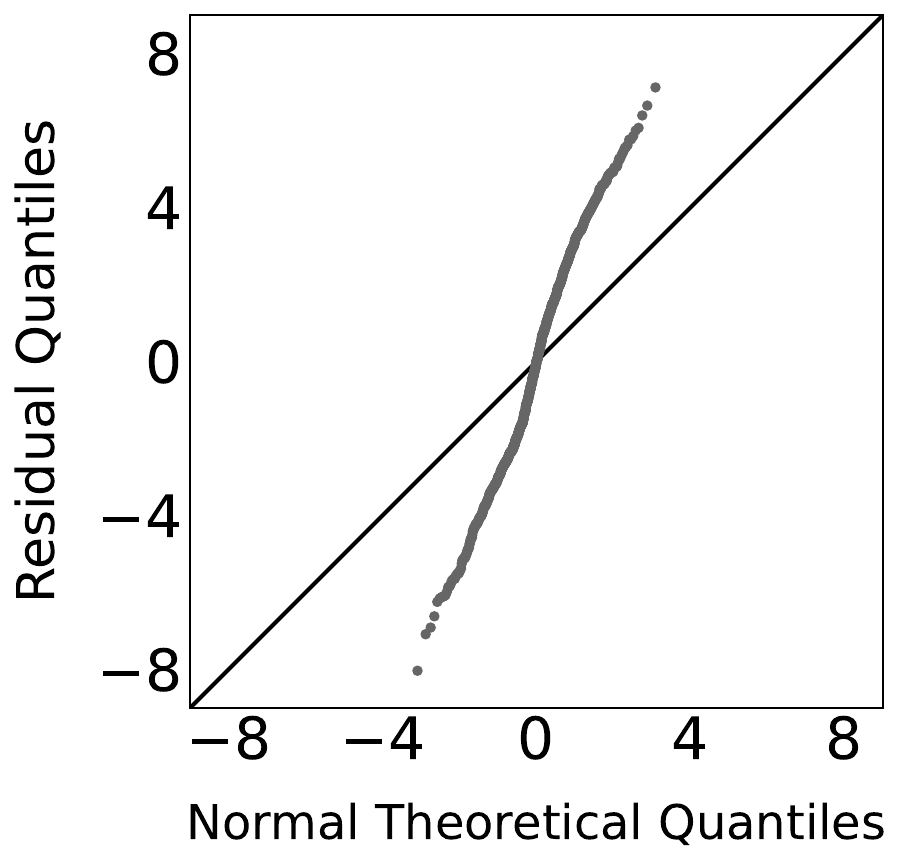}
    }
    \hspace{0.5mm}
    \subfigure[Q-Q plot of \textbf{imitative} models on \textit{members}.]
    {
    \includegraphics[width=0.22\linewidth]{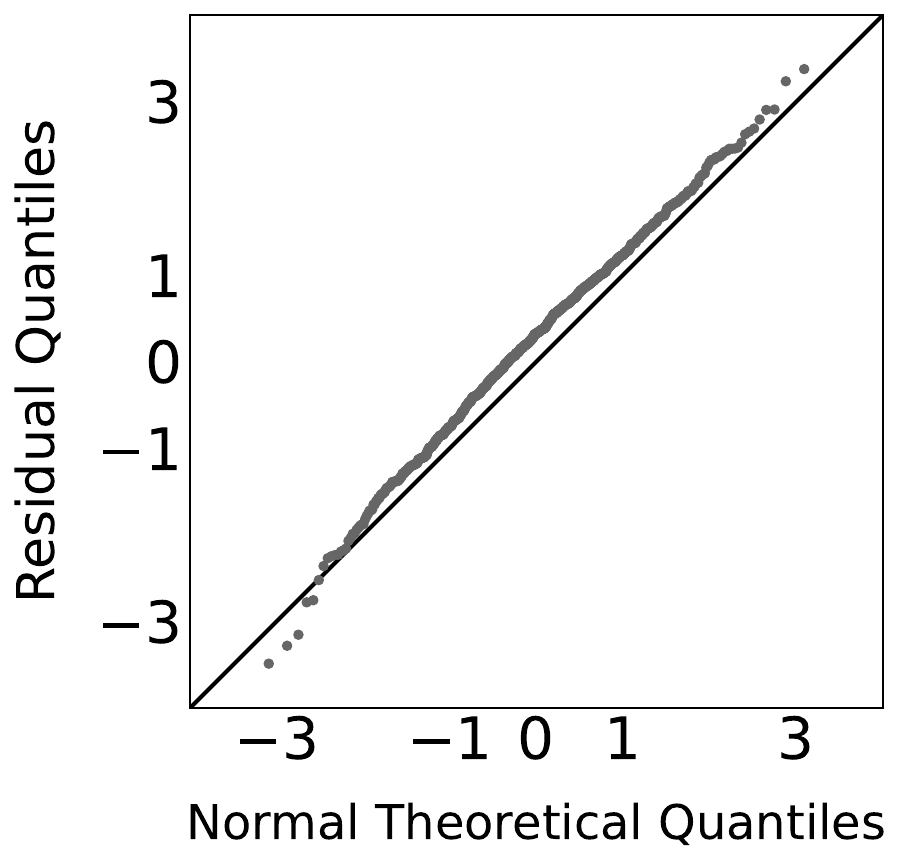}
    }
    \hspace{0.5mm}
    \subfigure[Q-Q plot of \textbf{shadow} models on \textit{non-members}.]
    {
    \includegraphics[width=0.22\linewidth]{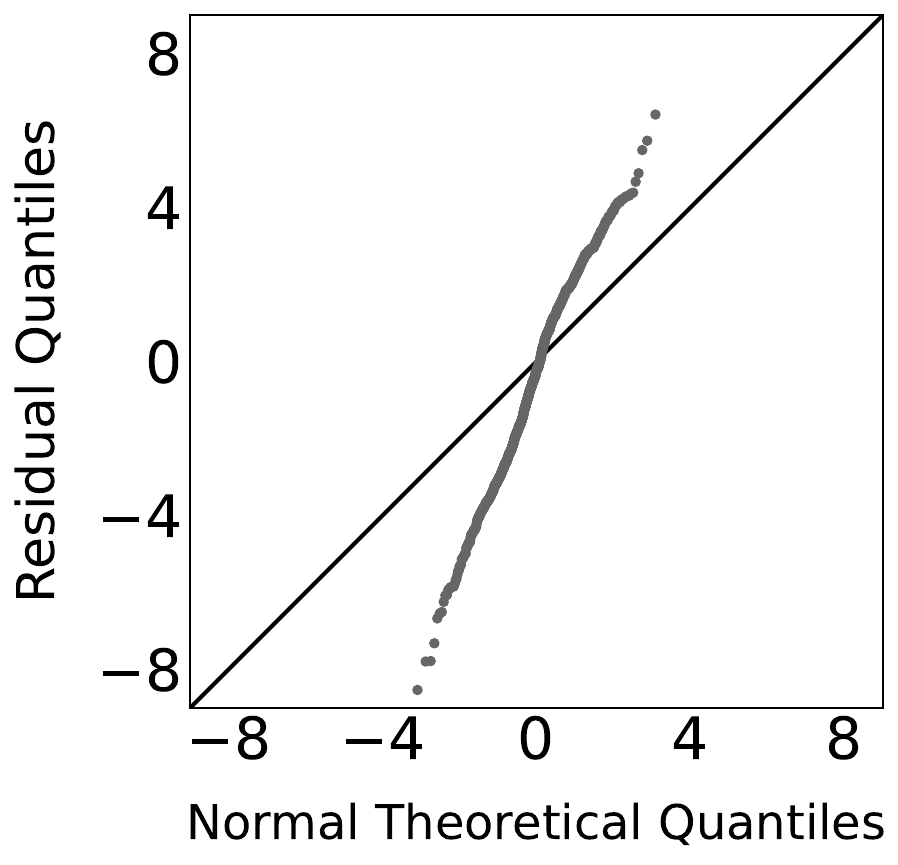}
    }
    \hspace{0.5mm}
    \subfigure[Q-Q plot of \textbf{imitative} models on \textit{non-members}.]
    {
    \includegraphics[width=0.22\linewidth]{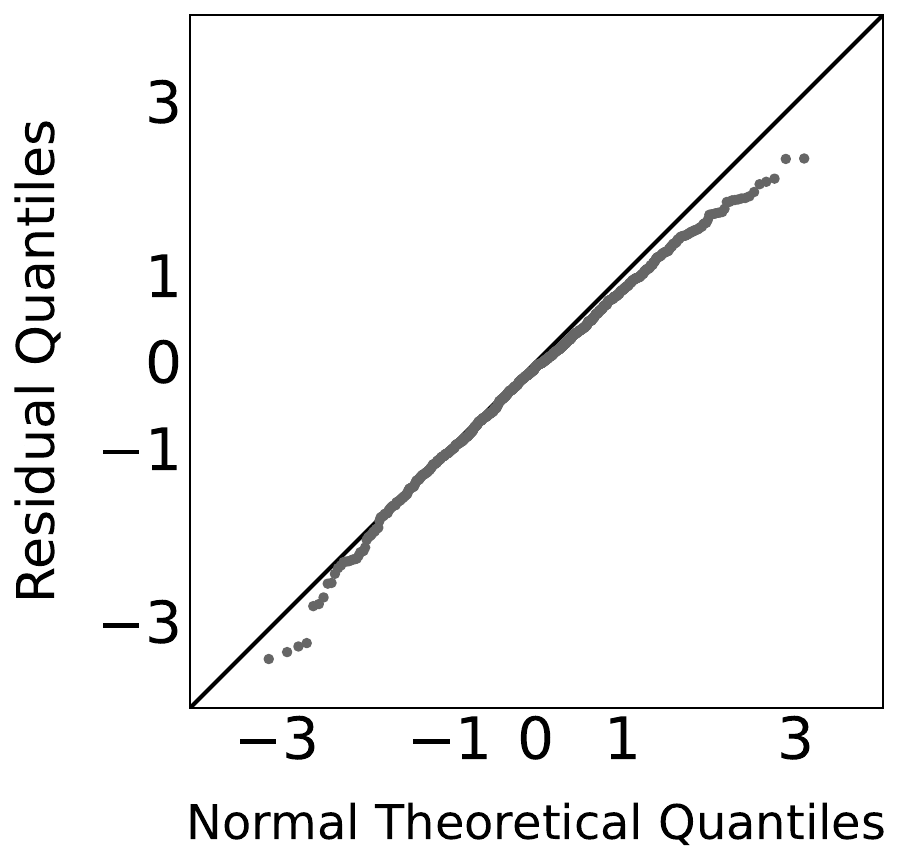}
    }
    \caption{Quantile-Quantile (Q-Q) plots of shadow models vs.\ imitative models on CIFAR-100. }
    \label{fig:z_score_qq_cifar100}
\end{figure*}

\begin{figure*}[t]
    \centering
    \subfigure[Q-Q plot of \textbf{shadow} models on \textit{members}.]
    {
    \includegraphics[width=0.22\linewidth]{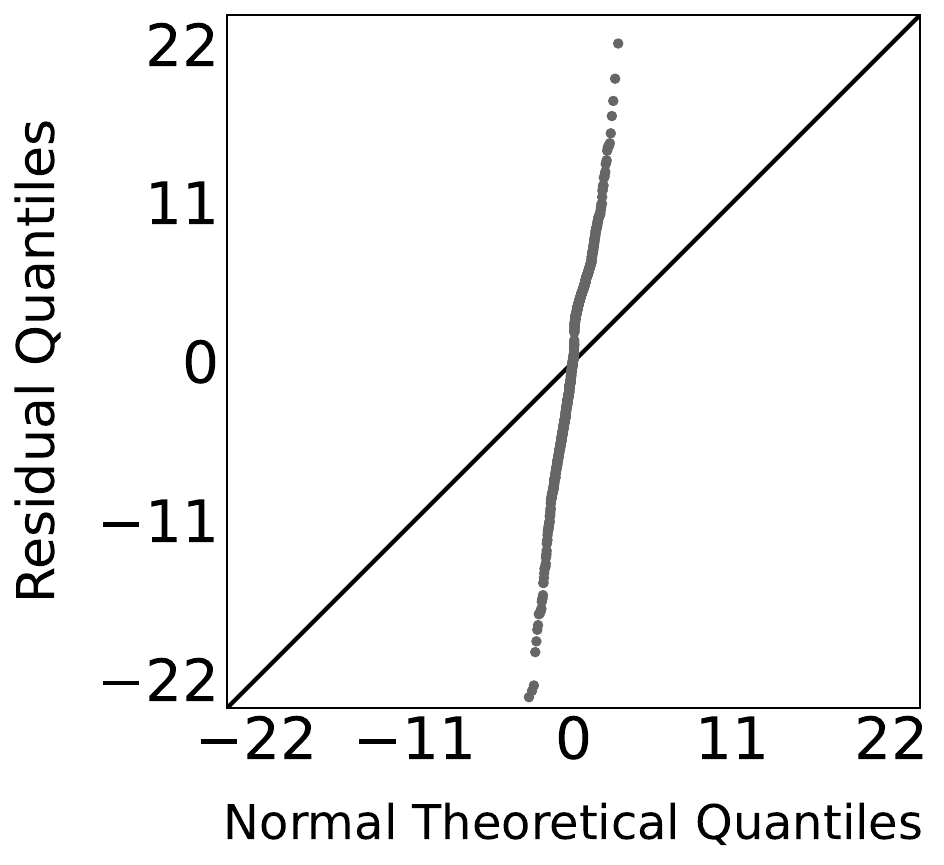}
    }
    \hspace{0.5mm}
    \subfigure[Q-Q plot of \textbf{imitative} models on \textit{members}.]
    {
    \includegraphics[width=0.22\linewidth]{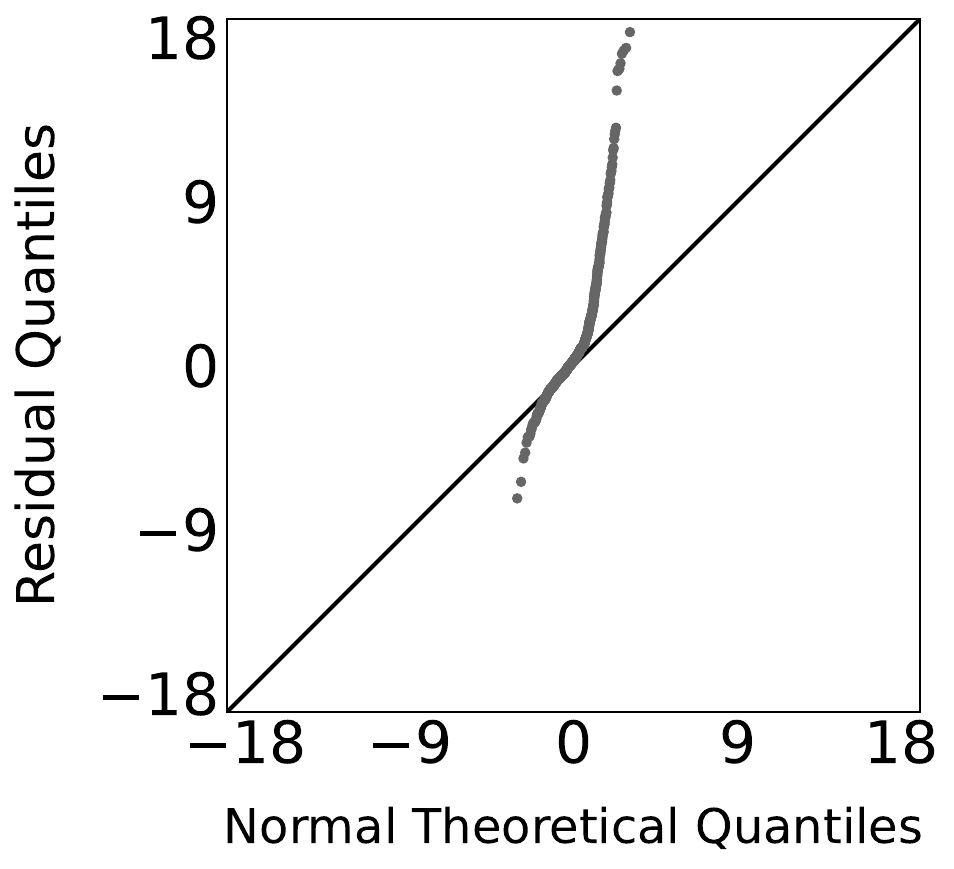}
    }
    \hspace{0.5mm}
    \subfigure[Q-Q plot of \textbf{shadow} models on \textit{non-members}.]
    {
    \includegraphics[width=0.22\linewidth]{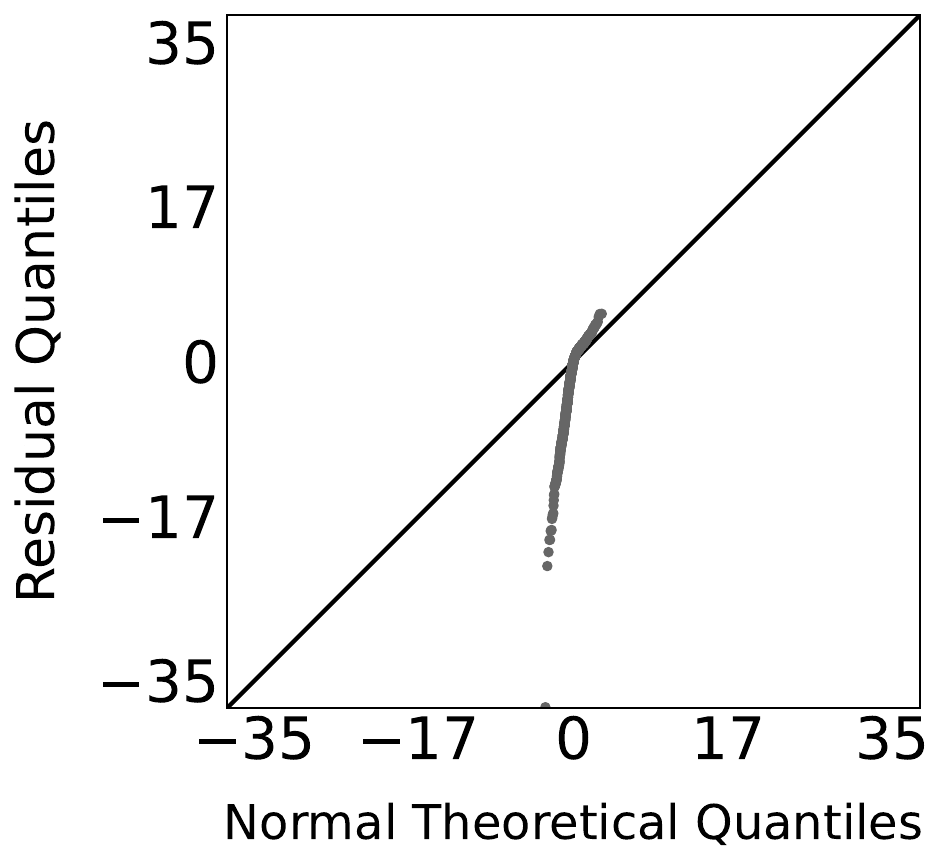}
    }
    \hspace{0.5mm}
    \subfigure[Q-Q plot of \textbf{imitative} models on \textit{non-members}.]
    {
    \includegraphics[width=0.22\linewidth]{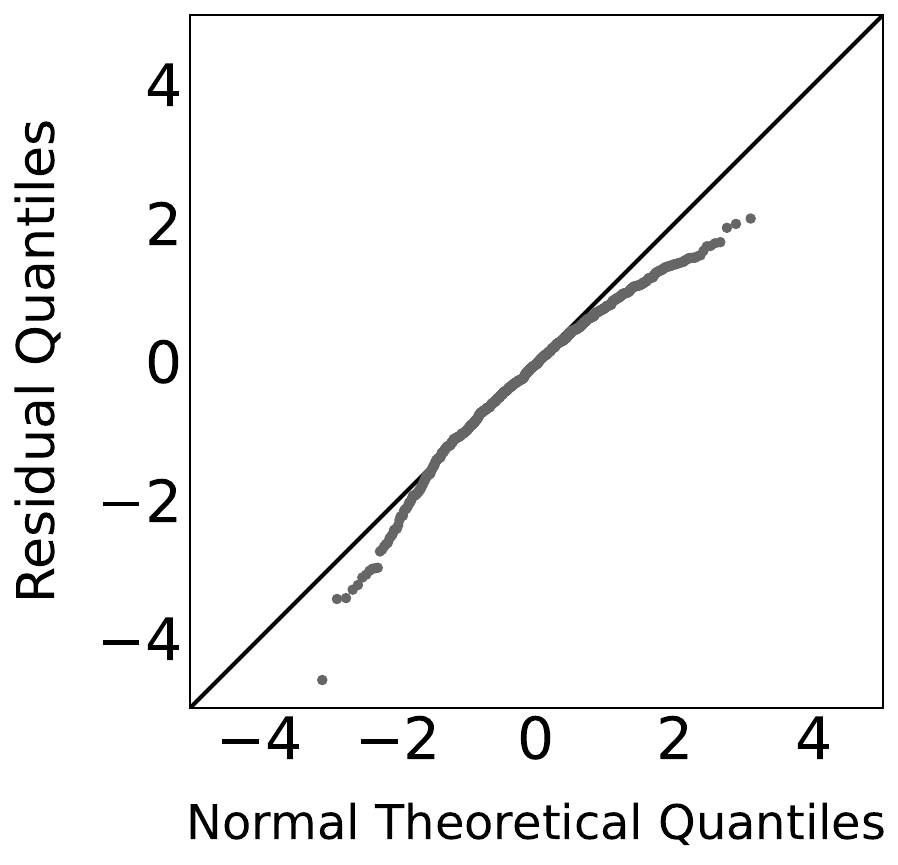}
    }
    \caption{Quantile-Quantile (Q-Q) plots of shadow models vs.\ imitative models on Purchase. }
    \label{fig:z_score_qq_purchase}
\end{figure*}

\begin{figure*}[t]
    \centering
    \subfigure[Q-Q plot of \textbf{shadow} models on \textit{members}.]
    {
    \includegraphics[width=0.22\linewidth]{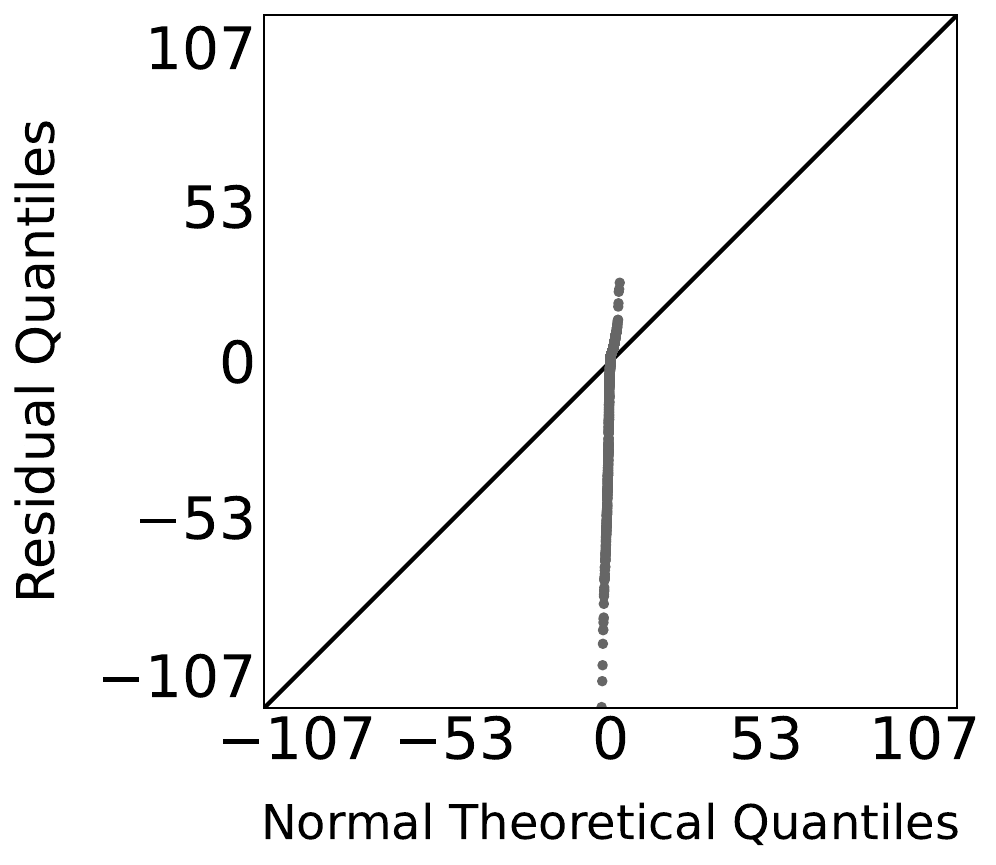}
    }
    \hspace{0.5mm}
    \subfigure[Q-Q plot of \textbf{imitative} models on \textit{members}.]
    {
    \includegraphics[width=0.22\linewidth]{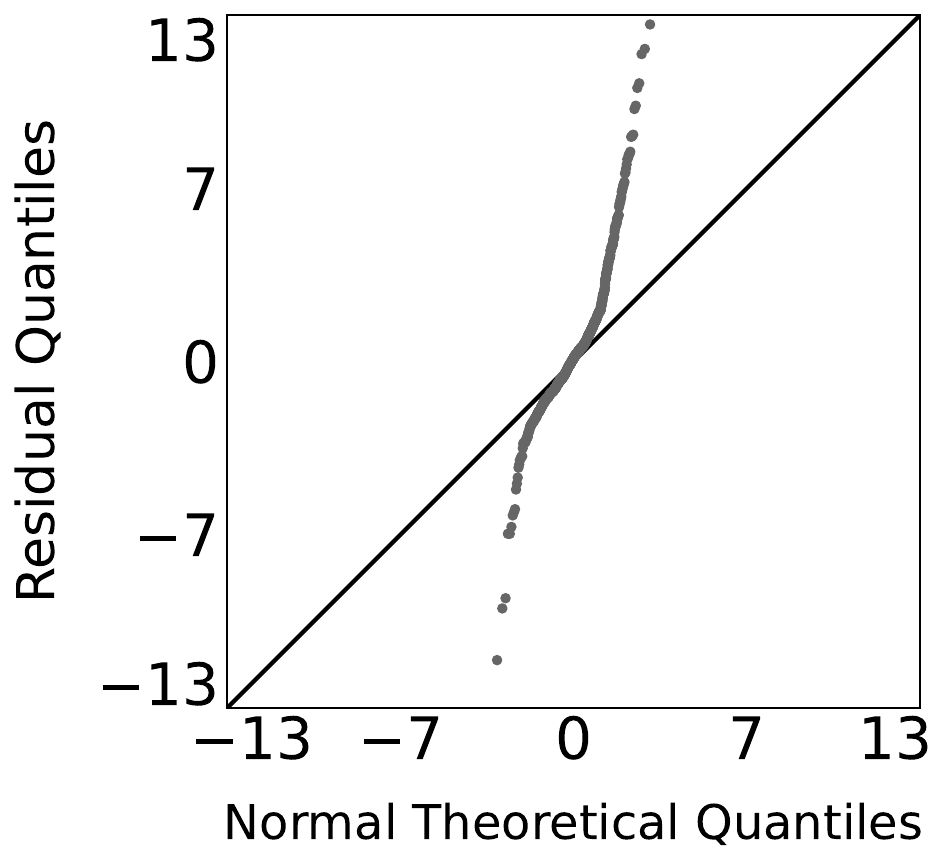}
    }
    \hspace{0.5mm}
    \subfigure[Q-Q plot of \textbf{shadow} models on \textit{non-members}.]
    {
    \includegraphics[width=0.22\linewidth]{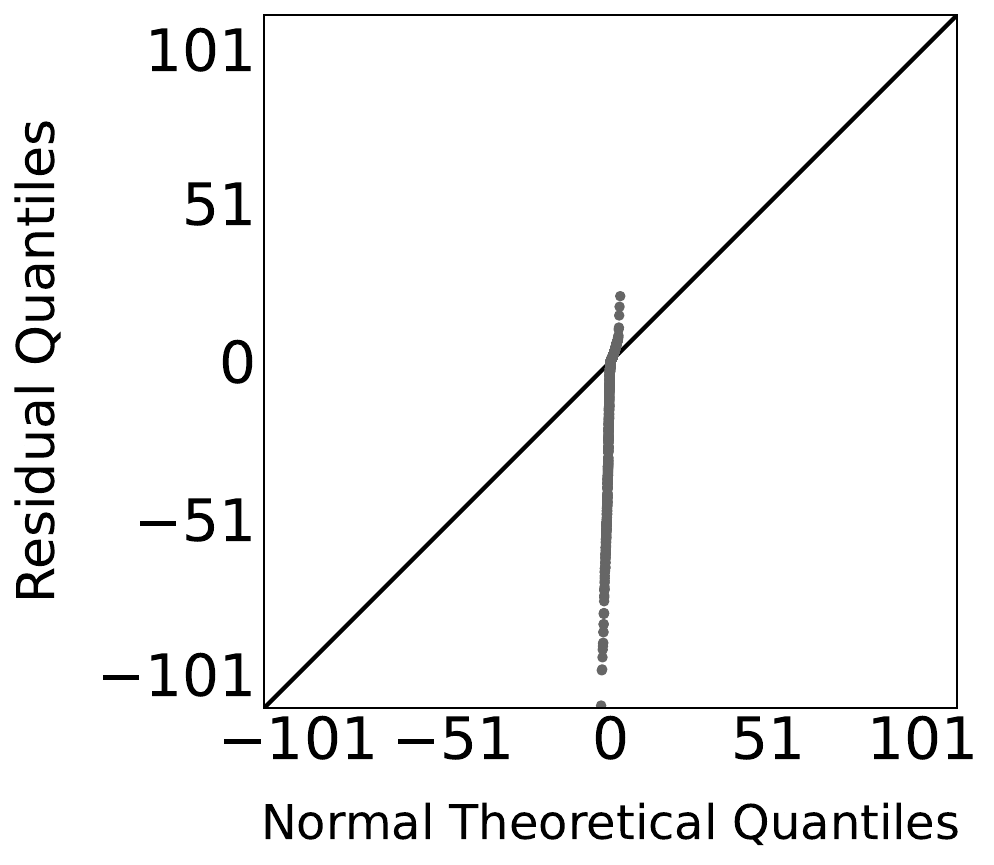}
    }
    \hspace{0.5mm}
    \subfigure[Q-Q plot of \textbf{imitative} models on \textit{non-members}.]
    {
    \includegraphics[width=0.22\linewidth]{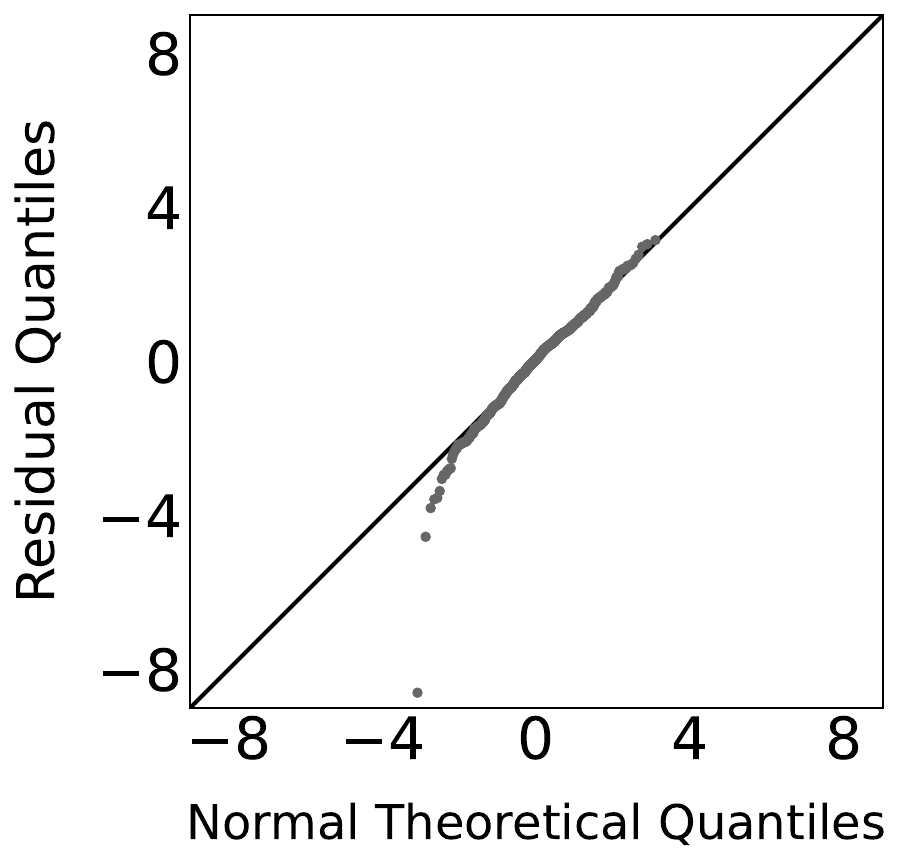}
    }
    \caption{Quantile-Quantile (Q-Q) plots of shadow models vs.\ imitative models on Texas. }
    \label{fig:z_score_qq_texas}
\end{figure*}

\begin{figure*}[t]
    \centering
    \subfigure[MNIST]
    {
    \includegraphics[width=0.23\linewidth]{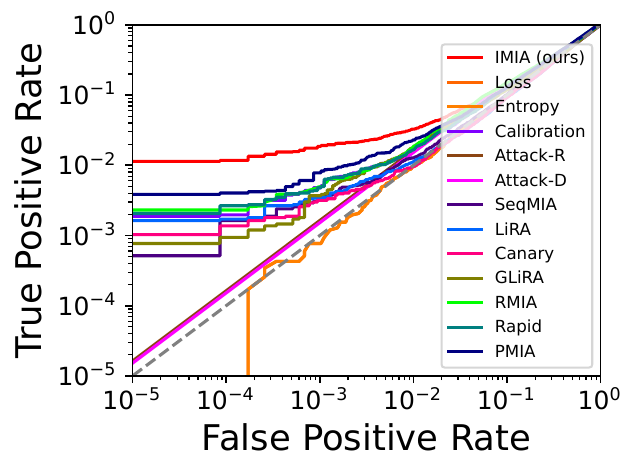}
    }
    \subfigure[Fashion-MNIST]
    {
    \includegraphics[width=0.23\linewidth]{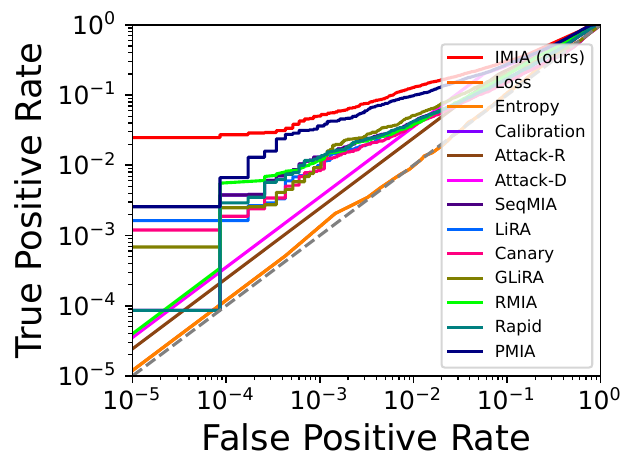}
    }
    \subfigure[CIFAR-10]
    {
    \includegraphics[width=0.23\linewidth]{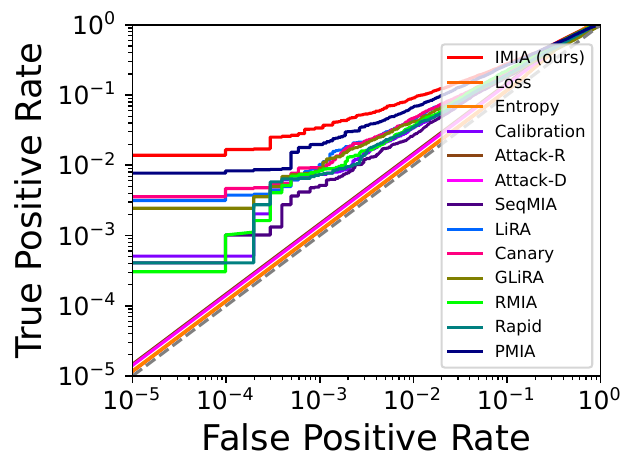}
    }
    \subfigure[CIFAR-100]
    {
    \includegraphics[width=0.23\linewidth]{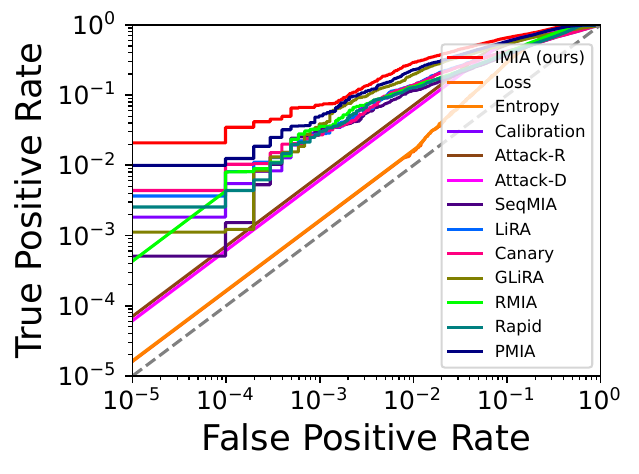}
    }
    \caption{The ROC curves of attack results on ResNet models trained on four image datasets in the \textit{non-adaptive} setting.}
    \label{fig:imia_curve_resnet}
\end{figure*}

\subsection{Effectiveness of Imitative Training}
\label{appendix:effectivness_imiative}

The results of normalized residual distributions on MNIST, FMNIST, CIFAR-10, Purchase, and Texas are in~\Cref{fig:z_score_hist_mnist},~\Cref{fig:z_score_hist_fmnist},~\Cref{fig:z_score_hist_cifar10},~\Cref{fig:z_score_hist_purchase}, and~\Cref{fig:z_score_hist_texas}, respectively. 
As illustrated, the normalized residuals from imitative models consistently align more closely with the standard normal distribution than those from shadow models. 
We also present two additional experiments to further demonstrate the effectiveness of imitative training.

\mypara{Distribution Comparison via Quantile-Quantile Plot}
To quantitatively assess how closely the normalized residual distribution approximates the standard normal distribution, we employ the quantile-quantile (Q-Q) plot, a standard method for comparing distributional discrepancies. 
The results for both shadow and imitative models on four image and two non-image datasets are shown in~\Cref{fig:z_score_qq_mnist},~\Cref{fig:z_score_qq_fmnist},~\Cref{fig:z_score_qq_cifar10},~\Cref{fig:z_score_qq_cifar100},~\Cref{fig:z_score_qq_purchase}, and~\Cref{fig:z_score_qq_texas}.
As shown, most residuals of imitative models align closely with the theoretical normal distribution (diagonal), indicating a near-perfect fit. 
In contrast, shadow models exhibit noticeable deviations from the diagonal, highlighting their poorer alignment with the target model's behavior.

\begin{table}[t]
\centering
\caption{Averaged likelihood ratio comparing imitative training to shadow training. A higher ratio (greater than one) indicates that imitative models better capture the target model's true behaviors compared to shadow models.}
\label{tab:fitness_likelihood_ratio}
\resizebox{0.47\textwidth}{!}{
\begin{tabular}{l|c|c|c|c|c|c}
\toprule
 & \textbf{MNIST} & \textbf{FMNIST} & \textbf{C-10} & \textbf{C-100} & \textbf{Purchase} & \textbf{Texas}\\
\midrule
Members & 1.16 & 1.18 & 1.31 & 1.52 & 1.64 & 1.52 \\ 
Non-members & 1.38 & 1.52 & 1.60 & 1.92 & 1.97 & 1.88 \\
\bottomrule
\end{tabular}}
\end{table}

\mypara{Likelihood-Based Analysis}
Likelihood is a widely used metric for assessing the fitness of a given value to a distribution. 
Here, we use it to quantify how closely the target model’s true confidence scores align with the distributions learned by imitative models and shadow models. 
Specifically, for both members and non-members, we compute the likelihood of the target model’s confidence score under the estimated Gaussian distribution of imitative and shadow models.
We then define the \textbf{Averaged Likelihood Ratio} as the mean likelihood of imitative models divided by that of shadow models, averaged over all members and non-members. 
A ratio greater than one indicates that the imitative model’s distribution more accurately approximates the target model’s behavior than the shadow model’s distribution.
As shown in~\Cref{tab:fitness_likelihood_ratio}, this ratio is consistently above one across all datasets, providing strong quantitative evidence that imitative training creates more faithful and accurate mimics of the target model's behavior.

\subsection{Additional Ablation Studies}
\label{appendix:ablation}

\begin{table}[t]
    \centering
    \caption{Impact of query budget on attack performance (\ie TPR @ 0\%FPR). Querying the model with multiple augmented views enhances attack performance, yielding significant gains with just two queries.}
    \label{tab:query_budget}
    \resizebox{0.42\textwidth}{!}{
    \begin{tabular}{ccccc}
    \toprule
    \multirow{2}{*}{Queries} & \multicolumn{2}{c}{Non-Adaptive Setting} & \multicolumn{2}{c}{Adaptive Setting} \\
    \cmidrule(lr){2-3} \cmidrule(lr){4-5} 
     & CIFAR-10 & CIFAR-100 & CIFAR-10 & CIFAR-100 \\
    \midrule
    1  & 0.73\% & 1.07\% & 1.53\% & 5.54\% \\
    2  & 1.22\% & 1.89\% & 2.06\% & 7.69\% \\
    9  & 1.43\% & 2.04\% & 2.31\% & 8.42\% \\
    18 & 1.45\% & 2.10\% & 2.33\% & 8.52\% \\
    36 & 1.45\% & 2.11\% & 2.33\% & 8.55\% \\
    \bottomrule
    \end{tabular}}
\end{table}

\mypara{Impact of Query Budget}
We evaluate the impact of varying query budgets on CIFAR-10 and CIFAR-100. As shown in~\Cref{tab:query_budget}, augmenting random queries (e.g., horizontal flips and shifts by up to $\pm 4$ pixels) improves attack performance. Notably, significant gains occur with just two queries, and performance stabilizes around 18 queries. This observation aligns with findings in prior work~\cite{sp22lira}.

\mypara{Impact of Mismatched Model Architecture}
We evaluate the robustness of \mymethod when the adversary is unaware of the exact architecture of the target model.
Specifically, we train imitative models using different architectures, and report the results in~\Cref{fig:impact_arch}.
As expected, the attack performs best when the imitative models match the target model’s architecture. 
However, using a different architecture results in only a slight drop in performance.
We attribute this to the imitative training paradigm.
By forcing the imitative models to mimic the target's behavior, the training process ensures they develop similar predictive patterns, regardless of their underlying structural differences.

\mypara{Attack with Distribution Shift}
In realistic attack scenarios, the adversary's data is often not perfectly aligned with the target's data. 
We follow prior work~\cite{sp22lira,ccs24rapid} and conduct the following experiments to evaluate \mymethod in such cases:
\begin{itemize}
    \item $\mathbb{D}_\text{target} = \mathbb{D}_\text{attack}$. 
    Both the target and imitative models are trained using disjoint subsets of the CIFAR-10 dataset. 
    This follows the non-adaptive setting in our main experiments.
    \item $\mathbb{D}_\text{target} \neq \mathbb{D}_\text{attack}$. 
    The target model is trained on CIFAR-10, while the imitative models use the ImageNet~\cite{cvpr09imagenet} portion of the CINIC-10~\cite{arxiv18cinic}, creating a distribution shift between the target’s data and the adversary’s data.
\end{itemize}

The results in~\Cref{fig:impact_shift} show that this distribution shift leads to a noticeable performance decrease, particularly for TPR at 0.1\% FPR.
This can be attributed to compounded approximation errors: when the adversary's data differs significantly from the target's, the imitative models and proxies become less effective at capturing membership-related behaviors of the target model.
Nonetheless, even in this challenging setting, \mymethod maintains strong performance on balanced accuracy.



\begin{table}[t]
\centering
\caption{Comparison of attack performance (\ie TPR @ 0\% FPR) between LiRA and \mymethod using CMIA~\cite{ndss26cpmia}. The runtime is measured (in hours) with 9 cascading iterations of both LiRA and \mymethod using an A100 GPU.}
\label{tab:cascading_imia}
\resizebox{0.49\textwidth}{!}{
\begin{tabular}{c|cc|cc|cc|cc}
\toprule
\multirow{2}{*}{\begin{tabular}{c}\# Cascading \\ iterations\end{tabular}} & \multicolumn{2}{c|}{MNIST} & \multicolumn{2}{c|}{FMNIST} & \multicolumn{2}{c|}{CIFAR-10} & \multicolumn{2}{c}{CIFAR-100} \\
\cmidrule(lr){2-3} \cmidrule(lr){4-5} \cmidrule(lr){6-7} \cmidrule(lr){8-9}
 & LiRA & \mymethod & LiRA & \mymethod & LiRA & \mymethod & LiRA & \mymethod \\
\midrule
1 & 0.80\% & 1.33\% & 3.85\% & 4.62\% & 1.30\% & 2.33\% & 5.41\% & 8.52\% \\
3 & 1.05\% & 1.42\% & 4.29\% & 4.86\% & 2.05\% & 2.55\% & 7.49\% & 10.65\% \\
6 & 1.46\% & 1.53\% & 4.72\% & 4.92\% & 2.26\% & 2.59\% & 9.30\% & 11.42\% \\
9 & 1.53\% & 1.65\% & 4.90\% & 4.98\% & 2.53\% & 2.60\% & 10.62\% & 11.45\% \\
\midrule
Runtime (hrs) & 128.8 & 5.8 & 129.8 & 5.9 & 1281.2 & 60.5 & 2562.3 & 120.9 \\
\bottomrule
\end{tabular}}
\end{table}

\subsection{Incorporating with CMIA}
CMIA~\cite{ndss26cpmia} is a recently proposed attack-agnostic framework that enhances the performance of shadow-based MIAs by iteratively training models on carefully selected datasets.
We incorporate this framework with \mymethod and compare with LiRA.
The result in~\Cref{tab:cascading_imia} shows that the performance improves with more cascading iterations, particularly during the initial few stages, which is consistent with observations from the original paper.
We observe that CMIA provides a much larger boost to LiRA than to \mymethod; after 9 iterations, LiRA eventually reaches performance levels comparable to \mymethod.
We think this is because CMIA trains more reliable shadow models by constructing datasets that resemble the target's training data.
This is similar to the goal of imitative training, where we aim to train target-informed imitative models.
As a result, CMIA’s impact is more noticeable for target-agnostic attacks like LiRA.
However, as shown in the table, applying CMIA with LiRA incurs significant computational overhead (over $20\times$ higher than \mymethod) and is only feasible in the adaptive setting, which limits its practicality.




\subsection{Impact of Relaxing Assumptions}
\label{appendix:non_assumpmtion}

Our experiments adopt the standard assumptions used in the MIA literature. 
Here, we examine the impact of relaxing these assumptions to demonstrate the robustness of \mymethod.

\mypara{Equal-size Assumption} 
While the standard evaluation constructs the query set with an equal number of members and non-members, our attack, like existing MIAs, does not rely on this assumption. 
First, the training of imitative models is independent of the query set's composition; the adversary randomly selects instances to train imitative \textit{out} models and selects pivot instances to train imitative \textit{in} models regardless of the query set distribution. 
Second, our primary evaluation metric, TPR at 0\% FPR, measures the highest true positive rate while permitting no false positives.
This metric remains valid regardless of how the query set is balanced.

\begin{table}[t]
\centering
\caption{Attack performance under different overlap ratios between the adversary’s dataset and the query set.}
\label{tab:disjoint}
\resizebox{0.47\textwidth}{!}{
\begin{tabular}{c|cc|cc|cc}
\toprule
\multirow{2}{*}{\begin{tabular}{c}\# Overlap \\ ratio\end{tabular}} & \multicolumn{2}{c|}{\textbf{TPR @ 0\%FPR}} & \multicolumn{2}{c|}{\textbf{TPR @ 0.1\%FPR}} & \multicolumn{2}{c}{\textbf{Balanced Accuracy}} \\
 & MNIST & C-10 & MNIST & C-10 & MNIST & C-10 \\
 \midrule
0 & 1.01\% & 1.45\% & 1.86\% & 3.42\% & 54.14\% & 61.08\% \\ 
1/10 & 1.04\% & 1.44\% & 1.79\% & 3.46\% & 54.12\% & 61.04\% \\ 
1/5 & 1.03\% & 1.43\% & 1.84\% & 3.48\% & 54.18\% & 61.10\% \\ 
1/3 & 1.08\% & 1.57\% & 1.85\% & 3.52\% & 54.15\% & 61.20\% \\ 
\bottomrule
\end{tabular}}
\end{table}

\mypara{Disjoint Assumption}
In the non-adaptive setting, prior work~\cite{sp22lira, ndss26cpmia} assumes that the adversary’s dataset and the query set are disjoint, \ie $D_\text{adv}^{\text{non-adapt}} \cap D_\text{query} =\emptyset$. 
This assumption may not hold in realistic scenarios, where high-probability instances can appear in both datasets.
\mymethod handles this overlap straightforwardly. 
If a query instance $(x,y)$ appears in the adversary's dataset $D_\text{adv}^{\text{non-adapt}}$, we simply discard the imitative \textit{out} models trained with $(x,y)$ and use only those where $(x,y)$ was held out.
In expectation, this requires twice as many imitative models to maintain the same number of imitative \textit{out} models per query instance.

We use this approach to evaluate our attack under varying degrees of overlap between $D_\text{adv}^{\text{non-adapt}}$ and $D_\text{query}$. 
The results in \Cref{tab:disjoint} show that attack performance remains stable across different overlap ratios. 
We also note that this strategy is highly conservative: when substantial overlap exists, one could leverage both the imitative \textit{in} and imitative \textit{out} behaviors for overlapped instances (same as the adaptive version of \mymethod) to achieve even stronger performance.
Overall, these experiments demonstrate that our attack does not depend on the disjoint assumption and continues to perform reliably when it is relaxed.

\subsection{Experiments on Other Models}
\label{appendix:imia_other_models}
We extend our evaluation to three model architectures: VGG16, DenseNet121, and MobileNetV2.
For the non-adaptive setting, the detailed results are provided in~\Cref{tab:imia_nonadapt_vgg,tab:imia_nonadapt_densenet,tab:imia_nonadapt_mobilenet}, with corresponding ROC curves shown in~\Cref{fig:imia_curve_resnet}.
Similarly, for the adaptive setting, the results are reported in~\Cref{tab:imia_adapt_vgg,tab:imia_adapt_densenet,tab:imia_adapt_mobilenet}.

\begin{table*}[t]
\centering
\caption{Performance comparison of \textit{non-adaptive} attacks on VGG16 across four image datasets.
}
\label{tab:imia_nonadapt_vgg}
\resizebox{0.98\textwidth}{!}{
\begin{tabular}{l|*{12}{c}}
\toprule
\multirow{2}{*}{\textbf{Method}} & \multicolumn{4}{c}{\textbf{TPR @ 0\% FPR}} & \multicolumn{4}{c}{\textbf{TPR @ 0.1\% FPR}} & \multicolumn{4}{c}{\textbf{Balanced Accuracy}} \\
\cmidrule(lr){
2-5}\cmidrule(lr){6-9}\cmidrule(lr){10-13}
&  MNIST & FMNIST & C-10 & C-100 & MNIST & FMNIST & C-10 & C-100 & MNIST & FMNIST & C-10 & C-100 \\
\midrule
LOSS & 0.00\% & 0.00\% & 0.00\% & 0.00\% & 0.00\% & 0.00\% & 0.03\% & 0.14\% & 52.06\% & 60.14\% & \underline{67.38\%} & 88.94\% \\
Entropy & 0.00\% & 0.00\% & 0.00\% & 0.02\% & 0.00\% & 0.00\% & 0.02\% & 0.19\% & 52.03\% & 60.08\% & 67.07\% & 89.01\% \\
Calibration & 0.01\% & 0.00\% & 0.91\% & 1.10\% & 0.35\% & 0.10\% & 2.06\% & 5.15\% & 51.20\% & 54.57\% & 60.13\% & 71.17\% \\
Attack-R & 0.00\% & 0.00\% & 0.00\% & 0.00\% & 0.00\% & 0.00\% & 0.00\% & 0.00\% & 52.01\% & 57.64\% & 66.90\% & 83.45\% \\
Attack-D & 0.00\% & 0.00\% & 0.00\% & 0.00\% & 0.00\% & 0.00\% & 0.00\% & 0.00\% & 52.42\% & 58.14\% & 65.64\% & 86.86\% \\
SeqMIA & 0.00\% & 0.00\% & 0.14\% & 0.47\% & 0.13\% & 0.06\% & 1.83\% & 4.78\% & 51.39\% & 57.78\% & 62.85\% & 77.02\%  \\
LiRA & 0.00\% & 0.00\% & 1.31\% & 0.26\% & 0.06\% & 0.09\% & 2.33\% & 6.77\% & 50.06\% & 52.01\% & 57.59\% & 74.90\% \\
Canary & 0.00\% & 0.00\% & 1.35\% & 0.30\% & 0.05\% & 0.08\% & 2.31\% & 6.72\% & 50.01\% & 52.03\% & 57.86\% & 74.98\% \\
GLiRA & 0.03\% & 0.00\% & 1.10\% & 0.56\% & 0.14\% & 0.11\% & 2.50\% & 5.71\% & 50.05\% & 58.93\% & 59.95\% & 80.09\% \\
RMIA & 0.03\% & 0.02\% & 1.31\% & 1.14\% & 0.38\% & 0.10\% & 2.93\% & 6.00\% & 52.82\% & 58.37\% & 63.16\% & 78.89\% \\
RAPID & 0.00\% & 0.03\% & 0.96\% & 1.24\% & 0.37\% & 0.08\% & 2.04\% & 5.52\% & \underline{52.91\%} & 58.25\% & 64.04\% & 79.33\% \\
PMIA & \underline{0.05\%} & \underline{0.04\%} & \underline{1.44\%} & \underline{2.55\%} & \underline{0.39\%} & \underline{0.12\%} & \underline{5.01\%} & \underline{12.20\%} & 52.07\% & \underline{60.51\%} & 67.16\% & \underline{89.03\%} \\
\midrule
\mymethod & \textbf{0.39\%} & \textbf{0.34\%} & \textbf{2.04\%} & \textbf{5.18\%} & \textbf{1.40\%} & \textbf{3.80\%} & \textbf{6.13\%} & \textbf{19.05\%} & \textbf{53.16\%} & \textbf{60.91\%} & \textbf{67.89\%} & \textbf{89.97\%} \\
\cellcolor[gray]{0.9}\%Imp.  & \cellcolor[gray]{0.9}680.00\% & \cellcolor[gray]{0.9}750.00\% & \cellcolor[gray]{0.9}41.67\% & \cellcolor[gray]{0.9}103.14\% & \cellcolor[gray]
{0.9}258.97\% & \cellcolor[gray]{0.9}3066.67\% & \cellcolor[gray]{0.9}22.36\% & \cellcolor[gray]{0.9}56.15\% & \cellcolor[gray]{0.9}0.47\% & \cellcolor[gray]{0.9}0.66\% & \cellcolor[gray]{0.9}0.76\% & \cellcolor[gray]{0.9}1.06\%  \\
\bottomrule
\end{tabular}}
\end{table*}
\begin{table*}[t]
\centering
\caption{Performance comparison of \textit{non-adaptive} attacks on DenseNet121 across four image datasets.
}
\label{tab:imia_nonadapt_densenet}
\resizebox{0.98\textwidth}{!}{
\begin{tabular}{l|*{12}{c}}
\toprule
\multirow{2}{*}{\textbf{Method}} & \multicolumn{4}{c}{\textbf{TPR @ 0\% FPR}} & \multicolumn{4}{c}{\textbf{TPR @ 0.1\% FPR}} & \multicolumn{4}{c}{\textbf{Balanced Accuracy}} \\
\cmidrule(lr){
2-5}\cmidrule(lr){6-9}\cmidrule(lr){10-13}
&  MNIST & FMNIST & C-10 & C-100 & MNIST & FMNIST & C-10 & C-100 & MNIST & FMNIST & C-10 & C-100 \\
\midrule
LOSS & 0.00\% & 0.00\% & 0.00\% & 0.04\% & 0.02\% & 0.03\% & 0.00\% & 0.10\% & 52.27\% & 60.05\% & 68.31\% & 87.90\% \\
Entropy & 0.00\% & 0.00\% & 0.00\% & 0.00\% & 0.03\% & 0.03\% & 0.00\% & 0.09\% & 52.27\% & 60.11\% & 68.04\% & 88.33\% \\
Calibration & 0.05\% & 0.01\% & 0.31\% & 1.74\% & 0.29\% & 0.97\% & 2.31\% & 3.99\% & 52.42\% & 55.27\% & 60.09\% & 69.45\% \\
Attack-R & 0.00\% & 0.00\% & 0.00\% & 0.00\% & 0.00\% & 0.00\% & 0.00\% & 0.00\% & 52.36\% & 58.08\% & 66.74\% & 85.89\% \\
Attack-D & 0.00\% & 0.00\% & 0.00\% & 0.00\% & 0.00\% & 0.00\% & 0.00\% & 0.00\% & 52.57\% & 58.07\% & 67.36\% & 83.91\% \\
SeqMIA & 0.01\% & 0.00\% & 0.29\% & 1.26\% & 0.02\% & 0.04\% & 1.04\% & 2.75\% & 52.28\% & 59.68\% & 63.86\% & 82.26\%  \\
LiRA & 0.09\% & 0.08\% & 0.81\% & 3.21\% & 0.33\% & 1.02\% & 2.69\% & 15.23\% & 50.49\% & 52.24\% & 59.38\% & 78.81\% \\
Canary & 0.08\% & 0.04\% & 0.79\% & 3.36\% & 0.27\% & 0.98\% & 2.50\% & 14.79\% & 51.68\% & 52.74\% & 59.82\% & 79.18\% \\
GLiRA & 0.02\% & 0.09\% & 0.39\% & 1.53\% & 0.25\% & 0.52\% & 4.45\% & 9.86\% & 50.51\% & 54.05\% & 62.34\% & 82.84\% \\
RMIA & 0.00\% & 0.00\% & 0.45\% & 1.78\% & 0.36\% & 1.23\% & 2.51\% & 4.39\% & 52.29\% & 58.09\% & 63.15\% & 74.58\% \\
RAPID & 0.02\% & 0.03\% & 0.37\% & 0.96\% & 0.29\% & 0.95\% & 2.18\% & 4.28\% & 51.37\% & 57.67\% & 62.19\% & 72.20\% \\
PMIA & \underline{0.23\%} & \underline{0.11\%} & \underline{1.44\%} & \underline{6.03\%} & \underline{0.38\%} & \underline{2.31\%} & \underline{5.37\%} & \underline{18.78\%} & \underline{52.59\%} & \underline{60.14\%} & \underline{68.52\%} & \underline{88.61\%} \\
\midrule
\mymethod & \textbf{1.01\%} & \textbf{1.93\%} & \textbf{4.51\%} & \textbf{6.59\%} & \textbf{1.27\%} & \textbf{7.67\%} & \textbf{8.42\%} & \textbf{21.56\%} & \textbf{52.62\%} & \textbf{60.46\%} & \textbf{70.85\%} &  \textbf{88.93\%}\\
\cellcolor[gray]{0.9}\%Imp.  & \cellcolor[gray]{0.9}339.13\% & \cellcolor[gray]{0.9}1654.55\% & \cellcolor[gray]{0.9}213.19\% & \cellcolor[gray]{0.9}9.29\% & \cellcolor[gray]
{0.9}234.21\% & \cellcolor[gray]{0.9}232.03\% & \cellcolor[gray]{0.9}56.79\% & \cellcolor[gray]{0.9}14.80\% & \cellcolor[gray]{0.9}0.06\% & \cellcolor[gray]{0.9}0.53\% & \cellcolor[gray]{0.9}3.40\% & \cellcolor[gray]{0.9}0.36\%  \\
\bottomrule
\end{tabular}}
\end{table*}
\begin{table*}[t]
\centering
\caption{Performance comparison of \textit{non-adaptive} attacks on MobileNetV2 across four image datasets.
}
\label{tab:imia_nonadapt_mobilenet}
\resizebox{0.98\textwidth}{!}{
\begin{tabular}{l|*{12}{c}}
\toprule
\multirow{2}{*}{\textbf{Method}} & \multicolumn{4}{c}{\textbf{TPR @ 0\% FPR}} & \multicolumn{4}{c}{\textbf{TPR @ 0.1\% FPR}} & \multicolumn{4}{c}{\textbf{Balanced Accuracy}} \\
\cmidrule(lr){
2-5}\cmidrule(lr){6-9}\cmidrule(lr){10-13}
&  MNIST & FMNIST & C-10 & C-100 & MNIST & FMNIST & C-10 & C-100 & MNIST & FMNIST & C-10 & C-100 \\
\midrule
LOSS & 0.00\% & 0.00\% & 0.00\% & 0.01\% & 0.01\% & 0.02\% & 0.00\% & 0.10\% & 55.80\% & 60.92\% & 70.01\% & 89.35\% \\
Entropy & 0.00\% & 0.00\% & 0.00\% & 0.02\% & 0.01\% & 0.02\% & 0.00\% & 0.10\% & 55.82\% & 60.50\% & 70.46\% & 89.01\% \\
Calibration & 0.15\% & 0.00\% & 0.44\% & 1.51\% & 0.39\% & 1.01\% & 2.28\% & 6.53\% & 54.03\% & 55.92\% & 60.68\% & 70.28\% \\
Attack-R & 0.00\% & 0.00\% & 0.00\% & 0.00\% & 0.00\% & 0.00\% & 0.00\% & 0.00\% & 53.19\% & 58.41\% & 67.32\% & 85.31\% \\
Attack-D & 0.00\% & 0.00\% & 0.00\% & 0.00\% & 0.00\% & 0.00\% & 0.00\% & 0.00\% & 53.84\% & 59.36\% & 67.12\% & 86.98\% \\
SeqMIA & 0.02\% & 0.00\% & 0.08\% & 0.42\% & 0.08\% & 0.12\% & 0.94\% & 1.47\% & 53.12\% & 55.49\% & 68.38\% & 75.40\%  \\
LiRA & 0.04\% & 0.01\% & 0.10\% & 3.17\% & 0.14\% & 0.77\% & 3.01\% & 15.34\% & 50.17\% & 52.32\% & 58.91\% & 80.61\% \\
Canary & 0.03\% & 0.01\% & 0.12\% & 3.20\% & 0.11\% & 0.72\% & 3.12\% & 15.42\% & 51.02\% & 52.39\% & 59.32\% & 81.95\% \\
GLiRA & 0.00\% & 0.11\% & 0.45\% & 0.20\% & 0.07\% & 0.13\% & 3.27\% & 7.06\% & 50.08\% & 54.43\% & 63.61\% & 84.47\% \\
RMIA & 0.11\% & 0.00\% & 0.40\% & 1.89\% & 0.40\% & 1.12\% & 2.79\% & 7.83\% & 55.76\% & 59.30\% & 65.11\% & 74.76\% \\
RAPID & 0.06\% & 0.00\% & 0.23\% & 1.73\% & 0.21\% & 0.99\% & 2.05\% & 6.94\% & 55.94\% & 60.41\% &70.02\% & 87.63\% \\
PMIA & \underline{0.23\%} & \underline{0.15\%} & \underline{0.46\%} & \underline{4.38\%} & \underline{0.96\%} & \underline{2.67\%} & \underline{4.97\%} & \underline{19.42\%} & \underline{56.31\%} & \underline{61.02\%} & \underline{70.62\%} & \underline{90.02\%} \\
\midrule
\mymethod & \textbf{1.21\%} & \textbf{2.80\%} & \textbf{1.97\%} & \textbf{6.34\%} & \textbf{2.41\%} & \textbf{5.00\%} & \textbf{6.48\%} & \textbf{23.57\%} & \textbf{56.55\%} & \textbf{61.15\%} & \textbf{72.27\%} & \textbf{90.19\%} \\
\cellcolor[gray]{0.9}\%Imp.  & \cellcolor[gray]{0.9}426.09\% & \cellcolor[gray]{0.9}1766.67\% & \cellcolor[gray]{0.9}328.46\% & \cellcolor[gray]{0.9}44.75\% & \cellcolor[gray]
{0.9}151.04\% & \cellcolor[gray]{0.9}87.27\% & \cellcolor[gray]{0.9}30.38\% & \cellcolor[gray]{0.9}21.37\% & \cellcolor[gray]{0.9}0.43\% & \cellcolor[gray]{0.9}0.21\% & \cellcolor[gray]{0.9}2.34\% & \cellcolor[gray]{0.9}0.19\%  \\
\bottomrule
\end{tabular}}
\end{table*}

\begin{table*}[t]
\centering
\caption{Performance comparison of \textit{adaptive} attacks on VGG16 across four image datasets.
}
\label{tab:imia_adapt_vgg}
\resizebox{0.98\textwidth}{!}{
\begin{tabular}{l|*{12}{c}}
\toprule
\multirow{2}{*}{\textbf{Method}} & \multicolumn{4}{c}{\textbf{TPR @ 0\% FPR}} & \multicolumn{4}{c}{\textbf{TPR @ 0.1\% FPR}} & \multicolumn{4}{c}{\textbf{Balanced Accuracy}} \\
\cmidrule(lr){2-5}\cmidrule(lr){6-9}\cmidrule(lr){10-13}
&  MNIST & FMNIST & C-10 & C-100 & MNIST & FMNIST & C-10 & C-100 & MNIST & FMNIST & C-10 & C-100 \\
\midrule
Calibration & 0.03\% & 0.04\% & 0.89\% & 6.51\% & 0.07\% & 0.06\% & 3.64\% & 13.63\% & 51.85\% & 54.74\% & 59.64\% & 72.85\% \\
Attack-R & 0.00\% & 0.00\% & 0.00\% & 0.00\% & 0.00\% & 0.00\% & 0.00\% & 0.00\% & 52.41\% & 58.35\% & 68.43\% & 86.89\% \\
Attack-D & 0.00\% & 0.00\% & 0.00\% & 0.00\% & 0.00\% & 0.00\% & 0.00\% & 0.00\% & 52.69\% & 58.84\% & 65.72\% & 86.99\% \\
SeqMIA & 0.00\% & 0.01\% & 0.73\% & 2.64\% & 0.02\% & 0.07\% & 3.32\% & 8.47\% & 51.92\% & 58.91\% & 66.52\% & 73.80\%  \\
LiRA & 0.00\% & 0.00\% & 3.25\% & 17.74\% & 0.08\% & 0.29\% & 9.92\% & \underline{43.30\%} & \underline{53.71\%} & 61.03\% & 69.98\% & \underline{92.13\%} \\
Canary & 0.01\% & 0.01\% &  3.27\% & 17.04\% & 0.07\% & 0.26\% & \underline{9.96\%} & 42.96\% & 53.70\% & \underline{61.05\%} & \underline{69.99\%} & 92.08\%\\
RMIA & \underline{0.05\%} & \underline{0.16\%} & \underline{4.03\%} & \underline{18.93\%} & \underline{0.13\%} & \underline{1.25\%} & 9.01\% & 35.21\% & 53.04\% & 60.92\% & 68.06\% & 89.04\% \\
RAPID & 0.04\% & 0.02\% & 3.01\% & 13.74\% & 0.09\% & 0.19\% & 9.23\% & 30.46\% & 53.11\% & 61.04\% & 68.82\% & 88.25\% \\
\midrule
\mymethod & \textbf{0.09\%} & \textbf{0.72\%} & \textbf{5.08\%} & \textbf{25.28\%} & \textbf{1.56\%} & \textbf{4.28\%} & \textbf{10.57\%} & \textbf{47.54\%} & \textbf{53.86\%} & \textbf{61.16\%} & \textbf{70.31\%} & \textbf{92.53\%} \\
\cellcolor[gray]{0.9}\%Imp.  & \cellcolor[gray]{0.9}80.00\% & \cellcolor[gray]{0.9}350.00\% & \cellcolor[gray]{0.9}26.05\% & \cellcolor[gray]{0.9}33.54\% & \cellcolor[gray]{0.9}1100.00\% & \cellcolor[gray]{0.9}242.40\% & \cellcolor[gray]{0.9}6.12\% & \cellcolor[gray]{0.9}9.79\% & \cellcolor[gray]{0.9}0.28\% & \cellcolor[gray]{0.9}0.18\% & \cellcolor[gray]{0.9}0.46\% & \cellcolor[gray]{0.9}0.43\%  \\
\bottomrule
\end{tabular}}
\end{table*}
\begin{table*}[t]
\centering
\caption{Performance comparison of \textit{adaptive} attacks on DenseNet121 across four image datasets.
}
\label{tab:imia_adapt_densenet}
\resizebox{0.98\textwidth}{!}{
\begin{tabular}{l|*{12}{c}}
\toprule
\multirow{2}{*}{\textbf{Method}} & \multicolumn{4}{c}{\textbf{TPR @ 0\% FPR}} & \multicolumn{4}{c}{\textbf{TPR @ 0.1\% FPR}} & \multicolumn{4}{c}{\textbf{Balanced Accuracy}} \\
\cmidrule(lr){
2-5}\cmidrule(lr){6-9}\cmidrule(lr){10-13}
&  MNIST & FMNIST & C-10 & C-100 & MNIST & FMNIST & C-10 & C-100 & MNIST & FMNIST & C-10 & C-100 \\
\midrule
Calibration & 0.21\% & 0.80\% & 0.81\% & 5.51\% & 0.71\% & 2.08\% & 4.99\% & 14.09\% & 52.18\% & 54.55\% & 59.26\% & 71.26\% \\
Attack-R & 0.00\% & 0.00\% & 0.00\% & 0.00\% & 0.00\% & 0.00\% & 0.00\% & 0.00\% & 52.14\% & 58.79\% & 68.90\% & 86.89\% \\
Attack-D & 0.00\% & 0.00\% & 0.00\% & 0.00\% & 0.00\% & 0.00\% & 0.00\% & 0.00\% & 52.44\% & 58.26\% & 67.04\% & 86.37\% \\
SeqMIA & 0.05\% & 0.23\% & 0.45\% & 3.59\% & 0.18\% & 1.70\% & 3.71\% & 6.34\% & 51.33\% & 55.42\% & 60.17\% & 77.62\%  \\
LiRA & \underline{1.08\%} & \underline{2.09\%} & 5.73\% & \underline{21.49\%} & \underline{2.17\%} & \underline{8.46\%} & 13.03\% & \underline{47.96\%} & \underline{53.21\%} & 61.05\% & \underline{72.36\%} & \underline{92.18\%} \\
Canary & 1.05\% & 1.96\% & \underline{5.76\%} & 20.64\% & 2.09\% & 8.40\% & \underline{13.22\%} & 46.61\% & 53.20\% & \underline{61.15\%} & 71.53\% & 91.98\% \\
RMIA & 0.41\% & 1.94\% & 5.09\% & 20.96\% & 1.31\% & 4.07\%  & 9.73\% & 33.07\% & 52.07\% & 60.79\% & 69.65\% & 88.25\% \\
RAPID & 0.46\% & 1.86\% & 4.27\% & 18.60\% & 1.03\% & 4.71\% & 10.63\% & 35.69\% & 52.03\% & 60.36\% & 70.41\% & 89.31\% \\
\midrule
\mymethod & \textbf{1.28\%} & \textbf{2.16\%} & \textbf{6.00\%} & \textbf{23.39\%} & \textbf{2.48\%} & \textbf{8.60\%} & \textbf{15.89\%} & \textbf{49.25\%} & \textbf{53.25\%} & \textbf{61.69\%} & \textbf{72.69\%} & \textbf{92.68\%} \\
\cellcolor[gray]{0.9}\%Imp.  & \cellcolor[gray]{0.9}18.52\% & \cellcolor[gray]{0.9}3.35\% & \cellcolor[gray]{0.9}4.17\% & \cellcolor[gray]{0.9}8.84\% & \cellcolor[gray]
{0.9}14.29\% & \cellcolor[gray]{0.9}1.65\% & \cellcolor[gray]{0.9}20.20\% & \cellcolor[gray]{0.9}2.69\% & \cellcolor[gray]{0.9}0.08\% & \cellcolor[gray]{0.9}0.88\% & \cellcolor[gray]{0.9}0.46\% & \cellcolor[gray]{0.9}0.54\%  \\
\bottomrule
\end{tabular}}
\end{table*}
\begin{table*}[t]
\centering
\caption{Performance comparison of \textit{adaptive} attacks on MobileNetV2 across four image datasets.
}
\label{tab:imia_adapt_mobilenet}
\resizebox{0.98\textwidth}{!}{
\begin{tabular}{l|*{12}{c}}
\toprule
\multirow{2}{*}{\textbf{Method}} & \multicolumn{4}{c}{\textbf{TPR @ 0\% FPR}} & \multicolumn{4}{c}{\textbf{TPR @ 0.1\% FPR}} & \multicolumn{4}{c}{\textbf{Balanced Accuracy}} \\
\cmidrule(lr){
2-5}\cmidrule(lr){6-9}\cmidrule(lr){10-13}
&  MNIST & FMNIST & C-10 & C-100 & MNIST & FMNIST & C-10 & C-100 & MNIST & FMNIST & C-10 & C-100 \\
\midrule
Calibration & 0.12\% & 0.47\% & 0.55\% & 9.52\% & 0.65\% & 1.49\% & 3.52\% & 16.74\% & 54.26\% & 55.82\% & 59.31\% & 71.45\% \\
Attack-R & 0.00\% & 0.00\% & 0.00\% & 0.00\% & 0.00\% & 0.00\% & 0.00\% & 0.00\% & 54.04\% & 58.75\% & 69.03\% & 87.87\% \\
Attack-D & 0.00\% & 0.00\% & 0.00\% & 0.00\% & 0.00\% & 0.00\% & 0.00\% & 0.00\% & 54.07\% & 59.03\% & 67.05\% & 87.62\% \\
SeqMIA & 0.06\% & 0.15\% & 0.26\% & 5.18\% & 0.22\% & 0.43\% & 0.63\% & 10.28\% & 54.72\% & 57.37\% & 59.62\% & 77.38\%  \\
LiRA & \underline{0.86\%} & \underline{2.51\%} & 4.23\% & \underline{31.19\%} & \underline{2.24\%} & \underline{7.61\%} & \underline{11.12\%} & \underline{50.17\%} & 57.10\% & \underline{62.47\%} & 73.01\% & \underline{92.69\%} \\
Canary & 0.79\% & 2.50\% & \underline{4.32\%} & 30.03\% & 2.06\% & 7.55\% & 10.59\% & 44.29\% & \underline{57.55\%} & 62.34\% & \underline{73.05\%} & 91.08\% \\
RMIA & 0.14\% & 1.86\% & 3.05\% & 27.08\% & 1.36\% & 3.53\% &10.92\% & 39.71\% & 57.08\% & 62.42\% & 70.07\% & 89.36\% \\
RAPID & 0.09\% & 1.05\% & 2.95\% & 20.53\% & 0.97\% & 2.06\% & 9.51\% & 25.02\% & 56.12\% & 61.63\% & 71.51\% & 90.60\% \\
\midrule
\mymethod & \textbf{0.98\%} & \textbf{2.89\%} & \textbf{4.98\%} & \textbf{34.46\%} & \textbf{2.94\%} & \textbf{7.89\%} & \textbf{13.92\%} & \textbf{54.55\%} & \textbf{57.94\%} & \textbf{62.97\%} & \textbf{73.92\%} & \textbf{93.15\%} \\
\cellcolor[gray]{0.9}\%Imp.  & \cellcolor[gray]{0.9}13.95\% & \cellcolor[gray]{0.9}15.14\% & \cellcolor[gray]{0.9}15.28\% & \cellcolor[gray]{0.9}10.48\% & \cellcolor[gray]
{0.9}31.25\% & \cellcolor[gray]{0.9}3.68\% & \cellcolor[gray]{0.9}25.18\% & \cellcolor[gray]{0.9}8.73\% & \cellcolor[gray]{0.9}0.68\% & \cellcolor[gray]{0.9}0.80\% & \cellcolor[gray]{0.9}1.19\% & \cellcolor[gray]{0.9}0.50\%  \\
\bottomrule
\end{tabular}}
\end{table*}

\end{document}